\documentclass[aps,prl,reprint,superscriptaddress,floatfix,longbibliography]{revtex4-2}

\usepackage{amsmath,amssymb,bm}
\usepackage{graphicx}
\usepackage{xcolor}
\usepackage{hyperref}
\hypersetup{hidelinks}

\newcommand{\dd}{\mathrm{d}}

\newcommand{\prlheading}[1]{\par\smallskip\noindent\textit{#1---}\nobreak\ignorespaces}

\begin{document}

\title{Sector-Resolved Winding Selection Rules for Structured-Light-Driven dc Currents}

\author{Tomohiro Tamaya}
\email{tamaya@g.ecc.u-tokyo.ac.jp}
\affiliation{Department of Nuclear Engineering and Management, Graduate School of Engineering, The University of Tokyo, Tokyo 113-8656, Japan}

\author{Kenichi L. Ishikawa}
\affiliation{Department of Nuclear Engineering and Management, Graduate School of Engineering, The University of Tokyo, Tokyo 113-8656, Japan}
\affiliation{Photon Science Center, Graduate School of Engineering, The University of Tokyo, Tokyo 113-8656, Japan}
\affiliation{Research Institute for Photon Science and Laser Technology, The University of Tokyo, Tokyo 113-0033, Japan}
\affiliation{Institute for Attosecond Laser Facility, The University of Tokyo, Tokyo 113-0033, Japan}

\date{\today}

\begin{abstract}
Structured light can generate electronic dc currents with azimuthal
winding, yet the rules governing their winding order \(m\) remain
unclear.
Here we identify sector-resolved winding selection rules using
graphene as a clean two-dimensional platform.
By decomposing the current response into local and gradient sectors at
the current-operator level, we show that the winding order is
determined not by the optical orbital angular momentum \(\ell\) alone,
but by the angular structure of the corresponding current operators
together with projection onto the azimuthal direction.
For scalar Laguerre--Gaussian beams, linear polarization yields
\(m=|\ell\pm1|\) in the local sector and
\(m=|\ell|,\ |\ell\pm2|\) in the gradient sector, whereas circular
polarization with helicity \(\sigma=\pm1\) selects
\(m=|\ell-\sigma|\) in the local sector and
\(m=|\ell+2\sigma|\) in the gradient sector.
Numerical time-evolution calculations verify these rules and further
show that helicity can select an \(m=0\) branch, producing an
azimuthally uniform circulating current whose radial profile
determines the axial magnetic field \(B_z(z)\).
Our results provide a sector-resolved organizing principle for
classifying and controlling structured-light-driven dc currents with
tailored winding structures.
\end{abstract}

\maketitle

\prlheading{Introduction}
Structured light offers spatial control over optical phase,
polarization, and intensity, enabling access to electronic responses
beyond the plane-wave limit%
~\cite{RubinszteinDunlop2017RoadmapStructuredLight,
Forbes2021StructuredLight,Litchinitser2012StructuredMatter}.
A representative example is a Laguerre--Gaussian (LG) beam, whose
azimuthal phase factor \(e^{i\ell\phi}\) carries optical orbital
angular momentum (OAM) \(\ell\)%
~\cite{Allen1992OAMLG,Yao2011OAMReview,
AndrewsBabiker2012AngularMomentumLight,
Beijersbergen1993ModeConverter}.
Such beams have been widely explored for controlling the
angular-momentum content of nonlinear optical responses, including
high-harmonic generation%
~\cite{Dholakia1996SHGOAM,Courtial1997SHGConservationOAM,
Zurch2012StrongFieldSingularBeams,Gariepy2014HHGOAM,
Gauthier2017TunableOAMHHG,Rego2019TimeVaryingOAM,
Buono2022NonlinearStructuredLight,Gunyaga2025SpatialStructureSHG}.

Beyond nonlinear optical emission, structured light can also drive
dc currents%
~\cite{Quinteiro2009TwistedLightCurrents,
Watzel2016CentrifugalPhotogalvanic,
Gunyaga2023StructuredLightPhotocurrents,
Ji2020OAMPhotocurrent}.
The azimuthal component of the induced dc current can exhibit a
well-defined winding order, giving rise to a real-space current
texture and an associated magnetic signature.
Previous studies have reported OAM-dependent dc currents and examined
contributions arising from spatial variations in optical intensity,
polarization, and phase~\cite{Gunyaga2023StructuredLightPhotocurrents,Ji2020OAMPhotocurrent,Gunyaga2026}.
However, a systematic rule connecting these optical structures to the
azimuthal winding order \(m\) of the induced dc current remains
unclear.

Graphene, whose excitation by twisted light has been studied
previously%
~\cite{Farias2013GrapheneTwistedLight}, provides a clean
two-dimensional platform for addressing this question.
Its inversion symmetry strongly constrains dc current responses under
spatially uniform optical driving%
~\cite{Mikhailov2007NonlinearGraphene,
Cheng2014ThirdOrderGraphene,Glazov2014NonlinearGraphene}.
Structured light, by contrast, introduces field gradients and
azimuthal angular factors absent in the plane-wave limit, enabling
spatially structured dc currents without requiring additional
material-specific symmetry breaking.

In this Letter, we formulate sector-resolved winding selection rules
for structured-light-driven dc currents and verify them numerically
in graphene.
We show that the winding order follows from the angular structure of
the local and gradient current operators together with the geometric
projection onto the azimuthal direction, rather than from the optical
OAM \(\ell\) alone.
For scalar LG beams, linear polarization yields
\(m=\lvert\ell\pm1\rvert\) in the local sector and
\(m=\lvert\ell\rvert,\lvert\ell\pm2\rvert\) in the gradient sector,
whereas circular polarization with helicity \(\sigma=\pm1\) selects
\(m=\lvert\ell-\sigma\rvert\) in the local sector and
\(m=\lvert\ell+2\sigma\rvert\) in the gradient sector.
Although the gradient sector contains channels of distinct microscopic
origin, these channels share a common angular structure at first order
in the field-gradient expansion and therefore obey a common
leading-order winding rule.
In particular, helicity can select an \(m=0\) branch for suitable
\((\ell,\sigma)\), producing an azimuthally uniform circulating
current whose radial profile determines the on-axis magnetic field
\(B_z(z)\).

\prlheading{Local and gradient sectors and winding observable}
We decompose the structured-light-driven dc current into local and
gradient sectors and introduce the winding observable used below.
To make this separation explicit, we write the microscopic electronic
coordinate as $\mathbf{r}=\bm{X}_0+\boldsymbol{\xi}$, where
$\bm{X}_0$ denotes the coarse-grained reference position with polar
coordinates $(R_0,\phi_0)$, and $\boldsymbol{\xi}$ denotes the microscopic
intra-orbital coordinate. We then expand the slowly varying optical
vector potential about $\bm{X}_0$, following the long-wavelength
expansion underlying multipolar light--matter coupling
\cite{Mahon2019,Cheng2017QuadrupoleGraphene}.
\begin{equation}
A_j(\bm{X}_0+\bm{\xi},t)
\simeq
A_j(\bm{X}_0,t)
+\xi_\alpha \partial_\alpha A_j(\bm{X}_0,t),
\label{eq:local_expansion}
\end{equation}
where $\alpha,j=x,y$, and repeated Cartesian indices are summed.
Here, $\partial_\alpha$ acts on the coarse-grained spatial dependence
of the structured-light field, rather than on the microscopic coordinate
$\boldsymbol{\xi}$.
The first term in Eq.~\eqref{eq:local_expansion} defines the local
sector and depends only on the local value of the optical field,
whereas the second term defines the gradient sector and contains the
first spatial gradient of the field.

This local/gradient separation is not confined to the optical-field
expansion but is carried through to the current operator. 
For an interaction Hamiltonian
\(H=\int \dd^2X\,\mathcal{H}[A,\partial A]\), where
\(\mathcal{H}\) may depend on \(A_i\) and
\(\partial_j A_i\), the corresponding current-density operator is
obtained by functional differentiation with respect to the vector
potential:
\begin{equation}
J_i
=
-c\frac{\delta H}{\delta A_i}
=
-c\left[
\frac{\partial \mathcal{H}}{\partial A_i}
-
\partial_j\!\left(
\frac{\partial \mathcal{H}}
{\partial(\partial_j A_i)}
\right)
\right],
\label{eq:current_operator}
\end{equation}
where repeated spatial indices are summed.
We classify each current contribution according to the Hamiltonian
term from which it originates: terms containing the local field
\(A_i\) but no spatial gradient of the field define the local sector,
whereas terms containing the first spatial gradient
\(\partial_j A_i\) define the gradient sector.
The sector label therefore refers to the originating Hamiltonian term,
rather than to the two individual terms in
Eq.~\eqref{eq:current_operator}.
The detailed derivation, microscopic current operators, and numerical
time-evolution scheme are given in Secs.~I and II of the Supplemental
Material~\cite{SM}.

In the numerical calculations, we solve the time evolution of the
driven electronic state at each reference position \(\bm{X}_0\) and
evaluate the expectation value of the corresponding current operator.
For the winding analysis, we evaluate the azimuthal current along a
circle of radius \(R_0\), with \(\bm{X}_0=(R_0,\phi)\).
After applying the window-normalized temporal Fourier transform defined
in the Supplemental Material~\cite{SM} to \(J_\theta(\phi;R_0,t)\), we denote the
frequency-domain current by
\(\mathcal{J}_\theta(\phi;R_0,\omega)\) and define its dc component as \(J_\theta^{\mathrm{dc}}(\phi;R_0)
\equiv
\mathcal{J}_\theta(\phi;R_0,\omega=0)\).
Here and below, dc denotes the zero-frequency component of the
pulse-induced current rather than a current that persists indefinitely
after the pulse.
The corresponding signed azimuthal Fourier coefficient is defined as
\begin{equation}
\Gamma_\mu(R_0)=R_0
\int_0^{2\pi}
J_\theta^{\mathrm{dc}}(\phi;R_0)
e^{-i\mu\phi}
\dd\phi,
\label{eq:winding_coefficient}
\end{equation}
where \(\mu\in\mathbb{Z}\) is the signed azimuthal harmonic index.
Because \(J_\theta^{\mathrm{dc}}(\phi;R_0)\) is real,
\(\Gamma_{-\mu}(R_0)=\Gamma_\mu^*(R_0)\).
We therefore define the nonnegative winding order
\(m\equiv|\mu|\) and write the corresponding winding power as
\(|\Gamma_m(R_0)|^2\).
The \(m=0\) component represents the azimuthally uniform part of the
circulating current. Its amplitude is given by the azimuthal average
\begin{equation}
J_\theta^{\mathrm{dc}}(R_0)
=
\frac{\Gamma_{m=0}(R_0)}{2\pi R_0}.
\label{eq:averaged_current}
\end{equation}
The structured-light field, numerical parameters, and extraction
procedure are described in Secs.~III A and III B of the Supplemental
Material~\cite{SM}.
Having defined the winding observable, we now derive the winding
orders allowed in the local and gradient sectors.

\begin{figure*}[t]
\centering
\includegraphics[width=\textwidth]{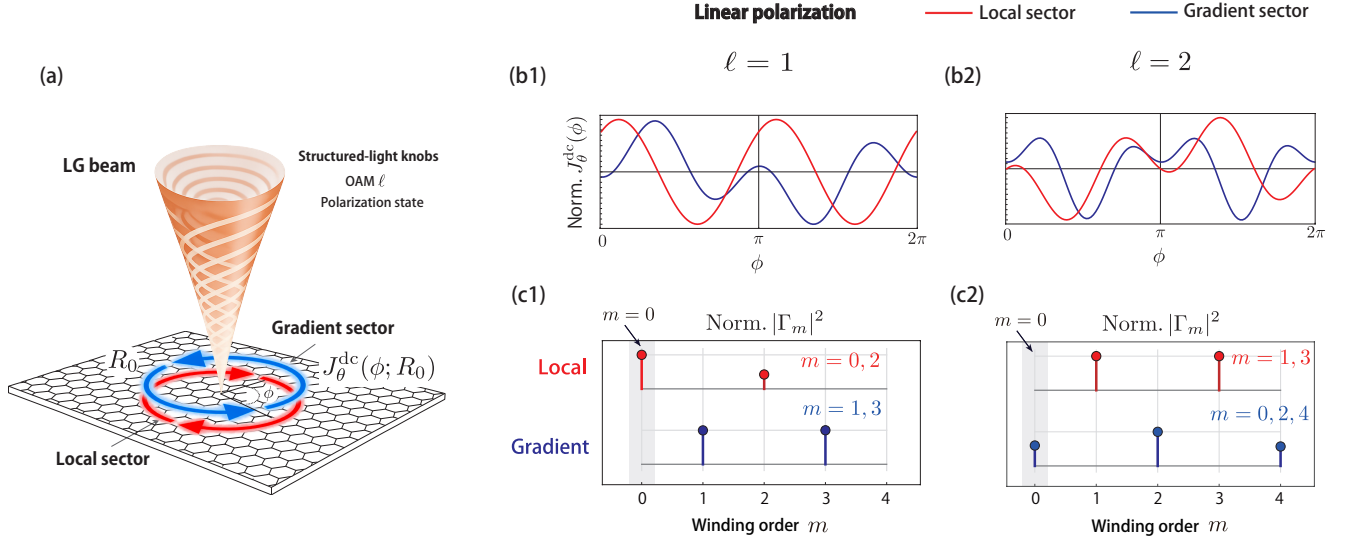}
\caption{
Sector-resolved winding diagnostic and linear-polarization selection.
(a) Schematic of the winding-spectrum diagnostic for the dc azimuthal current
$J_\theta^{\mathrm{dc}}(\phi;R_0)$ on a graphene sheet, evaluated
along an observation circle of radius $R_0$.
(b1,b2) Azimuthal dependence of $J_\theta^{\mathrm{dc}}(\phi;R_0)$
for linearly polarized scalar LG beams at $R_0=R_\ell^{\mathrm{max}}$,
where $R_\ell^{\mathrm{max}}$ is the radius of maximum LG intensity,
for $\ell=1$ and $\ell=2$.
For visualization in (b1,b2), the azimuthally uniform $m=0$ component,
i.e., the angular average of $J_\theta^{\mathrm{dc}}(\phi;R_0)$,
has been subtracted.
Red and blue curves denote local- and gradient-sector contributions, respectively.
(c1,c2) Corresponding normalized winding spectra $|\Gamma_m(R_0)|^2$,
computed from the unsubtracted $J_\theta^{\mathrm{dc}}(\phi;R_0)$,
verifying $m=|\ell\pm1|$ in the local sector and
$m=|\ell|,\ |\ell\pm2|$ in the gradient sector.
The gray region marks the $m=0$ branch.
In (b1,b2) and (c1,c2), the local- and gradient-sector results are
normalized independently within each panel; their relative heights
therefore do not represent the absolute amplitude ratio between the
two sectors.
}
\label{fig:linear_winding_selection}
\end{figure*}

\begin{figure*}[t]
\centering
\includegraphics[width=\textwidth]{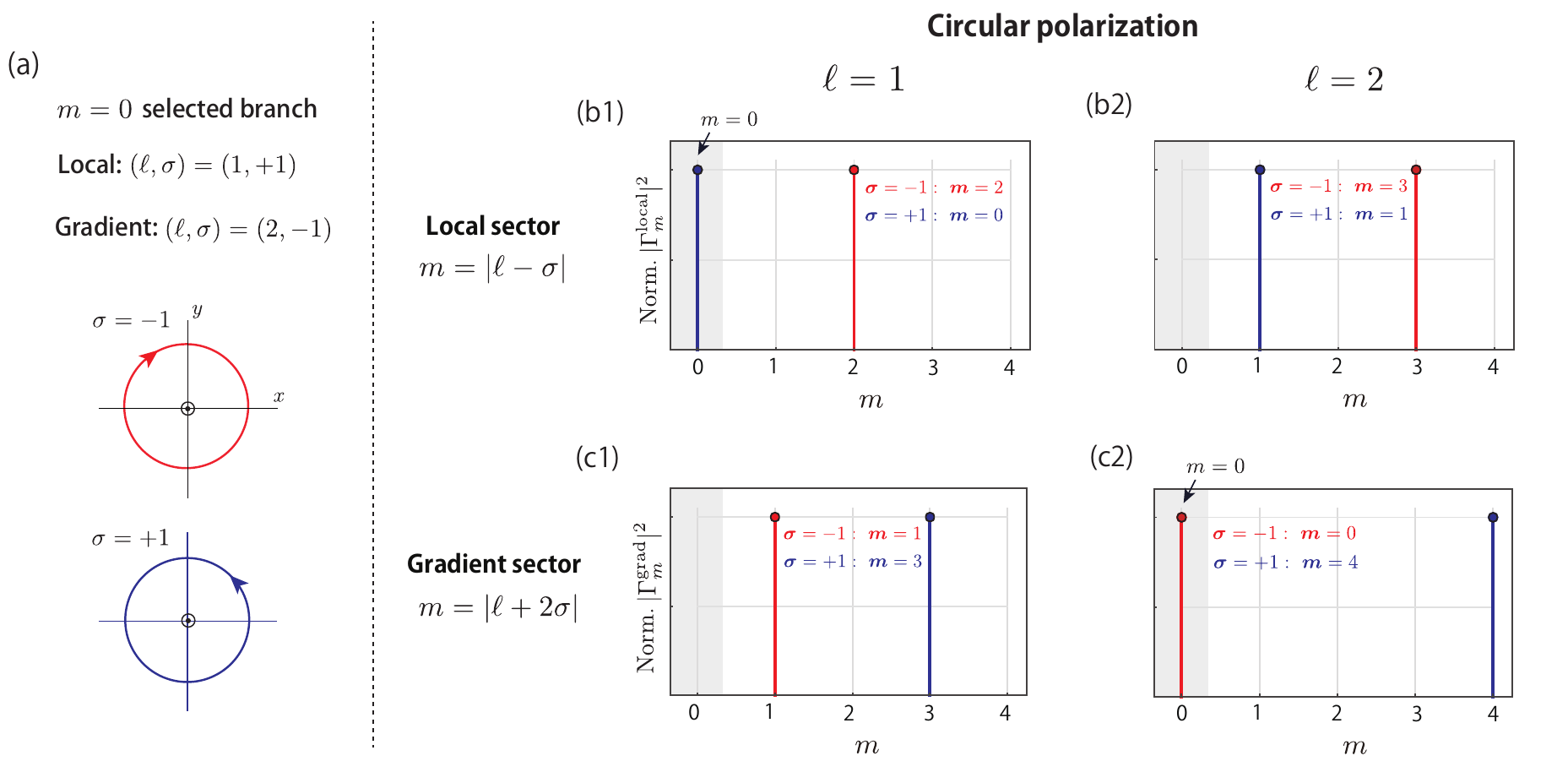}
\caption{
Helicity-controlled branch selection for circularly polarized scalar LG beams.
(a) Helicity convention and sector-dependent combinations selecting the
$m=0$ circulating-current branch.
(b1,b2) Local-sector winding spectra and (c1,c2) gradient-sector winding spectra
for $\ell=1,2$.
Red and blue sticks denote $\sigma=-1$ and $\sigma=+1$, respectively,
and the gray region marks the $m=0$ branch.
Each panel is normalized independently to its maximum value.
}
\label{fig:helicity_branch_selection}
\end{figure*}

\prlheading{Sector-resolved winding selection rules}
The observable winding order follows from a two-step angular mapping.
First, the local or gradient character of the originating Hamiltonian
term constrains the allowed signed azimuthal harmonics $N$ of the
Cartesian current, where $N$ labels the harmonics obtained from the
time-dependent electronic dynamics.
Second, projection onto the azimuthal direction shifts these harmonics
by one unit, yielding $\mu=N\pm1$.
Thus, the sector dependence enters through the allowed pre-projection
harmonics $N$, while the mapping from $N$ to $\mu$ is purely geometric.
Consequently, the winding order is determined jointly by the sector
and the optical OAM $\ell$, rather than by $\ell$ alone.

For a scalar LG beam, the vector potential can be written
schematically as
\(\bm{A}(\bm{X},t)
\propto
u_{\ell}(R)e^{i\ell\phi}\bm{e}_{\mathrm{pol}}
+\mathrm{c.c.}\),
where \(u_{\ell}(R)\) is the radial envelope and
\(\bm{e}_{\mathrm{pol}}\) is a spatially uniform polarization vector.
For the \(e^{i\ell\phi}\) component, local-sector current
contributions involve the local field but no spatial derivative and
therefore retain the harmonic \(N=\ell\).
By contrast, gradient-sector contributions involve one spatial
derivative of the field, which introduces an additional one-unit
angular shift.
Indeed,
\(\partial_X\pm i\partial_Y
=e^{\pm i\phi}
(\partial_R\pm iR^{-1}\partial_\phi)\),
showing that a Cartesian derivative shifts an azimuthal harmonic by
\(\pm1\).
The allowed pre-projection harmonics in the gradient sector are
therefore \(N=\ell\pm1\), with the complex-conjugate field supplying
the corresponding negative harmonics.

The subsequent projection onto the azimuthal direction is given by
\(J_\theta=\bm{e}_{\theta}\cdot\bm{J}\), where
\(\bm{e}_{\theta}=(-\sin\phi,\cos\phi)\).
Because the Cartesian components of \(\bm{e}_{\theta}\) are linear
combinations of \(e^{\pm i\phi}\), this projection shifts each
pre-projection harmonic by one unit, yielding \(\mu=N\pm1\).
The polarization fixes the relative combination of the Cartesian
current components and thereby determines which of the allowed signed
branches contribute to \(J_\theta\).

Linear polarization contains both helicity components with equal
weight, allowing both helicity-resolved branches to contribute.
In the local sector, the pre-projection harmonic \(N=\ell\) combines
with the projection-induced shift \(\pm1\), yielding
\(\mu=\ell\pm1\).
In the gradient sector, the derivative-induced shift \(\pm1\) combines
with the projection-induced shift \(\pm1\).
The same-sign combinations yield \(\mu=\ell\pm2\), whereas the
opposite-sign combinations yield \(\mu=\ell\).
The resulting winding selection rules are
\begin{equation}
\begin{aligned}
\mathrm{local}:&\qquad
m=|\ell\pm1|,\\
\mathrm{gradient}:&\qquad
m=|\ell|,\ |\ell\pm2|,
\end{aligned}
\label{eq:selection_linear}
\end{equation}
where \(m=|\mu|\) is the nonnegative winding order displayed in the
spectra.

For circular polarization, the driving field has a definite helicity
\(\sigma=\pm1\), which selects the corresponding helicity-resolved
component of the current.
With the convention used here, the local sector yields the signed
harmonic
\(\mu_{\mathrm{loc}}=\ell-\sigma\).
In the gradient sector, the derivative-induced and
projection-induced one-unit shifts have the same sign in the selected
helicity component, yielding
\(\mu_{\mathrm{grad}}=\ell+2\sigma\).
The resulting winding-order selection rules are
\begin{equation}
\begin{aligned}
\mathrm{local}:&\quad
m=\lvert\ell-\sigma\rvert,\\
\mathrm{gradient}:&\quad
m=\lvert\ell+2\sigma\rvert.
\end{aligned}
\label{eq:selection_circular}
\end{equation}

Equations~\eqref{eq:selection_linear} and
\eqref{eq:selection_circular} summarize the resulting sector-resolved
winding selection rules.
The distinction between the local and gradient sectors originates
from their different sets of allowed pre-projection harmonics, while
the common azimuthal projection maps these harmonics onto the
observable winding orders.
Although the gradient sector contains microscopically distinct
channels, they obey the same leading winding rule because they share
the same angular structure to first order in the field-gradient
expansion.
A detailed derivation of the angular factors and a numerical
verification of the pre-projection harmonics are given in
Sec.~III D of the Supplemental Material~\cite{SM}.

The selection rules above are derived for scalar LG beams, for which
the azimuthal dependence relevant to the winding selection is carried
by the phase factor $e^{i\ell\phi}$, while the polarization vector is
spatially uniform.
More general structured beams, particularly those with spatially
varying polarization, can introduce additional angular factors and
thereby modify the allowed pre-projection harmonics $N$.
The same organizing principle nevertheless applies: the optical angular
structure and the sector-resolved operator structure determine the
allowed $N$, which are then mapped onto the observable winding orders
by the common azimuthal projection.

We next test this organizing principle in graphene using full-lattice
time-evolution calculations.
Although the full lattice allows $C_6$-related sidebands, the leading
winding branches remain those predicted by the isotropic Dirac-cone
analysis in the parameter regime considered here.
We also verified that explicitly including the finite in-plane optical
momentum and the associated Bloch-momentum transfer does not change
the selected winding orders, with relative amplitude changes below
$0.02\%$~\cite{SM}.

\begin{figure*}[t]
\centering
\includegraphics[width=\textwidth]{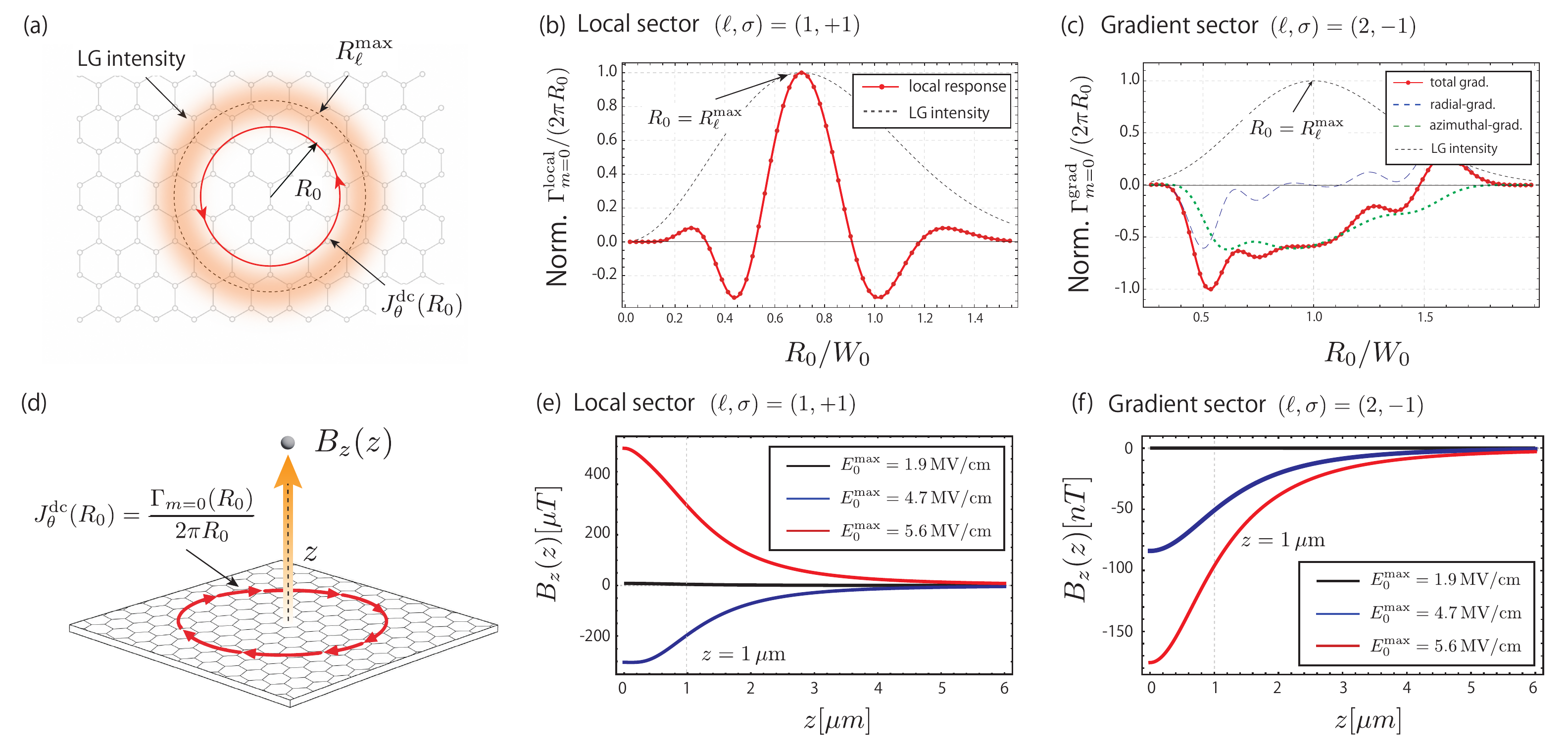}
\caption{
Real-space profiles and magnetic readout of sector-selected \(m=0\)
currents.
(a) Schematic extraction of the radial profile of the azimuthal current
\(J_\theta^{\mathrm{dc}}(R_0)=\Gamma_{m=0}(R_0)/(2\pi R_0)\)
from the selected \(m=0\) component, overlaid on the radial intensity
profile of the LG beam.
(b,c) Radial profiles of the selected \(m=0\) current for the local
branch \((\ell,\sigma)=(1,+1)\) and the gradient branch
\((\ell,\sigma)=(2,-1)\), respectively.
Dashed gray curves show the corresponding LG intensity profiles.
In (c), the gradient response is decomposed into total,
radial-gradient, and azimuthal-gradient contributions.
(d) Schematic magnetic readout of the circulating sheet current.
(e,f) Corresponding on-axis fields \(B_z(z)\), calculated using the
Biot--Savart law for different peak field amplitudes
\(E_0^{\max}\).
The fields in (e) and (f) are plotted in \(\mu\mathrm{T}\) and
\(\mathrm{nT}\), respectively, reflecting the different absolute
scales of the local and gradient sectors.
The current profiles in (b) and (c) are normalized independently
within each panel; their amplitudes therefore cannot be compared
between the local and gradient sectors.
}
\label{fig:radial_profile_m0}
\end{figure*}

The numerical calculations are performed for scalar LG beams in
Figs.~\ref{fig:linear_winding_selection} and
\ref{fig:helicity_branch_selection} at the radius of maximum LG
intensity, \(R_0=R_\ell^{\mathrm{max}}\).
At this radius, the gradient-sector response is dominated by the
azimuthal-gradient contribution, while the radial-gradient
contribution is examined separately in
Fig.~\ref{fig:radial_profile_m0}.

\prlheading{Linear-polarization winding selection}
Figure~\ref{fig:linear_winding_selection} verifies the
sector-resolved selection rules for linearly polarized scalar LG
beams. The schematic in
Fig.~\ref{fig:linear_winding_selection}(a) defines the
azimuthal-current diagnostic on an observation circle of radius
\(R_0\). Figures~\ref{fig:linear_winding_selection}(b1) and (b2) show
\(J_\theta^{\mathrm{dc}}(\phi;R_0)\) at
\(R_0=R_\ell^{\mathrm{max}}\) for \(\ell=1\) and \(\ell=2\),
respectively. The corresponding spectra in
Figs.~\ref{fig:linear_winding_selection}(c1) and (c2) exhibit the
predicted local-sector peaks at \(m=|\ell\pm1|\) and gradient-sector
peaks at \(m=|\ell|,\ |\ell\pm2|\), confirming
Eq.~\eqref{eq:selection_linear}.
Channel-resolved spectra for the local
\(\bm{A}\cdot\bm{p}\), first-gradient
\(\bm{A}\cdot\bm{p}\), and first-gradient \(A^2\) channels
are provided in Sec.~III E.1 of the Supplemental
Material~\cite{SM}.

\prlheading{Helicity-controlled branch selection}
Circular polarization fixes the helicity \(\sigma\), so
Eq.~\eqref{eq:selection_circular} selects a definite winding branch
within each sector rather than a superposition of branches.
Figure~\ref{fig:helicity_branch_selection}(a) summarizes the helicity
convention and the sector-dependent combinations that yield the
\(m=0\) circulating-current branch.
The local and gradient selection rules give \(m=0\) for
\((\ell,\sigma)=(1,+1)\) and \((\ell,\sigma)=(2,-1)\),
respectively.

The winding spectra directly verify these predictions.
Figures~\ref{fig:helicity_branch_selection}(b1) and (b2) show the
local-sector spectra for \(\ell=1\) and \(\ell=2\), respectively.
Their peaks follow the local-sector rule \(m=|\ell-\sigma|\), with the
\(m=0\) branch appearing for \((\ell,\sigma)=(1,+1)\).
Figures~\ref{fig:helicity_branch_selection}(c1) and (c2) show the
corresponding gradient-sector spectra, whose peaks instead follow
\(m=|\ell+2\sigma|\), with the \(m=0\) branch selected for
\((\ell,\sigma)=(2,-1)\).
Thus, helicity reversal changes the selected winding order in a
sector-dependent manner. The same sector-resolved rules are confirmed for the individual microscopic current channels, including the distinct gradient-sector
channels, in Sec.~III E.2 of the Supplemental Material~\cite{SM}.

\prlheading{Real-space profiles and magnetic readout of selected
$m=0$ currents}
The helicity-selected \(m=0\) branches have a direct real-space
consequence: they form azimuthally uniform circulating currents and
therefore generate an axial magnetic field.
For the selected local and gradient branches, corresponding to
\((\ell,\sigma)=(1,+1)\) and \((2,-1)\), respectively, we evaluate
\(\Gamma_{m=0}(R_0)\), obtain the averaged azimuthal current
\(J_\theta^{\mathrm{dc}}(R_0)=\Gamma_{m=0}(R_0)/(2\pi R_0)\), and calculate
the corresponding axial magnetic field using
\begin{equation}
B_z(z)=
\frac{\mu_0}{2}
\int_0^\infty \dd R_0\,
J_\theta^{\mathrm{dc}}(R_0)
\frac{R_0^2}{(R_0^2+z^2)^{3/2}}.
\label{eq:magnetic_readout}
\end{equation}
Positive \(J_\theta^{\mathrm{dc}}\) is defined along the
increasing-\(\phi\) direction, counterclockwise as viewed from \(+z\).
Accordingly, the sign of \(\Gamma_{m=0}(R_0)\) specifies the local
circulation direction, whereas the sign of \(B_z(z)\) is determined
by the radially weighted integral in
Eq.~\eqref{eq:magnetic_readout}. Details of the magnetic-field
calculation and physical-unit calibration are given in Sec.~III G
of the Supplemental Material~\cite{SM}.

Figure~\ref{fig:radial_profile_m0}(a) illustrates the extraction of
the radial profile of the azimuthal current entering the magnetic readout, while
Figures~\ref{fig:radial_profile_m0}(b) and (c) show the selected
profiles in the local and gradient sectors. The local response is concentrated near the high-intensity region of the LG beam but is not simply proportional to the intensity and can
change sign, reflecting the nonlinear field-amplitude dependence of
the zero-frequency response.
The gradient response contains both radial- and azimuthal-gradient
contributions: the sign-changing radial-gradient contribution is partially canceled in the Biot--Savart integration, leaving the broader
azimuthal-gradient contribution dominant in the magnetic signal.
The selected winding order remains robust against changes in the
observation radius. Further
analyses of the field-amplitude dependence, observation-radius
robustness, and the corresponding intensity-gradient-like profile are
provided in Secs.~III E.3--III E.5 of the Supplemental
Material~\cite{SM}.

Figure~\ref{fig:radial_profile_m0}(d) illustrates the magnetic
readout. Only the $m=0$ component contributes to the on-axis $B_z$,
whereas the $m\neq0$ contributions cancel under azimuthal averaging.
Figures~\ref{fig:radial_profile_m0}(e) and (f) show the corresponding
on-axis magnetic fields for several peak amplitudes \(E_0^{\max}\).
The magnetic fields depend nonlinearly on \(E_0^{\max}\), and the
local-sector field changes sign as \(E_0^{\max}\) is varied.
At \(E_0^{\max}=5.6\,\mathrm{MV/cm}\) and
\(z=1\,\mu\mathrm{m}\), we obtain approximately
\(B_z=280\,\mu\mathrm{T}\) and \(B_z=-120\,\mathrm{nT}\) for the
local and gradient branches, respectively.
This difference reflects the smaller gradient-sector current and the
additional spatial-gradient scale entering its response.
The conversion to physical units is calibrated against the universal
optical conductivity of graphene
\cite{Nair2008FineStructureGraphene,Mak2008OpticalConductivityGraphene,
Kuzmenko2008UniversalOpticalConductance,
Stauber2008OpticalConductivityGrapheneVisible};
the resulting absolute values should therefore be regarded as
order-of-magnitude estimates.
The winding selection rules and the conditions for selecting the
\(m=0\) branch are independent of this calibration.

Thus, the winding classification has a direct real-space
consequence: OAM- and helicity-controlled selection of
an $m=0$ branch produces an azimuthally uniform dc
circulating current and thereby generates a finite on-axis
$B_z$.

\prlheading{Conclusion}
We have formulated sector-resolved winding selection rules for
structured-light-driven dc currents and verified them using
full-lattice time-evolution calculations for graphene.
For scalar LG beams, linear polarization yields
\(m=|\ell\pm1|\) in the local sector and
\(m=|\ell|,\ |\ell\pm2|\) in the gradient sector, whereas circular
polarization selects \(m=|\ell-\sigma|\) in the local sector and
\(m=|\ell+2\sigma|\) in the gradient sector.
These rules follow from the angular structure of the local and
gradient current operators together with projection onto the
azimuthal direction, rather than from the optical OAM \(\ell\) alone.
Accordingly, gradient-sector channels with distinct microscopic
origins obey the same leading winding rule.

Helicity-controlled selection of an \(m=0\) branch produces an
azimuthally uniform circulating dc current and thereby generates a
finite on-axis magnetic field \(B_z(z)\).
Together, these results establish a sector-resolved organizing
principle for tailoring dc-current winding with optical OAM and
helicity while providing a direct magnetic-field readout of the
selected winding branch.

\begin{acknowledgments}
T.T. acknowledges funding from JST PRESTO under Grant No.~JPMJPR2107.
This research was supported by the MEXT Quantum Leap Flagship
Program (MEXT Q-LEAP) under Grant No.~JPMXS0118067246 and by JSPS
KAKENHI under Grant Nos.~JP24H00427 and JP25H00704.
This work was also partially supported by the JST K Program under
Grant No.~JPMJKP24M1 and by the RIKEN TRIP initiative
(HIKARI-COOL Tokyo).
%This work was supported by [funding information].
\end{acknowledgments}

% Add references to references.bib, or replace this block with a thebibliography environment.
\bibliographystyle{apsrev4-2}
\bibliography{references}

\end{document}

\setcounter{figure}{0}
\renewcommand{\thefigure}{S\arabic{figure}}

\title{\texorpdfstring{Supplemental Material for\\
``Sector-Resolved Winding Selection Rules for Structured-Light-Driven dc Currents''}
{Supplemental Material for Sector-Resolved Winding Selection Rules for Structured-Light-Driven dc Currents}}

\author{Tomohiro Tamaya}
\email{\TAMAYAEMAIL}
\affiliation{\NEM}

\author{Kenichi L. Ishikawa}
\affiliation{\NEM}
\affiliation{\PSC}
\affiliation{\UTripl}
\affiliation{\ALFA}

\date{\today}

\maketitle
This Supplemental Material provides further details of the theoretical
framework and numerical calculations underlying the results presented
in the main text.
In Sec.~\ref{sec:microscopic_hamiltonian_time_evolution}, we derive the
microscopic light--matter interaction Hamiltonian for graphene driven
by a spatially inhomogeneous optical field and formulate the equations
of motion used in the numerical calculations.
The main calculations employ a local long-wavelength,
Bloch-momentum-diagonal approximation, in which the residual
site-to-site phase variation of the structured-light field is neglected.
Because this phase variation corresponds to a finite local in-plane
optical wave vector $\bm{q}_{\parallel}$, we also formulate a
finite-wave-vector extension that restores the associated momentum
transfer $\pm\bm{q}_{\parallel}$ between Bloch states.
Using this framework, we assess the validity of the
momentum-diagonal approximation through a direct numerical comparison
between the two formulations.
In Sec.~\ref{sec:current_channels}, we derive the current operators and
define the sector-resolved channel decomposition.
Finally, in Sec.~\ref{sec:winding_selection_rules}, we specify the
structured-light field and numerical parameters, define the azimuthal
current and winding spectrum, derive the sector-resolved winding
selection rules, and present additional numerical results including
the finite-$\bm{q}_{\parallel}$ validation.

\section{Details of the microscopic Hamiltonian and numerical time evolution}
\label{sec:microscopic_hamiltonian_time_evolution}
In this section, we derive the microscopic light--matter Hamiltonian
and formulate the equations of motion used in the numerical
calculations presented in the main text.
We consider monolayer graphene irradiated by structured light.
Although the numerical calculations employ a Laguerre--Gaussian (LG)
beam, the formulation can be applied more generally to spatially
inhomogeneous optical fields.
Throughout this work, spin degeneracy is assumed.
Starting from the minimal-coupling Hamiltonian, we perform a
site-centered spatial-gradient expansion and project the resulting
interaction onto the graphene Bloch basis.
The local momentum-diagonal formulation used in the main calculations
is derived in
Secs.~\ref{sec:minimal_coupling}--\ref{sec:bloch_hamiltonian}.
In Secs.~\ref{sec:finite_q_hamiltonian} and
\ref{sec:finite_q_bloch_projection}, we then extend this formulation
by restoring the leading site-to-site optical phase associated with a
finite local in-plane optical wave vector
$\bm{q}_{\parallel}$, in order to test the validity of the
momentum-diagonal approximation.
The equations of motion for both the momentum-diagonal and
finite-$\bm{q}_{\parallel}$ formulations are given in
Sec.~\ref{sec:time_evolution}.
The effects of finite-$\bm{q}_{\parallel}$ coupling are examined
numerically in
Sec.~\ref{sec:finite_q_numerical_validation}.

\subsection{Minimal-coupling Hamiltonian}
\label{sec:minimal_coupling}

In this subsection, we derive the light--matter interaction Hamiltonian in a general form. 
In the absence of an external electromagnetic field, the single-particle Hamiltonian of an electron in a crystalline solid is written as
\begin{equation*}
    \hat{h}_0
    =
    \frac{\hat{\bm{p}}^{\,2}}{2m_0}
    +
    \sum_i V\!\left(\bm{r}-\bm{R}_i\right),
\end{equation*}
where \(\hat{\bm{p}}=-i\hbar\bm{\nabla}\) is the momentum operator, \(m_0\) is the bare electron mass, \(\bm{r}\) is the electron coordinate, \(\bm{R}_i\) denotes the position of the \(i\)th atom, and \(V(\bm{r}-\bm{R}_i)\) is the potential generated by the corresponding nucleus or ionic core.
Here, \(e\) is defined with its sign such that the minimal-coupling prescription takes the form\cite{CohenTannoudji1989PhotonsAtoms,Mahan2000ManyParticle}
\begin{equation*}
    \hat{\bm{p}}
    \longrightarrow
    \hat{\bm{p}}
    -
    \frac{e}{c}\bm{A}(\bm{r},t),
\end{equation*}
where \(c\) is the speed of light and \(\bm{A}(\bm{r},t)\) is the vector potential of the optical field. 
We neglect the scalar-potential contribution and retain the general spatial dependence of the vector potential. 
The Hamiltonian in the presence of the optical field is therefore given by
\begin{equation}
    \hat{h}
    =
    \frac{1}{2m_0}
    \left[
        \hat{\bm{p}}
        -
        \frac{e}{c}\bm{A}(\bm{r},t)
    \right]^2
    +
    \sum_i V\!\left(\bm{r}-\bm{R}_i\right).
    \label{eq:minimal_coupling_H}
\end{equation}
Expanding Eq.~\eqref{eq:minimal_coupling_H}, while keeping the operator ordering between \(\hat{\bm{p}}\) and the spatially dependent vector potential \(\bm{A}(\bm{r},t)\), we obtain
\begin{equation*}
    \hat{h}
    =
    \hat{h}_0
    -
    \frac{e}{2m_0c}
    \left[
        \hat{\bm{p}}\cdot\bm{A}(\bm{r},t)
        +
        \bm{A}(\bm{r},t)\cdot\hat{\bm{p}}
    \right]
    +
    \frac{e^2}{2m_0c^2}
    \bm{A}^2(\bm{r},t).
\end{equation*}
For a spatially dependent vector potential, the two terms in the square brackets are not generally identical. 
Acting on an arbitrary wave function, one has
\begin{equation*}
    \hat{\bm{p}}\cdot\bm{A}(\bm{r},t)
    =
    \bm{A}(\bm{r},t)\cdot\hat{\bm{p}}
    -
    i\hbar\,\bm{\nabla}\cdot\bm{A}(\bm{r},t).
\end{equation*}
In the Coulomb gauge\cite{CohenTannoudji1989PhotonsAtoms,Mahan2000ManyParticle,Jackson1998ClassicalElectrodynamics},
\begin{equation*}
    \bm{\nabla}\cdot\bm{A}(\bm{r},t)=0,
\end{equation*}
this relation reduces to
\begin{equation*}
    \hat{\bm{p}}\cdot\bm{A}(\bm{r},t)
    =
    \bm{A}(\bm{r},t)\cdot\hat{\bm{p}}.
\end{equation*}
The Hamiltonian can then be decomposed as
\begin{equation*}
    \hat{h}
    =
    \hat{h}_0
    +
    \hat{h}_{I}^{A}
    +
    \hat{h}_{I}^{B},
\end{equation*}
where
\begin{align*}
    \hat{h}_{I}^{A}
    &=
    -\frac{e}{m_0c}
    \bm{A}(\bm{r},t)\cdot\hat{\bm{p}},
    \\
    \hat{h}_{I}^{B}
    &=
    \frac{e^2}{2m_0c^2}
    \bm{A}^2(\bm{r},t).
\end{align*}
Here, \(\hat{h}_{I}^{A}\) and \(\hat{h}_{I}^{B}\) denote the light--matter interaction terms that are linear and quadratic in the vector potential, respectively. 
In the following subsection, we expand these interaction terms in spatial gradients of the structured-light field.

\subsection{Gradient expansion of the structured-light field}
\label{sec:gradient_expansion}

For a spatially uniform optical field, the vector potential can be
approximated as $\bm{A}(\bm{r},t)\simeq\bm{A}(t)$, and the
translational symmetry of the crystal is preserved.
The Bloch wave vector $\bm{k}$ then remains a good quantum number.
By contrast, structured light has spatially varying amplitude, phase,
and polarization, and its spatial dependence generally breaks the
global translational symmetry of the crystal.
Consequently, $\bm{k}$ is no longer an exact quantum number.
To treat this problem approximately, we adopt a local long-wavelength
description.
At each coarse-grained position $\bm{X}_0$, we define a locally
translationally invariant electronic problem by evaluating the optical
field and its spatial gradients at $\bm{X}_0$.
Within this local approximation, different Bloch wave vectors are not
mixed, so that $\bm{k}$ is treated as a good quantum number.
The spatial dependence of the optical field is thus retained through
the parameter $\bm X_0$, while the electronic states are described in
the Bloch basis.

In the present work, we assume that the structured-light field varies
sufficiently slowly over the spatial extent of the atomic orbitals to
allow a site-centered spatial-gradient expansion.
We denote the coarse-grained in-plane position by
$\bm{X}_0=(X_0,Y_0)$.
When used as the argument of the three-dimensional optical field,
$\bm{X}_0$ is understood as the point $(X_0,Y_0,0)$ in the graphene
plane.
Accordingly, $z$ derivatives are taken before evaluating the
three-dimensional field at the graphene plane, $z=0$.
The position of the $n$th lattice site is written as
\begin{equation*}
    \bm{R}_n^{\bm{X}_0}
    =
    \bm{X}_0+\bm{\rho}_n,
\end{equation*}
where $\bm{\rho}_n$ denotes the displacement of the $n$th lattice site
from the coarse-grained position $\bm{X}_0$.
The real-space coordinate of an electron in the vicinity of this site
is then written as
\begin{equation*}
    \bm{r}
    =
    \bm{R}_n^{\bm{X}_0}+\bm{\xi}
    =
    \bm{X}_0+\bm{\rho}_n+\bm{\xi},
\end{equation*}
where $\bm{\xi}$ denotes the microscopic coordinate measured from the
lattice-site center.
Because the optical field varies slowly over the spatial extent of the
atomic orbitals, each component of the vector potential can be expanded
to first order about the lattice-site position as
\begin{equation}
    A_j\!\left(
        \bm{R}_n^{\bm{X}_0}+\bm{\xi},t
    \right)
    \simeq
    A_j\!\left(
        \bm{R}_n^{\bm{X}_0},t
    \right)
    +
    \sum_{\lambda=x,y,z}
    \xi_\lambda\,
    \partial_\lambda
    A_j\!\left(
        \bm{R}_n^{\bm{X}_0},t
    \right),
    \qquad
    j=x,y,z.
    \label{eq:vector_potential_gradient_expansion}
\end{equation}
Here, $\partial_\lambda$ acts on the spatial dependence of the
structured-light field.
For the in-plane directions, we use
$\partial_x\equiv\partial_X$ and
$\partial_y\equiv\partial_Y$.
This site-centered expansion is consistent with established
multipole formulations of optical response in crystalline systems
\cite{Mahon2019,MahonSipe2020} and with microscopic treatments of
spatial dispersion that include field-gradient couplings
\cite{DuffSipe2022}.
Terms of second and higher order in this spatial-gradient expansion
are neglected because the characteristic microscopic length associated
with $\bm{\xi}$ is much smaller than the spatial-variation scale of the
structured-light field.

For the local momentum-diagonal formulation used in the main
calculations, we further neglect the residual site-to-site variation
of the optical field about each coarse-grained position $\bm{X}_0$
and approximate
\begin{align}
    A_j\!\left(
        \bm{R}_n^{\bm{X}_0},t
    \right)
    &\simeq
    A_j(\bm{X}_0,t),
    \nonumber\\
    \partial_\lambda A_j\!\left(
        \bm{R}_n^{\bm{X}_0},t
    \right)
    &\simeq
    \partial_\lambda A_j(\bm{X}_0,t),
    \qquad
    j,\lambda=x,y,z.
    \label{eq:local_field_approximation}
\end{align}
Equation~\eqref{eq:vector_potential_gradient_expansion} then reduces to
\begin{equation}
    A_j\!\left(
        \bm{R}_n^{\bm{X}_0}+\bm{\xi},t
    \right)
    \simeq
    A_j(\bm{X}_0,t)
    +
    \sum_{\lambda=x,y,z}
    \xi_\lambda\,
    \partial_\lambda A_j(\bm{X}_0,t),
    \qquad
    j=x,y,z.
    \label{eq:local_vector_potential_gradient_expansion}
\end{equation}
The zeroth-order term in
Eq.~\eqref{eq:local_vector_potential_gradient_expansion}
describes the coupling to the local value of the optical field,
whereas the terms proportional to $\bm{\xi}$ describe the leading
variation of the field over the microscopic extent of the electronic
orbital.
These first-order terms provide the microscopic origin of the
gradient sector introduced below.

The site-centered gradient expansion should be distinguished from the
additional local approximation in
Eq.~\eqref{eq:local_field_approximation}.
The former retains the dependence on the microscopic intra-site
coordinate $\bm{\xi}$ to first order, whereas the latter neglects the
residual site-to-site variation through $\bm{\rho}_n$ by replacing the
field and its spatial derivatives at each lattice site with their
values at the common coarse-grained position $\bm{X}_0$.
As a result, the light--matter interaction becomes diagonal in the
Bloch wave vector $\bm{k}$, while its dependence on $\bm{X}_0$ is
retained.
In Secs.~\ref{sec:finite_q_hamiltonian} and
\ref{sec:finite_q_bloch_projection}, we partially relax this local
approximation by restoring the leading site-to-site optical phase
associated with a finite local in-plane optical wave vector
$\bm{q}_{\parallel}$.

Substituting Eq.~\eqref{eq:local_vector_potential_gradient_expansion}
into the light--matter interaction Hamiltonian,
\begin{equation*}
    \hat{h}_I
    =
    \hat{h}_I^{A}
    +
    \hat{h}_I^{B},
\end{equation*}
and retaining terms up to first order in the spatial gradients, the
interaction term linear in the vector potential becomes
\begin{align}
    \hat{h}_I^{A}
    \simeq{}&
    -\frac{e}{m_0c}
    \sum_{j=x,y,z}
    A_j(\bm{X}_0,t)\hat{p}_j
    \nonumber\\
    &-
    \frac{e}{m_0c}
    \sum_{j=x,y,z}
    \sum_{\lambda=x,y,z}
    \xi_\lambda\,
    \partial_\lambda A_j(\bm{X}_0,t)
    \hat{p}_j
    \nonumber\\
    \equiv{}&
    \hat{h}_I^{A1}
    +
    \hat{h}_I^{A2}.
    \label{eq:3D_A_interaction}
\end{align}
Similarly, the interaction term quadratic in the vector potential is
written as
\begin{align}
    \hat{h}_I^{B}
    \simeq{}&
    \frac{e^2}{2m_0c^2}
    \sum_{j=x,y,z}
    A_j^2(\bm{X}_0,t)
    \nonumber\\
    &+
    \frac{e^2}{m_0c^2}
    \sum_{j=x,y,z}
    \sum_{\lambda=x,y,z}
    A_j(\bm{X}_0,t)
    \xi_\lambda\,
    \partial_\lambda A_j(\bm{X}_0,t)
    \nonumber\\
    \equiv{}&
    \hat{h}_I^{B1}
    +
    \hat{h}_I^{B2}.
    \label{eq:3D_B_interaction}
\end{align}
The terms $\hat{h}_I^{A1}$ and $\hat{h}_I^{B1}$ are the
zero-gradient contributions determined by the local value of the
optical field at $\bm{X}_0$.
We refer to these terms collectively as the \textit{local sector}.
By contrast, $\hat{h}_I^{A2}$ and $\hat{h}_I^{B2}$ are the
first-gradient contributions arising from the spatial inhomogeneity
of the structured-light field and are collectively referred to as the
\textit{gradient sector}.
In the following subsection, these interaction terms are projected
onto the graphene Bloch basis and expressed in second-quantized form.

\subsection{Second-quantized Hamiltonian in the graphene Bloch basis}
\label{sec:bloch_hamiltonian}

\begin{figure}[t]
    \centering
    \includegraphics[width=0.70\textwidth]{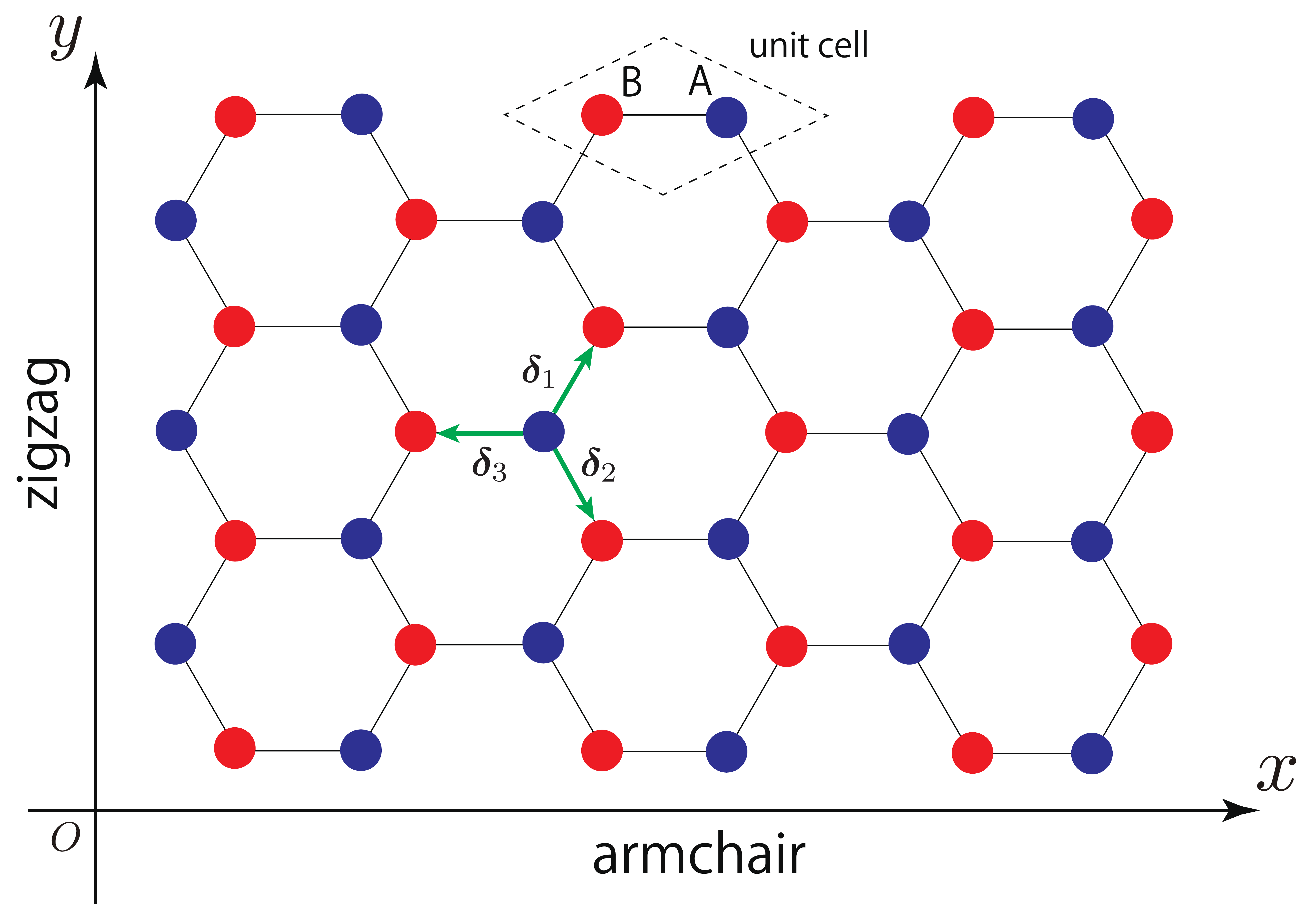}
    \caption{
        Lattice structure of graphene and the coordinate system used in this work.
        The dashed rhombus denotes a unit cell containing two carbon atoms
        belonging to the $A$ and $B$ sublattices.
        The bond vectors $\bm{\delta}_i$ $(i=1,2,3)$ are defined as the vectors
        directed from an $A$-sublattice site to its three nearest-neighbor
        $B$-sublattice sites.
        The $x$ and $y$ axes are chosen along the armchair and zigzag directions,
        respectively.
    }
    \label{fig:graphene_lattice}
\end{figure}

In this subsection, we express the light--matter interaction Hamiltonian
derived in the preceding subsection in second-quantized form using the
Bloch basis of graphene
\cite{Wallace1947BandTheoryGraphite,CastroNeto2009ElectronicPropertiesGraphene}.
We describe the $\pi$-electron system of graphene within a tight-binding
model constructed from the carbon $2p_z$ orbitals.
The unit cell of graphene contains two inequivalent carbon atoms,
conventionally denoted by the $A$ and $B$ sublattices.
The localized carbon $2p_z$ orbital centered at the origin is denoted
by $\phi_{2p_z}(\bm{\xi})$, where $\bm{\xi}$ is the microscopic
coordinate introduced in the preceding subsection.
We denote the three nearest-neighbor bond vectors directed from an
$A$-sublattice site to the neighboring $B$-sublattice sites by
$\bm{\delta}_i$ $(i=1,2,3)$
(Fig.~\ref{fig:graphene_lattice}).
An orbital centered at a site displaced by $\bm{\delta}_i$ is then
represented by $\phi_{2p_z}(\bm{\xi}-\bm{\delta}_i)$.

The reciprocal lattice of graphene is generated by two primitive
reciprocal lattice vectors, which may be written in the present
coordinate system as
\begin{equation}
    \bm{b}_1
    =
    \left(
        \frac{2\pi}{3a},
        \frac{2\pi}{\sqrt{3}a}
    \right),
    \qquad
    \bm{b}_2
    =
    \left(
        \frac{2\pi}{3a},
        -\frac{2\pi}{\sqrt{3}a}
    \right).
    \label{eq:reciprocal_lattice_vectors}
\end{equation}
These reciprocal lattice vectors define the primitive reciprocal cell
used in the following formulation.
In the following projection of the light--matter interaction, onsite
and nearest-neighbor matrix elements are examined separately, and the
leading nonvanishing contribution is retained within each interaction
sector.

\subsubsection{Graphene Bloch basis and unperturbed tight-binding Hamiltonian}
\label{sec:unperturbed_tb_hamiltonian}

In this subsection, we illustrate the Bloch projection and
second-quantization procedure for the field-free single-particle
Hamiltonian of graphene,
\begin{equation*}
    \hat{h}_0
    =
    \frac{\hat{\bm{p}}^{\,2}}{2m_0}
    +
    \sum_i V\!\left(\bm{r}-\bm{R}_i\right).
\end{equation*}
The same procedure will subsequently be applied to the light--matter
interaction terms.
For the field-free electronic structure, we adopt the standard
nearest-neighbor tight-binding approximation, neglecting the ordinary
overlap matrix between orbitals on different sites.
The electronic states are constructed from carbon $2p_z$ orbitals
localized at the lattice sites.
Let $\bm{T}_n$ denote the Bravais lattice vector of the $n$th unit
cell and $\bm{\tau}_{\alpha}$ ($\alpha=A,B$) the position of
sublattice $\alpha$ within the unit cell.
The lattice geometry and notation are summarized in
Fig.~\ref{fig:graphene_lattice}.
We denote the localized orbital at
$\bm{T}_n+\bm{\tau}_{\alpha}$ by
$\lvert n,\alpha\rangle$, with
\begin{equation*}
    \langle \bm{r}\vert n,\alpha\rangle
    =
    \phi_{2p_z}
    \!\left(
        \bm{r}-\bm{T}_n-\bm{\tau}_{\alpha}
    \right).
\end{equation*}
The argument of $\phi_{2p_z}$ corresponds to the microscopic
site-centered coordinate $\bm{\xi}$ introduced in
Sec.~\ref{sec:gradient_expansion}.
Because translational symmetry is exact for $\hat{h}_0$, the
corresponding Bloch basis is
\begin{equation*}
    \lvert \bm{k},\alpha\rangle
    =
    \frac{1}{\sqrt{N}}
    \sum_n
    e^{i\bm{k}\cdot
    \left(\bm{T}_n+\bm{\tau}_{\alpha}\right)}
    \lvert n,\alpha\rangle,
\end{equation*}
where $N$ is the number of unit cells.
We denote the annihilation operators for the $A$ and $B$ sublattices
by $\hat{a}_{\bm{k}}$ and $\hat{b}_{\bm{k}}$, respectively.

In this basis, the second-quantized field-free Hamiltonian is
\begin{equation*}
    \hat{H}_0
    =
    \sum_{\bm{k},\bm{k}'}
    \sum_{\alpha,\beta=A,B}
    \langle \bm{k},\alpha
    \vert
    \hat{h}_0
    \vert
    \bm{k}',\beta\rangle
    \hat{c}_{\bm{k},\alpha}^{\dagger}
    \hat{c}_{\bm{k}',\beta},
\end{equation*}
with
$\hat{c}_{\bm{k},A}=\hat{a}_{\bm{k}}$ and
$\hat{c}_{\bm{k},B}=\hat{b}_{\bm{k}}$.
Taking the common onsite energy of the $A$ and $B$ sublattices
as the zero of energy and retaining only nearest-neighbor hopping, the
Hamiltonian reduces to
\begin{equation}
    \hat{H}_0
    =
    \gamma
    \sum_{\bm{k}}
    \left[
        f(\bm{k})
        \hat{a}_{\bm{k}}^{\dagger}
        \hat{b}_{\bm{k}}
        +
        \mathrm{H.c.}
    \right],
    \label{eq:graphene_tb_hamiltonian}
\end{equation}
or equivalently,
\begin{equation*}
    \hat{H}_0
    =
    \sum_{\bm{k}}
    \begin{pmatrix}
        \hat{a}_{\bm{k}}^{\dagger} &
        \hat{b}_{\bm{k}}^{\dagger}
    \end{pmatrix}
    \begin{pmatrix}
        0 & \gamma f(\bm{k}) \\
        \gamma f^{*}(\bm{k}) & 0
    \end{pmatrix}
    \begin{pmatrix}
        \hat{a}_{\bm{k}} \\
        \hat{b}_{\bm{k}}
    \end{pmatrix}.
\end{equation*}
Here, $\gamma$ is the nearest-neighbor hopping matrix element.
Choosing the origin of the microscopic coordinate $\bm{\xi}$ at an
$A$-sublattice site, it can be written as
\begin{equation*}
    \gamma
    =
    \langle n,A\vert\hat{h}_0\vert n_i,B\rangle
    =
    \int d^3\xi\,
    \phi_{2p_z}^{*}(\bm{\xi})
    \hat{h}_0
    \phi_{2p_z}\!\left(\bm{\xi}-\bm{\delta}_i\right).
\end{equation*}
where $\lvert n_i,B\rangle$ denotes the $B$-sublattice orbital
connected to $\lvert n,A\rangle$ by the nearest-neighbor bond
$\bm{\delta}_i$ ($i=1,2,3$), as shown in
Fig.~\ref{fig:graphene_lattice}.
These three bonds are symmetry-equivalent, so that $\gamma$ is
independent of $i$.
The corresponding graphene structure factor is
\begin{equation*}
    f(\bm{k})
    =
    \sum_{i=1}^{3}
    e^{i\bm{k}\cdot\bm{\delta}_i}.
\end{equation*}

\subsubsection{\texorpdfstring{$\bm{A}\cdot\bm{p}$}{A dot p}-type interaction}
\label{sec:A_dot_p_interaction}

We next consider the interaction term linear in the vector potential.
Starting from the three-dimensional first-gradient interaction in
Eq.~\eqref{eq:3D_A_interaction}, we project the microscopic
Hamiltonian onto the graphene $\pi$-orbital subspace and determine
which out-of-plane contributions survive.
Terms containing either $\hat{p}_z$ or $\hat{\xi}_z$, but not both,
vanish by mirror symmetry about the graphene plane.
The remaining contribution involving
$\hat{\xi}_z\hat{p}_z$ requires separate consideration, as discussed below.
After accounting for this contribution, the effective graphene
$\pi$-band interaction is restricted to the in-plane components and
can be written as
\begin{equation*}
    \hat{h}_I^{A}
    =
    -\frac{e}{m_0c}
    \left[
        A_j(\bm{X}_0,t)\hat{p}_j
        +
        \xi_{\lambda}
        \partial_{X_{\lambda}}A_j(\bm{X}_0,t)
        \hat{p}_j
    \right],
\end{equation*}
where repeated indices $j,\lambda=x,y$ are summed over.
Here, $\xi_{\lambda}$ denotes the microscopic coordinate measured from
the lattice-site center, whereas $\partial_{X_{\lambda}}$ denotes a
spatial derivative acting on the structured-light field with respect
to the coarse-grained coordinate.
We use the compact notation
$X_x\equiv X$ and $X_y\equiv Y$, so that
$\partial_{X_x}\equiv\partial_X$ and
$\partial_{X_y}\equiv\partial_Y$.

We now project this effective interaction onto the
sublattice-resolved Bloch basis of graphene.
In general, the corresponding second-quantized interaction Hamiltonian
is given by
\begin{equation*}
    \hat{H}_I^{A}
    =
    \sum_{\bm{k},\bm{k}'}
    \sum_{\alpha,\beta=A,B}
    \langle
        \bm{k},\alpha
        \vert
        \hat{h}_I^{A}
        \vert
        \bm{k}',\beta
    \rangle
    \hat{c}_{\bm{k},\alpha}^{\dagger}
    \hat{c}_{\bm{k}',\beta},
\end{equation*}
where
$\hat{c}_{\bm{k},A}=\hat{a}_{\bm{k}}$ and
$\hat{c}_{\bm{k},B}=\hat{b}_{\bm{k}}$.
Within the local momentum-diagonal approximation,
$A_j(\bm{R}_n^{\bm{X}_0},t)$ and its spatial derivatives are
approximated by their values at the common coarse-grained position
$\bm{X}_0$, as in Eq.~\eqref{eq:local_field_approximation}.
The optical-field factors therefore carry no lattice-site index $n$.
Upon projection onto the Bloch basis, the sum over lattice sites then
produces
\begin{equation*}
    \frac{1}{N}
    \sum_n
    e^{i(\bm{k}'-\bm{k})\cdot\bm{T}_n}
    =
    \delta_{\bm{k},\bm{k}'},
\end{equation*}
where $\bm{T}_n$ is the Bravais lattice vector introduced above.
Thus, within this approximation, the light--matter interaction is
diagonal in the Bloch wave vector $\bm{k}$.
Retaining the nearest-neighbor matrix elements between the $A$ and
$B$ sublattices, we obtain
\begin{equation*}
    \hat{H}_I^{A}
    =
    \sum_{\bm{k}}
    \sum_{j=x,y}
    d_{\bm{k}}^{j}
    A_j(\bm{X}_0,t)
    \hat{a}_{\bm{k}}^{\dagger}
    \hat{b}_{\bm{k}}
    +
    \sum_{\bm{k}}
    \sum_{\lambda,j=x,y}
    d_{\bm{k}}^{\lambda j}
    \partial_{X_{\lambda}}A_j(\bm{X}_0,t)
    \hat{a}_{\bm{k}}^{\dagger}
    \hat{b}_{\bm{k}}
    +
    \mathrm{H.c.}
    \equiv
    \hat{H}_I^{A1}
    +
    \hat{H}_I^{A2}.
\end{equation*}
Here, the matrix elements are defined as
\begin{equation*}
    d_{\bm{k}}^{j}
    =
    \frac{i\hbar e}{m_0c}
    \sum_{i=1}^{3}
    e^{i\bm{k}\cdot\bm{\delta}_i}
    \int d^3\xi\,
    \phi_{2p_z}^{*}(\bm{\xi})
    \partial_{\xi_j}
    \phi_{2p_z}(\bm{\xi}-\bm{\delta}_i),
\end{equation*}
and
\begin{equation*}
    d_{\bm{k}}^{\lambda j}
    =
    \frac{i\hbar e}{m_0c}
    \sum_{i=1}^{3}
    e^{i\bm{k}\cdot\bm{\delta}_i}
    \int d^3\xi\,
    \phi_{2p_z}^{*}(\bm{\xi})
    \xi_{\lambda}\,
    \partial_{\xi_j}
    \phi_{2p_z}(\bm{\xi}-\bm{\delta}_i),
\end{equation*}
where $\partial_{\xi_j}$ denotes differentiation with respect to the
microscopic orbital coordinate $\xi_j$.
It should therefore be distinguished from
$\partial_{X_{\lambda}}$, which acts on the spatial dependence of the
structured-light field.

For a possible onsite contribution to the local
$\bm{A}\cdot\bm{p}$ interaction, the parity of the localized
$2p_z$ orbital gives
\begin{equation*}
    \int d^3\xi\,
    \phi_{2p_z}^{*}(\bm{\xi})
    \partial_{\xi_j}\phi_{2p_z}(\bm{\xi})
    =
    0,
    \qquad j=x,y,z,
\end{equation*}
since the integrand is odd under
$\xi_j\rightarrow-\xi_j$.
Thus, the onsite contribution to $\hat{H}_I^{A1}$ vanishes.
For the onsite first-gradient interaction, the three-dimensional
microscopic matrix element must be considered before restricting the
effective Hamiltonian to the graphene plane.
Integration by parts gives
\begin{equation*}
    \int d^3\xi\,
    \phi_{2p_z}^{*}(\bm{\xi})\,
    \xi_{\lambda}
    \partial_{\xi_j}
    \phi_{2p_z}(\bm{\xi})
    =
    -\frac{1}{2}\delta_{\lambda j},
    \qquad
    \lambda,j=x,y,z.
\end{equation*}
The complete onsite first-gradient contribution is therefore
proportional to
\begin{equation*}
    \partial_X A_x
    +
    \partial_Y A_y
    +
    \partial_z A_z
    =
    \bm{\nabla}\cdot\bm{A}
    =
    0,
\end{equation*}
and hence vanishes in the Coulomb gauge.
For the nearest-neighbor contribution, we retain the in-plane matrix
elements that provide the leading terms of the effective $\pi$-band
interaction. The remaining $\hat{\xi}_z\hat{p}_z$ matrix element
represents an out-of-plane overlap-sensitive correction and is not
included in the present minimal projection. Combining these results
with the mirror-symmetry argument above, the effective interaction is
restricted to $\lambda,j=x,y$.

\subsubsection{\texorpdfstring{$\bm{A}^2$}{A2}-type interaction}
\label{sec:A2_interaction}

We next consider the interaction term quadratic in the vector
potential.
Starting from the three-dimensional first-gradient interaction in
Eq.~\eqref{eq:3D_B_interaction}, we project the microscopic
Hamiltonian onto the graphene $\pi$-orbital subspace and determine
which out-of-plane contributions can be neglected.
Explicitly, Eq.~\eqref{eq:3D_B_interaction} contains the structure
\begin{equation*}
    \sum_{j=x,y,z}
    \left[
        A_j^2
        +
        2A_j
        \sum_{\lambda=x,y,z}
        \xi_{\lambda}\partial_{\lambda}A_j
    \right].
\end{equation*}
The out-of-plane contributions can be omitted for two distinct
reasons.
First, all first-gradient terms proportional to $\xi_z$ vanish upon
projection onto the graphene $\pi$-orbital subspace by mirror symmetry.
For example, the nearest-neighbor matrix element satisfies
\begin{equation*}
    \int d^3\xi\,
    \phi_{2p_z}^{*}(\bm{\xi})\,
    \xi_z\,
    \phi_{2p_z}(\bm{\xi}-\bm{\delta}_i)
    =
    0.
\end{equation*}
Second, the remaining terms involving the longitudinal field component
$A_z$, namely $A_z^2$ and
$A_z\xi_{\lambda}\partial_{X_\lambda}A_z$ with
$\lambda=x,y$, are of higher paraxial order because
$A_z$ is much smaller than the in-plane components $A_x$ and $A_y$.
We therefore retain only the leading in-plane contribution,
\begin{equation*}
    \hat{h}_I^{B}
    =
    \frac{e^2}{2m_0c^2}
    \left[
        A_j^2(\bm{X}_0,t)
        +
        2A_j(\bm{X}_0,t)
        \hat{\xi}_{\lambda}
        \partial_{X_{\lambda}}A_j(\bm{X}_0,t)
    \right]
    \equiv
    \hat{h}_I^{B1}
    +
    \hat{h}_I^{B2}.
\end{equation*}
where repeated indices $j,\lambda=x,y$ are summed over.
The corresponding second-quantized Hamiltonian is
\begin{equation*}
    \hat{H}_I^{B}
    =
    \sum_{\bm{k},\bm{k}'}
    \sum_{\alpha,\beta=A,B}
    \langle
        \bm{k},\alpha
        \vert
        \hat{h}_I^{B}
        \vert
        \bm{k}',\beta
    \rangle
    \hat{c}_{\bm{k},\alpha}^{\dagger}
    \hat{c}_{\bm{k}',\beta}.
\end{equation*}
At fixed $\bm X_0$, the zero-gradient term
$\hat h_I^{B1}$ is independent of the microscopic coordinate and is
proportional to the identity operator in coordinate space.
Its onsite matrix element is unity and therefore provides the leading
contribution to the $B1$ sector. The off-diagonal nearest-neighbor
matrix element is proportional to the ordinary orbital overlap and is
omitted as a correction to this already nonzero onsite contribution.
For the first-gradient term $\hat h_I^{B2}$, by contrast, the relevant
microscopic operator is $\hat{\xi}_{\lambda}$. Its onsite matrix
element vanishes by parity, so that the nearest-neighbor A--B
first-moment matrix element provides the leading nonvanishing
contribution to the $B2$ sector and is retained. We thus obtain
\begin{equation*}
    \hat{H}_I^{B}
    =
    \sum_{\bm{k}}
    \left\{
        \frac{e^2A^2(\bm{X}_0,t)}{2m_0c^2}
        \left(
            \hat{a}_{\bm{k}}^{\dagger}\hat{a}_{\bm{k}}
            +
            \hat{b}_{\bm{k}}^{\dagger}\hat{b}_{\bm{k}}
        \right)
        +
        \sum_{\lambda=x,y}
        \partial_{X_{\lambda}}
        \left[
            \frac{A^2(\bm{X}_0,t)}{2}
        \right]
        \left[
            S_{\lambda}(\bm{k})
            \hat{a}_{\bm{k}}^{\dagger}\hat{b}_{\bm{k}}
            +
            \mathrm{H.c.}
        \right]
    \right\}
    \equiv
    \hat{H}_I^{B1}
    +
    \hat{H}_I^{B2}.
\end{equation*}
where
\begin{equation*}
    A^2(\bm{X}_0,t)
    =
    \sum_{j=x,y}
    A_j^2(\bm{X}_0,t),
\end{equation*}
denotes the squared in-plane vector potential retained in the
effective graphene Hamiltonian, and
\begin{equation*}
    S_{\lambda}(\bm{k})
    =
    \frac{e^2}{m_0c^2}
    \sum_{i=1}^{3}
    e^{i\bm{k}\cdot\bm{\delta}_i}
    \int d^3\xi\,
    \phi_{2p_z}^{*}(\bm{\xi})
    \xi_{\lambda}
    \phi_{2p_z}(\bm{\xi}-\bm{\delta}_i).
\end{equation*}
The matrix element entering $S_{\lambda}(\bm k)$ is the
nearest-neighbor first moment of the localized $2p_z$ orbitals.
For the onsite contribution, parity gives
\begin{equation*}
    \int d^3\xi\,
    \phi_{2p_z}^{*}(\bm{\xi})
    \xi_{\lambda}
    \phi_{2p_z}(\bm{\xi})
    =
    0,
    \qquad
    \lambda=x,y.
\end{equation*}
Thus, because the onsite first moment vanishes, the
nearest-neighbor first-moment matrix element provides the leading
nonvanishing contribution to $\hat H_I^{B2}$.

\subsubsection{Summary of the local and gradient Hamiltonian sectors}
\label{sec:hamiltonian_sector_summary}

Combining the results above, the Hamiltonian for graphene driven by a
spatially inhomogeneous optical field can be written as
\begin{equation*}
    \hat{H}(\bm{X}_0,t)
    =
    \hat{H}_0
    +
    \hat{H}_I^{A1}
    +
    \hat{H}_I^{A2}
    +
    \hat{H}_I^{B1}
    +
    \hat{H}_I^{B2}.
\end{equation*}
The individual contributions are
\begin{equation*}
    \hat{H}_0
    =
    \gamma
    \sum_{\bm{k}}
    \left[
        f(\bm{k})
        \hat{a}_{\bm{k}}^{\dagger}
        \hat{b}_{\bm{k}}
        +
        \mathrm{H.c.}
    \right],
\end{equation*}
\begin{equation}
    \hat{H}_I^{A1}
    =
    \sum_{\bm{k}}
    \sum_{j=x,y}
    \left[
        d_{\bm{k}}^{j}
        A_j(\bm{X}_0,t)
        \hat{a}_{\bm{k}}^{\dagger}
        \hat{b}_{\bm{k}}
        +
        \mathrm{H.c.}
    \right],
    \label{eq:summary_HA1}
\end{equation}
\begin{equation*}
    \hat{H}_I^{A2}
    =
    \sum_{\bm{k}}
    \sum_{\lambda,j=x,y}
    \left[
        d_{\bm{k}}^{\lambda j}
        \partial_{X_{\lambda}}A_j(\bm{X}_0,t)
        \hat{a}_{\bm{k}}^{\dagger}
        \hat{b}_{\bm{k}}
        +
        \mathrm{H.c.}
    \right],
\end{equation*}
\begin{equation*}
    \hat{H}_I^{B1}
    =
    \frac{e^2A^2(\bm{X}_0,t)}
         {2m_0c^2}
    \sum_{\bm{k}}
    \left(
        \hat{a}_{\bm{k}}^{\dagger}\hat{a}_{\bm{k}}
        +
        \hat{b}_{\bm{k}}^{\dagger}\hat{b}_{\bm{k}}
    \right),
\end{equation*}
and
\begin{equation*}
    \hat{H}_I^{B2}
    =
    \sum_{\bm{k}}
    \sum_{\lambda=x,y}
    \partial_{X_{\lambda}}
    \left[
        \frac{A^2(\bm{X}_0,t)}{2}
    \right]
    \left[
        S_{\lambda}(\bm{k})
        \hat{a}_{\bm{k}}^{\dagger}
        \hat{b}_{\bm{k}}
        +
        \mathrm{H.c.}
    \right].
\end{equation*}
Here,
\begin{equation*}
    A^2(\bm{X}_0,t)
    =
    \sum_{j=x,y}
    A_j^2(\bm{X}_0,t)
\end{equation*}
denotes the squared in-plane vector potential retained in the effective
graphene Hamiltonian.
The microscopic matrix elements entering the in-plane interaction
terms are
\begin{equation*}
    d_{\bm{k}}^{j}
    =
    \frac{i\hbar e}{m_0c}
    \sum_{i=1}^{3}
    e^{i\bm{k}\cdot\bm{\delta}_i}
    \int d^3\xi\,
    \phi_{2p_z}^{*}(\bm{\xi})
    \partial_{\xi_j}
    \phi_{2p_z}(\bm{\xi}-\bm{\delta}_i),
\end{equation*}
\begin{equation*}
    d_{\bm{k}}^{\lambda j}
    =
    \frac{i\hbar e}{m_0c}
    \sum_{i=1}^{3}
    e^{i\bm{k}\cdot\bm{\delta}_i}
    \int d^3\xi\,
    \phi_{2p_z}^{*}(\bm{\xi})
    \xi_{\lambda}
    \partial_{\xi_j}
    \phi_{2p_z}(\bm{\xi}-\bm{\delta}_i),
\end{equation*}
and
\begin{equation*}
    S_{\lambda}(\bm{k})
    =
    \frac{e^2}{m_0c^2}
    \sum_{i=1}^{3}
    e^{i\bm{k}\cdot\bm{\delta}_i}
    \int d^3\xi\,
    \phi_{2p_z}^{*}(\bm{\xi})
    \xi_{\lambda}
    \phi_{2p_z}(\bm{\xi}-\bm{\delta}_i).
\end{equation*}
Here, $j,\lambda=x,y$ label the effective in-plane couplings retained
after projection onto the graphene $\pi$-band subspace.
Accordingly, the total Hamiltonian can be decomposed into the local
and gradient sectors as
\begin{equation*}
    \hat{H}(\bm{X}_0,t)
    =
    \hat{H}_0
    +
    \hat{H}_{I,\mathrm{loc}}(\bm{X}_0,t)
    +
    \hat{H}_{I,\mathrm{grad}}(\bm{X}_0,t),
\end{equation*}
where
\begin{equation*}
    \hat{H}_{I,\mathrm{loc}}
    =
    \hat{H}_I^{A1}
    +
    \hat{H}_I^{B1},
    \qquad
    \hat{H}_{I,\mathrm{grad}}
    =
    \hat{H}_I^{A2}
    +
    \hat{H}_I^{B2}.
\end{equation*}

\subsection{Finite-in-plane-wave-vector light--matter Hamiltonian}
\label{sec:finite_q_hamiltonian}

The local long-wavelength formulation developed in
Secs.~\ref{sec:gradient_expansion} and \ref{sec:bloch_hamiltonian}
neglects the residual site-to-site variation of the optical field at
fixed $\bm{X}_0$, resulting in a momentum-diagonal light--matter
interaction.
To assess this approximation, we partially restore the site-to-site
optical phase associated with the local in-plane optical wave vector
$\bm{q}_{\parallel}$ generated by the transverse phase gradient of the
structured-light field.
Specifically, we retain the leading lattice-site phase factor
$e^{i\bm{q}_{\parallel}\cdot\bm{\rho}_n}$, which introduces momentum
transfer $\pm\bm{q}_{\parallel}$ between Bloch states.
We apply this finite-$\bm{q}_{\parallel}$ extension to the effective
in-plane graphene $\pi$-band interaction derived in
Sec.~\ref{sec:bloch_hamiltonian}, retaining the same treatment of the
out-of-plane contributions.
The resulting formulation is used to test the robustness of the
circular-polarization branch selection against finite-$\bm{q}_{\parallel}$
effects in Sec.~\ref{sec:finite_q_numerical_validation}.

To formulate this finite-$\bm{q}_{\parallel}$ extension explicitly,
we use the coordinate decomposition introduced in
Sec.~\ref{sec:gradient_expansion},
\begin{equation*}
    \bm{r}
    =
    \bm{R}_n^{\bm{X}_0}+\bm{\xi}
    =
    \bm{X}_0+\bm{\rho}_n+\bm{\xi},
\end{equation*}
where $\bm{\rho}_n$ and $\bm{\xi}$ denote the lattice-site displacement
from $\bm{X}_0$ and the microscopic site-centered coordinate,
respectively.
For a structured optical field with spatial phase $\Phi(\bm{X})$, we
define the local in-plane optical wave vector as
\begin{equation}
    \bm{q}_{\parallel}(\bm{X}_0)
    =
    \bm{\nabla}_{\parallel}\Phi(\bm{X}_0).
    \label{eq:local_q_parallel}
\end{equation}
Thus, $\bm{q}_{\parallel}$ is determined by the structured-light field
rather than introduced as an independent parameter; its explicit form
for the numerical validation is given in
Sec.~\ref{sec:finite_q_numerical_validation}.
We next specify how the finite-$\bm{q}_{\parallel}$ phase variation is
incorporated into the vector potential.
To track the opposite lattice-site phase factors, we decompose the
real-valued physical vector potential into two complex-conjugate
components,
\begin{equation*}
    A_j(\bm{X},t)
    =
    \mathcal{A}_j^{(+)}(\bm{X},t)
    +
    \mathcal{A}_j^{(-)}(\bm{X},t),
    \qquad
    \mathcal{A}_j^{(-)}(\bm{X},t)
    =
    \left[
        \mathcal{A}_j^{(+)}(\bm{X},t)
    \right]^*,
    \qquad
    j=x,y.
\end{equation*}
This decomposition is introduced only to track the
$\pm\bm{q}_{\parallel}$ lattice-site phase factors.
At fixed $\bm{X}_0$, the two complex-conjugate components acquire
opposite lattice-site phase factors, so that
\begin{equation}
    A_j(\bm{R}_n^{\bm{X}_0},t)
    \simeq
    \mathcal{A}_j^{(+)}(\bm{X}_0,t)
    e^{i\bm{q}_{\parallel}\cdot\bm{\rho}_n}
    +
    \mathcal{A}_j^{(-)}(\bm{X}_0,t)
    e^{-i\bm{q}_{\parallel}\cdot\bm{\rho}_n},
    \qquad
    j=x,y,
    \label{eq:finite_q_field}
\end{equation}
At each coarse-grained position $\bm{X}_0$,
$\bm{q}_{\parallel}(\bm{X}_0)$ is treated as a fixed local wave vector
in the lattice-site phase factors.
The corresponding spatial gradient is therefore represented as
\begin{equation}
    \partial_{\lambda}A_j(\bm{R}_n^{\bm{X}_0},t)
    \simeq
    \left[
        \partial_{X_\lambda}\mathcal{A}_j^{(+)}(\bm{X}_0,t)
    \right]
    e^{i\bm{q}_{\parallel}\cdot\bm{\rho}_n}
    +
    \left[
        \partial_{X_\lambda}\mathcal{A}_j^{(-)}(\bm{X}_0,t)
    \right]
    e^{-i\bm{q}_{\parallel}\cdot\bm{\rho}_n},
    \qquad
    \lambda,j=x,y.
    \label{eq:finite_q_gradient}
\end{equation}
The derivatives
$\partial_{X_\lambda}\mathcal{A}_j^{(\pm)}(\bm{X}_0,t)$
are evaluated from the full local complex structured-light field and
therefore contain the local variations of both its amplitude and
phase.
Substituting Eqs.~\eqref{eq:finite_q_field} and
\eqref{eq:finite_q_gradient} into the site-centered expansion in
Eq.~\eqref{eq:vector_potential_gradient_expansion}, we obtain
\begin{equation}
    A_j(\bm{R}_n^{\bm{X}_0}+\bm{\xi},t)
    \simeq
    \sum_{s=\pm1}
    e^{is\bm{q}_{\parallel}\cdot\bm{\rho}_n}
    \left[
        \mathcal{A}_j^{(s)}(\bm{X}_0,t)
        +
        \xi_x\partial_X\mathcal{A}_j^{(s)}(\bm{X}_0,t)
        +
        \xi_y\partial_Y\mathcal{A}_j^{(s)}(\bm{X}_0,t)
    \right],
    \qquad
    j=x,y.
    \label{eq:finite_q_site_expansion}
\end{equation}
Thus, the microscopic dependence on $\bm{\xi}$ is treated through the
same site-centered gradient expansion as in
Sec.~\ref{sec:gradient_expansion}, whereas the dependence on
$\bm{\rho}_n$, neglected in
Eq.~\eqref{eq:local_field_approximation}, is partially restored through
the lattice-site phase factors.

We first apply the finite-$\bm{q}_{\parallel}$ field representation
to the interaction linear in the vector potential.
For each component $s=\pm1$, substituting
Eq.~\eqref{eq:finite_q_site_expansion} into the effective in-plane
$\bm{A}\cdot\bm{p}$ interaction derived in
Sec.~\ref{sec:A_dot_p_interaction} gives
\begin{equation*}
    \hat{h}_{I,s\bm{q}_{\parallel}}^{A(s)}
    \simeq
    -\frac{e}{m_0c}
    e^{is\bm{q}_{\parallel}\cdot\bm{\rho}_n}
    \sum_{j=x,y}
    \left[
        \mathcal{A}_j^{(s)}(\bm{X}_0,t)
        +
        \sum_{\lambda=x,y}
        \hat{\xi}_{\lambda}
        \partial_{X_\lambda}\mathcal{A}_j^{(s)}(\bm{X}_0,t)
    \right]
    \hat{p}_j
    \equiv
    \hat{h}_{I,s\bm{q}_{\parallel}}^{A1(s)}
    +
    \hat{h}_{I,s\bm{q}_{\parallel}}^{A2(s)}.
\end{equation*}
The full interaction linear in the vector potential is therefore
\begin{equation*}
    \hat{h}_I^{A}
    \simeq
    \sum_{s=\pm1}
    \hat{h}_{I,s\bm{q}_{\parallel}}^{A(s)}.
\end{equation*}
Thus, both the $A1$ and $A2$ interactions carry
$\pm\bm{q}_{\parallel}$ lattice-site phase factors, which generate
momentum transfers $\pm\bm{q}_{\parallel}$ upon projection onto the
Bloch basis.

We next consider the interaction quadratic in the vector potential.
Because the quadratic interaction involves products of the two
components in Eq.~\eqref{eq:finite_q_field}, it acquires a different
lattice-site phase structure.
Defining
$\bm{\mathcal{A}}^{(s)}
=(\mathcal{A}_x^{(s)},\mathcal{A}_y^{(s)})$
with $s=\pm1$, and suppressing the field arguments for compactness,
the local quadratic factor becomes
\begin{equation}
    \sum_{j=x,y} A_j^2
    =
    \sum_{s=\pm1}
    \bm{\mathcal{A}}^{(s)}
    \cdot
    \bm{\mathcal{A}}^{(s)}
    e^{2is\bm{q}_{\parallel}\cdot\bm{\rho}_n}
    +
    2\bm{\mathcal{A}}^{(+)}
    \cdot
    \bm{\mathcal{A}}^{(-)}.
    \label{eq:finite_q_A2_phase}
\end{equation}
Similarly, using Eqs.~\eqref{eq:finite_q_field} and
\eqref{eq:finite_q_gradient}, the field-dependent factor entering the
first-gradient quadratic interaction becomes
\begin{equation}
    \sum_{j=x,y}
    A_j\partial_{\lambda}A_j
    =
    \sum_{s=\pm1}
    \left[
        \bm{\mathcal{A}}^{(s)}
        \cdot
        \partial_{X_\lambda}\bm{\mathcal{A}}^{(s)}
        e^{2is\bm{q}_{\parallel}\cdot\bm{\rho}_n}
        +
        \bm{\mathcal{A}}^{(s)}
        \cdot
        \partial_{X_\lambda}\bm{\mathcal{A}}^{(-s)}
    \right].
    \label{eq:finite_q_AgradA_phase}
\end{equation}
Thus, for general polarization, the quadratic interaction contains
lattice-site phase components corresponding to momentum transfers
$0$ and $\pm2\bm{q}_{\parallel}$.
We now specialize to the circularly polarized field used in the
finite-$\bm{q}_{\parallel}$ numerical validation.
For the $+\bm{q}_{\parallel}$ component, we write
\begin{equation*}
    \bm{\mathcal{A}}^{(+)}(\bm{X}_0,t)
    =
    \mathcal{A}(\bm{X}_0,t)\bm{e}_{\sigma},
    \qquad
    \bm{e}_{\sigma}
    =
    \frac{\bm{e}_x-i\sigma\bm{e}_y}{\sqrt{2}},
    \qquad
    \sigma=\pm1,
\end{equation*}
where $\sigma$ denotes the helicity and
$\bm{\mathcal{A}}^{(-)}
=[\bm{\mathcal{A}}^{(+)}]^*$.
Since
$\bm{e}_{\sigma}\cdot\bm{e}_{\sigma}=0$
and $\bm{e}_{\sigma}$ is independent of position,
\begin{equation*}
    \bm{\mathcal{A}}^{(s)}
    \cdot
    \bm{\mathcal{A}}^{(s)}
    =
    \bm{\mathcal{A}}^{(s)}
    \cdot
    \partial_{X_\lambda}\bm{\mathcal{A}}^{(s)}
    =0,
    \qquad
    s=\pm1.
    \label{eq:finite_q_circular_quadratic_vanish}
\end{equation*}
Hence the $\pm2\bm{q}_{\parallel}$ terms in
Eqs.~\eqref{eq:finite_q_A2_phase} and
\eqref{eq:finite_q_AgradA_phase} vanish in the circularly polarized
background Hamiltonian, leaving only the mixed $(+,-)$ terms,
which correspond to zero momentum transfer.
Substituting the surviving mixed terms into the quadratic interaction
derived in Sec.~\ref{sec:A2_interaction}, we obtain
\begin{equation*}
    \hat{h}_{I}^{B}
    \simeq
    \frac{e^2}{m_0c^2}
    \left\{
        \bm{\mathcal{A}}^{(+)}
        \cdot
        \bm{\mathcal{A}}^{(-)}
        +
        \sum_{\lambda=x,y}
        \hat{\xi}_{\lambda}
        \left[
            \bm{\mathcal{A}}^{(+)}
            \cdot
            \partial_{X_\lambda}\bm{\mathcal{A}}^{(-)}
            +
            \bm{\mathcal{A}}^{(-)}
            \cdot
            \partial_{X_\lambda}\bm{\mathcal{A}}^{(+)}
        \right]
    \right\}
    \equiv
    \hat{h}_{I}^{B1}
    +
    \hat{h}_{I}^{B2}.
    \label{eq:finite_q_B_hamiltonian}
\end{equation*}
No residual $\bm{\rho}_n$-dependent lattice-site phase factor remains
in either term.
Therefore, the background $B1$ and $B2$ Hamiltonians entering the
circularly polarized time evolution are momentum diagonal under the
Bloch-basis projection considered below.
This momentum-diagonal simplification applies only to the circularly
polarized background Hamiltonian used for the time evolution.
The corresponding finite-$\bm{q}_{\parallel}$ current operators are
constructed from the full quadratic interaction before imposing the
circular-polarization condition, as discussed in
Sec.~\ref{sec:finite_q_numerical_validation}.

\subsection{Projection of the finite-wave-vector Hamiltonian onto the graphene Bloch basis}
\label{sec:finite_q_bloch_projection}

We now project the finite-$\bm{q}_{\parallel}$ interaction introduced
in Sec.~\ref{sec:finite_q_hamiltonian} onto the graphene Bloch basis.
At each coarse-grained position $\bm{X}_0$,
$\bm{q}_{\parallel}(\bm{X}_0)$ is treated as a fixed local parameter.
The position of an $\alpha$-sublattice site in the $n$th unit cell is
$\bm{T}_n+\bm{\tau}_{\alpha}$, so that its displacement from
$\bm{X}_0$ is
\begin{equation*}
    \bm{\rho}_{n,\alpha}
    =
    \bm{T}_n
    +
    \bm{\tau}_{\alpha}
    -
    \bm{X}_0,
    \qquad
    \alpha=A,B.
\end{equation*}
Accordingly, the lattice-site phase factor associated with the
$+\bm{q}_{\parallel}$ component can be written as
\begin{equation*}
    e^{i\bm{q}_{\parallel}\cdot\bm{\rho}_{n,\alpha}}
    =
    e^{i\bm{q}_{\parallel}\cdot
    (\bm{\tau}_{\alpha}-\bm{X}_0)}
    e^{i\bm{q}_{\parallel}\cdot\bm{T}_n}.
\end{equation*}
Only the second factor depends on the unit-cell index $n$ and therefore
modifies the lattice sum in the Bloch-basis projection, generating the
finite Bloch-momentum transfer.
The first factor is independent of $n$ and therefore does not affect
the Bloch-momentum selection rule.

\subsubsection{\texorpdfstring{$\bm{A}\cdot\bm{p}$}{A dot p}-type interaction}
\label{sec:finite_q_A_projection}

We first consider the interaction linear in the vector potential.
For the component carrying the lattice-site phase
$e^{is\bm{q}_{\parallel}\cdot\bm{\rho}_n}$ with $s=\pm1$,
the part of the phase that depends on the unit-cell index $n$ is
\begin{equation*}
    e^{-i\bm{k}\cdot\bm{T}_n}
    e^{i\bm{k}'\cdot\bm{T}_n}
    e^{is\bm{q}_{\parallel}\cdot\bm{T}_n}.
\end{equation*}
The sum over unit cells therefore gives
\begin{equation*}
    \frac{1}{N}
    \sum_n
    e^{i(\bm{k}'-\bm{k}
    +s\bm{q}_{\parallel})\cdot\bm{T}_n}
    =
    \delta_{\bm{k}',\,\bm{k}-s\bm{q}_{\parallel}},
    \qquad
    s=\pm1,
\end{equation*}
where the momentum relation is understood modulo a reciprocal-lattice
vector.
Thus, the unit-cell-dependent phase factor determines the
Bloch-momentum transfer.
For the nearest-neighbor $A$--$B$ matrix element, imposing
$\bm{k}'=\bm{k}-s\bm{q}_{\parallel}$ cancels the corresponding
sublattice-dependent phase, while the common factor
$e^{-is\bm{q}_{\parallel}\cdot\bm{X}_0}$ remains.

Using the same nearest-neighbor matrix elements as in the
momentum-diagonal projection and relabeling the remaining momentum
index, the local $A1$ interaction becomes
\begin{equation}
    \hat{H}_{I}^{A1}
    =
    \sum_{s=\pm1}
    \sum_{\bm{k}}
    \sum_{j=x,y}
    \left[
        e^{-is\bm{q}_{\parallel}\cdot\bm{X}_0}
        d_{\bm{k}}^{j}
        \mathcal{A}_j^{(s)}(\bm{X}_0,t)
        \hat{a}_{\bm{k}+s\bm{q}_{\parallel}}^{\dagger}
        \hat{b}_{\bm{k}}
        +
        \mathrm{H.c.}
    \right].
    \label{eq:finite_q_HA1}
\end{equation}
The first-gradient $A2$ interaction carries the same lattice-site
phase and therefore becomes
\begin{equation}
    \hat{H}_{I}^{A2}
    =
    \sum_{s=\pm1}
    \sum_{\bm{k}}
    \sum_{\lambda,j=x,y}
    \left[
        e^{-is\bm{q}_{\parallel}\cdot\bm{X}_0}
        d_{\bm{k}}^{\lambda j}
        \partial_{X_\lambda}
        \mathcal{A}_j^{(s)}(\bm{X}_0,t)
        \hat{a}_{\bm{k}+s\bm{q}_{\parallel}}^{\dagger}
        \hat{b}_{\bm{k}}
        +
        \mathrm{H.c.}
    \right].
    \label{eq:finite_q_HA2}
\end{equation}
The coefficient functions
$d_{\bm{k}}^{j}$ and $d_{\bm{k}}^{\lambda j}$ are the same as those
introduced in the momentum-diagonal formulation in
Sec.~\ref{sec:A_dot_p_interaction}.
Thus, both $A1$ and $A2$ connect Bloch states according to
\begin{equation*}
    \hat{a}_{\bm{k}}^{\dagger}\hat{b}_{\bm{k}}
    \longrightarrow
    \hat{a}_{\bm{k}+s\bm{q}_{\parallel}}^{\dagger}
    \hat{b}_{\bm{k}},
    \qquad
    s=\pm1,
\end{equation*}
corresponding to momentum transfers
$\pm\bm{q}_{\parallel}$.
For the LG-beam geometry used in the finite-$\bm{q}_{\parallel}$
validation, $\bm{q}_{\parallel}(\bm{X}_0)$ is azimuthal whereas
$\bm{X}_0$ is radial, so that
$\bm{q}_{\parallel}(\bm{X}_0)\cdot\bm{X}_0=0$.
Therefore, the additional phase factor
$e^{-is\bm{q}_{\parallel}\cdot\bm{X}_0}$
is identically unity in the numerical calculations presented below,
although it is retained here in the general finite-
$\bm{q}_{\parallel}$ formulation.
In the limit $\bm{q}_{\parallel}\rightarrow\bm{0}$,
\begin{equation*}
    \mathcal{A}_j^{(+)}+\mathcal{A}_j^{(-)}=A_j,
    \qquad
    \partial_{X_\lambda}\mathcal{A}_j^{(+)}
    +
    \partial_{X_\lambda}\mathcal{A}_j^{(-)}
    =
    \partial_{X_\lambda}A_j,
\end{equation*}
and Eqs.~\eqref{eq:finite_q_HA1} and
\eqref{eq:finite_q_HA2} reduce to the momentum-diagonal expressions
derived in Sec.~\ref{sec:A_dot_p_interaction}.

\subsubsection{\texorpdfstring{$\bm{A}^2$}{A-squared}-type interaction}
\label{sec:finite_q_B_projection}

We next consider the interaction quadratic in the vector potential.
As shown in Sec.~\ref{sec:finite_q_hamiltonian}, for the circularly
polarized background field used in the finite-$\bm{q}_{\parallel}$
numerical time evolution, the $\pm2\bm{q}_{\parallel}$ components of
both the local and first-gradient quadratic interactions vanish from
the Hamiltonian.
Only the mixed $(+,-)$ components remain, carrying no residual
$\bm{\rho}_n$-dependent lattice-site phase.
The corresponding lattice sum is therefore
\begin{equation*}
    \frac{1}{N}
    \sum_n
    e^{i(\bm{k}'-\bm{k})\cdot\bm{T}_n}
    =
    \delta_{\bm{k},\bm{k}'}.
\end{equation*}
Thus, the quadratic background Hamiltonian entering the circularly
polarized time evolution is diagonal in the Bloch wave vector.
For compactness, we define
\begin{equation*}
    \mathcal{I}(\bm{X}_0,t)
    \equiv
    \bm{\mathcal{A}}^{(+)}(\bm{X}_0,t)
    \cdot
    \bm{\mathcal{A}}^{(-)}(\bm{X}_0,t),
\end{equation*}
so that
$A^2(\bm{X}_0,t)=2\mathcal{I}(\bm{X}_0,t)$
for circular polarization.
The local quadratic contribution then becomes
\begin{equation*}
    \hat{H}_{I,\bm{q}_{\parallel}}^{B1}
    =
    \frac{e^2}{m_0c^2}
    \mathcal{I}(\bm{X}_0,t)
    \sum_{\bm{k}}
    \left(
        \hat{a}_{\bm{k}}^{\dagger}\hat{a}_{\bm{k}}
        +
        \hat{b}_{\bm{k}}^{\dagger}\hat{b}_{\bm{k}}
    \right).
\end{equation*}
The first-gradient quadratic contribution is
\begin{equation*}
    \hat{H}_{I,\bm{q}_{\parallel}}^{B2}
    =
    \sum_{\bm{k}}
    \sum_{\lambda=x,y}
    \partial_{X_\lambda}
    \mathcal{I}(\bm{X}_0,t)
    \left[
        S_{\lambda}(\bm{k})
        \hat{a}_{\bm{k}}^{\dagger}
        \hat{b}_{\bm{k}}
        +
        \mathrm{H.c.}
    \right].
\end{equation*}
Here, $S_{\lambda}(\bm{k})$ is the same microscopic coefficient
function as that introduced in the momentum-diagonal formulation in
Sec.~\ref{sec:A2_interaction}.
Thus, both $\hat{H}_{I,\bm{q}_{\parallel}}^{B1}$ and
$\hat{H}_{I,\bm{q}_{\parallel}}^{B2}$ are momentum diagonal for the
circularly polarized background field considered here.
This conclusion applies only to the background Hamiltonian used in
the time evolution; as discussed in
Sec.~\ref{sec:finite_q_numerical_validation}, the corresponding
$B1$ and $B2$ current operators contain momentum-transfer components
$0$ and $\pm2\bm{q}_{\parallel}$.

\subsubsection{Summary of the finite-wave-vector Hamiltonian}
\label{sec:finite_q_hamiltonian_summary}

Combining the above results, the finite-wave-vector formulation
preserves the same local/gradient sector decomposition as the
momentum-diagonal formulation,
\begin{equation*}
    \hat{H}_{I,\mathrm{loc}}
    =
    \hat{H}_{I}^{A1}
    +
    \hat{H}_{I}^{B1},
    \qquad
    \hat{H}_{I,\mathrm{grad}}
    =
    \hat{H}_{I}^{A2}
    +
    \hat{H}_{I}^{B2}.
\end{equation*}
For the LG-beam geometry used in the finite-$\bm{q}_{\parallel}$
validation,
$\bm{q}_{\parallel}(\bm{X}_0)\cdot\bm{X}_0=0$, so that
$e^{-is\bm{q}_{\parallel}\cdot\bm{X}_0}=1$.
Accordingly, for the circularly polarized background field used in
the numerical time evolution, the four interaction sectors take the
explicit form
\begin{align*}
    \hat{H}_{I}^{A1}
    &=
    \sum_{s=\pm1}
    \sum_{\bm{k}}
    \sum_{j=x,y}
    \left[
        d_{\bm{k}}^{j}
        \mathcal{A}_{j}^{(s)}(\bm{X}_0,t)
        \hat{a}_{\bm{k}+s\bm{q}_{\parallel}}^{\dagger}
        \hat{b}_{\bm{k}}
        +\mathrm{H.c.}
    \right],
    \\
    \hat{H}_{I}^{A2}
    &=
    \sum_{s=\pm1}
    \sum_{\bm{k}}
    \sum_{\lambda,j=x,y}
    \left[
        d_{\bm{k}}^{\lambda j}
        \partial_{X_\lambda}\mathcal{A}_{j}^{(s)}(\bm{X}_0,t)
        \hat{a}_{\bm{k}+s\bm{q}_{\parallel}}^{\dagger}
        \hat{b}_{\bm{k}}
        +\mathrm{H.c.}
    \right],
    \\
    \hat{H}_{I}^{B1}
    &=
    \frac{e^{2}}{m_{0}c^{2}}
    \mathcal{I}(\bm{X}_0,t)
    \sum_{\bm{k}}
    \left(
        \hat{a}_{\bm{k}}^{\dagger}\hat{a}_{\bm{k}}
        +
        \hat{b}_{\bm{k}}^{\dagger}\hat{b}_{\bm{k}}
    \right),
    \\
    \hat{H}_{I}^{B2}
    &=
    \sum_{\bm{k}}
    \sum_{\lambda=x,y}
    \partial_{X_\lambda}\mathcal{I}(\bm{X}_0,t)
    \left[
        S_{\lambda}(\bm{k})
        \hat{a}_{\bm{k}}^{\dagger}\hat{b}_{\bm{k}}
        +\mathrm{H.c.}
    \right].
\end{align*}
Here,
\begin{equation*}
    \mathcal{I}(\bm{X}_0,t)
    =
    \bm{\mathcal{A}}^{(+)}(\bm{X}_0,t)
    \cdot
    \bm{\mathcal{A}}^{(-)}(\bm{X}_0,t),
    \qquad
    A^{2}(\bm{X}_0,t)
    =
    2\mathcal{I}(\bm{X}_0,t),
\end{equation*}
for circular polarization.
The momentum-transfer structure of the background Hamiltonian is
therefore
\begin{equation*}
    A1,\ A2:
    \qquad
    \Delta\bm{k}=\pm\bm{q}_{\parallel},
    \qquad
    B1,\ B2:
    \qquad
    \Delta\bm{k}=0.
\end{equation*}
For later reference, the corresponding current operators have the
momentum-transfer structure
\begin{equation*}
    A1,\ A2:
    \qquad
    \Delta\bm{k}=\pm\bm{q}_{\parallel},
    \qquad
    B1,\ B2:
    \qquad
    \Delta\bm{k}=0,\ \pm2\bm{q}_{\parallel}.
    \label{eq:finite_q_current_momentum_structure}
\end{equation*}
The origin of the additional $\pm2\bm{q}_{\parallel}$ current
components is discussed in
Sec.~\ref{sec:finite_q_numerical_validation}.
Within this finite-$\bm{q}_{\parallel}$ extension, the same microscopic
coefficient functions as in the momentum-diagonal formulation are
retained, and only the Bloch-momentum transfer associated with the
structured-light phase is restored.

\subsection{Equations of motion and numerical time evolution}
\label{sec:time_evolution}

We next formulate the equations of motion used in the numerical
calculations.
We first present the momentum-diagonal formulation employed in the
main calculations, in which each Bloch wave vector $\bm{k}$ evolves
independently.
We then extend the time-evolution scheme to the finite-
$\bm{q}_{\parallel}$ formulation, in which the restored lattice-site
phase couples Bloch states whose wave vectors differ by
$\pm\bm{q}_{\parallel}$.
This finite-$\bm{q}_{\parallel}$ evolution is used in the numerical
validation discussed in
Sec.~\ref{sec:finite_q_numerical_validation}.

\subsubsection{Momentum-diagonal formulation}
\label{sec:time_evolution_q0}

In the local momentum-diagonal formulation summarized in
Sec.~\ref{sec:hamiltonian_sector_summary}, the residual lattice-site
variation of the optical field is neglected at fixed $\bm{X}_0$.
The Bloch wave vector $\bm{k}$ therefore remains a good quantum number
at each $\bm{X}_0$, and the Hamiltonian decomposes into independent
momentum sectors,
\begin{equation*}
    \hat{H}(\bm{X}_0,t)
    =
    \sum_{\bm{k}}
    \hat{\Psi}_{\bm{k}}^{\dagger}
    h_{\bm{k}}(\bm{X}_0,t)
    \hat{\Psi}_{\bm{k}},
\end{equation*}
where
\begin{equation*}
    \hat{\Psi}_{\bm{k}}
    =
    \begin{pmatrix}
        \hat{a}_{\bm{k}}\\
        \hat{b}_{\bm{k}}
    \end{pmatrix},
\end{equation*}
and
\begin{equation*}
    h_{\bm{k}}(\bm{X}_0,t)
    =
    h_{0,\bm{k}}
    +
    h_{I,\bm{k}}^{A1}(\bm{X}_0,t)
    +
    h_{I,\bm{k}}^{B1}(\bm{X}_0,t)
    +
    h_{I,\bm{k}}^{A2}(\bm{X}_0,t)
    +
    h_{I,\bm{k}}^{B2}(\bm{X}_0,t).
\end{equation*}
The time-dependent Schr\"odinger equation can therefore be solved
independently for each pair $(\bm{X}_0,\bm{k})$,
\begin{equation*}
    i\hbar
    \frac{\partial}{\partial t}
    \lvert\psi_{\bm{k}}(\bm{X}_0,t)\rangle
    =
    h_{\bm{k}}(\bm{X}_0,t)
    \lvert\psi_{\bm{k}}(\bm{X}_0,t)\rangle.
    \label{eq:time_dependent_schrodinger_equation}
\end{equation*}
In the sublattice basis,
\begin{equation*}
    \lvert\psi_{\bm{k}}(\bm{X}_0,t)\rangle
    =
    \begin{pmatrix}
        \psi_{\bm{k}}^{A}(\bm{X}_0,t)\\
        \psi_{\bm{k}}^{B}(\bm{X}_0,t)
    \end{pmatrix}.
\end{equation*}
Before the optical field is applied, the system is initialized in the
valence-band eigenstate of the unperturbed Hamiltonian.
Under the convention in which the hopping parameter $\gamma$ in
Eq.~\eqref{eq:graphene_tb_hamiltonian} is positive, the initial state is
\begin{equation*}
    \lvert\psi_{\bm{k}}(\bm{X}_0,0)\rangle
    =
    \frac{1}{\sqrt{2}}
    \begin{pmatrix}
        -f(\bm{k})/\lvert f(\bm{k})\rvert\\
        1
    \end{pmatrix}.
\end{equation*}
This initial condition corresponds to charge-neutral graphene at zero
temperature, with the valence band fully occupied and the conduction
band empty.
No phenomenological dephasing or relaxation terms are included, so the
resulting dynamics correspond to the coherent limit.
For the momentum-diagonal calculation, the momentum integration is
performed over a primitive reciprocal cell parameterized as
\begin{equation*}
    \bm{k}
    =
    p_1\bm{b}_1+p_2\bm{b}_2,
    \qquad
    -\frac{1}{2}
    \leq
    p_1,p_2
    <
    \frac{1}{2},
\end{equation*}
where $\bm{b}_1$ and $\bm{b}_2$ are the primitive reciprocal-lattice
vectors defined in Eq.~\eqref{eq:reciprocal_lattice_vectors}.
For a uniform discretization containing $N_{\bm{k}}$ momentum points,
the normalized momentum sum approaches
\begin{equation*}
    \frac{1}{N_{\bm{k}}}
    \sum_{\bm{k}}F(\bm{k})
    \longrightarrow
    \frac{\mathcal{A}_{\mathrm{uc}}}{(2\pi)^2}
    \int_{\mathrm{cell}}d^2k\,F(\bm{k}),
\end{equation*}
where $\mathcal{A}_{\mathrm{uc}}$ is the area of the real-space
primitive unit cell.
The local current is evaluated using this normalized momentum-space
sum together with the prefactors associated with the corresponding
current operator.

\subsubsection{\texorpdfstring{Finite-$\bm{q}_{\parallel}$}{Finite-q-parallel} extension}
\label{sec:time_evolution_finite_q}

For finite $\bm{q}_{\parallel}$, the $A1$ and $A2$ interactions
derived in Sec.~\ref{sec:finite_q_A_projection} couple Bloch states
whose wave vectors differ by $\pm\bm{q}_{\parallel}$.
The time evolution therefore no longer separates into independent
$\bm{k}$ sectors.
At each fixed $\bm{X}_0$, the Bloch components connected through repeated
$\pm\bm{q}_{\parallel}$ momentum transfers are propagated simultaneously
under the coupled finite-$q_{\parallel}$ Hamiltonian.
The state containing the coupled Bloch components evolves according to
\begin{equation*}
    i\hbar
    \frac{\partial}{\partial t}
    \lvert\Psi(\bm{X}_0,t)\rangle
    =
    h_{\bm{q}_{\parallel}}(\bm{X}_0,t)
    \lvert\Psi(\bm{X}_0,t)\rangle,
\end{equation*}
where $h_{\bm{q}_{\parallel}}$ is the single-particle representation
of the finite-wave-vector Hamiltonian derived in
Sec.~\ref{sec:finite_q_bloch_projection}.
Before the optical field is applied, the initially occupied states are
the valence-band eigenstates of the field-free Hamiltonian, as in the
momentum-diagonal formulation.
The numerical implementation used for the finite-$\bm{q}_{\parallel}$
validation is described in
Sec.~\ref{sec:finite_q_numerical_validation}.

\section{Current operators and sector-resolved channel decomposition}
\label{sec:current_channels}

The current operator can be derived from charge conservation.
Following Noether's theorem, the conserved current associated with the
$U(1)$ phase symmetry can be identified with the source current coupled
to the electromagnetic gauge field
~\cite{Noether1971InvariantVariationProblems,Weinberg1995QFTVol1}.
In the present formulation, the external optical field enters the
Hamiltonian not only through the vector potential $A_i$ but also
through its first spatial gradients $\partial_j A_i$.
Therefore, in addition to the conventional contribution obtained by
differentiating with respect to $A_i$, the current contains
contributions originating from the gradient dependence of the optical
field.
The explicit Bloch-basis current operators derived in this section are
formulated within the momentum-diagonal approximation used in the main
calculations.
The corresponding finite-$\bm{q}_{\parallel}$ current evaluation is presented separately in Sec.~\ref{sec:finite_q_numerical_validation}.
In the preceding section, we decomposed the Hamiltonian for
structured-light-driven graphene as
\begin{equation}
    \hat{H}
    =
    \hat{H}_0
    +
    \hat{H}_I^{A1}
    +
    \hat{H}_I^{A2}
    +
    \hat{H}_I^{B1}
    +
    \hat{H}_I^{B2}.
    \label{eq:current_section_total_hamiltonian}
\end{equation}
Here, $\hat{H}_I^{A1}$ and $\hat{H}_I^{B1}$ correspond to the
zero-gradient contributions of the optical field, whereas
$\hat{H}_I^{A2}$ and $\hat{H}_I^{B2}$ arise from the first spatial
gradients of the field.
In this section, we derive the current operators associated with these
individual interaction terms.
It should be emphasized that the current channels defined below do not
represent responses obtained by evolving the electronic state with only
the corresponding interaction term.
Instead, the electronic state is always propagated under the full
Hamiltonian in Eq.~\eqref{eq:current_section_total_hamiltonian}, and
the expectation values of the individual current operators are
evaluated using this fully time-evolved state.
The channel decomposition therefore provides an operator-level
resolution of the total current into contributions associated with the
local and gradient sectors of the light--matter interaction.

\subsection{Noether's theorem and the continuity equation}
\label{sec:noether_continuity}

To derive the current operator in a form applicable to Hamiltonians that depend on both the vector potential and its spatial gradients, we consider the gauge variation of the action
\begin{equation*}
    S
    =
    \int dt \int d^3x\,
    \mathcal{L},
\end{equation*}
where the Lagrangian density depends on the electron field operators
$\psi$ and $\psi^\dagger$, the scalar potential $\phi$, the vector potential
$A_i$, and the first spatial gradients $\partial_j A_i$:
\begin{equation*}
    \mathcal{L}
    =
    \mathcal{L}
    \left[
        \psi,\psi^\dagger,\phi,A_i,\partial_j A_i
    \right].
\end{equation*}
We define a local $U(1)$ gauge transformation by using an arbitrary real function $\chi(\bm{x},t)$ as
\begin{equation*}    \psi'(\bm{x},t)
    =
    \exp
    \left[
        \frac{i e}{\hbar c}\chi(\bm{x},t)
    \right]
    \psi(\bm{x},t).
\end{equation*}
Consistently with the minimal-coupling convention
$\hat{\bm{p}}\rightarrow\hat{\bm{p}}-(e/c)\bm{A}$, the electromagnetic potentials transform as
\begin{equation*}    \phi'
    =
    \phi
    -
    \frac{1}{c}\partial_t\chi,
    \qquad
    A_i'
    =
    A_i
    +
    \partial_i\chi.
\end{equation*}
Thus, for an infinitesimal gauge transformation, we have
\begin{equation}    \delta\phi
    =
    -\frac{1}{c}\partial_t\chi,
    \qquad
    \delta A_i
    =
    \partial_i\chi,
    \qquad
    \delta(\partial_j A_i)
    =
    \partial_j\partial_i\chi.
    \label{eq:infinitesimal_gauge_variations}
\end{equation}
Gauge invariance of the action requires $\delta S=0$.
Using the Euler--Lagrange equations for the electron fields, the
terms associated with the variations of $\psi$ and $\psi^\dagger$
vanish, leaving the variations with respect to the external
electromagnetic potentials:
\begin{align*}    \delta S
    =
    \int dt \int d^3x
    \left[
        \frac{\partial\mathcal{L}}{\partial\phi}\delta\phi
        +
        \frac{\partial\mathcal{L}}{\partial A_i}\delta A_i
        +
        \frac{\partial\mathcal{L}}{\partial(\partial_j A_i)}
        \delta(\partial_j A_i)
    \right].
\end{align*}
Substituting Eq.~\eqref{eq:infinitesimal_gauge_variations} gives
\begin{align*}    \delta S
    =
    \int dt \int d^3x
    \left[
        -\frac{1}{c}
        \frac{\partial\mathcal{L}}{\partial\phi}
        \partial_t\chi
        +
        \frac{\partial\mathcal{L}}{\partial A_i}
        \partial_i\chi
        +
        \frac{\partial\mathcal{L}}{\partial(\partial_j A_i)}
        \partial_j\partial_i\chi
    \right].
\end{align*}
After integration by parts with respect to time and space, and assuming that $\chi$ and its derivatives vanish at the boundary, the surface terms can be neglected. We then find
\begin{align*}    \delta S
    =
    \int dt \int d^3x
    \left\{
        \frac{1}{c}
        \partial_t
        \left(
            \frac{\partial\mathcal{L}}{\partial\phi}
        \right)
        -
        \partial_i
        \left[
            \frac{\partial\mathcal{L}}{\partial A_i}
            -
            \partial_j
            \left(
                \frac{\partial\mathcal{L}}{\partial(\partial_j A_i)}
            \right)
        \right]
    \right\}
    \chi(\bm{x},t).
\end{align*}
Since the gauge function $\chi(\bm{x},t)$ is arbitrary, the coefficient of $\chi$ must vanish identically:
\begin{equation}    \frac{1}{c}
    \partial_t
    \left(
        \frac{\partial\mathcal{L}}{\partial\phi}
    \right)
    -
    \partial_i
    \left[
        \frac{\partial\mathcal{L}}{\partial A_i}
        -
        \partial_j
        \left(
            \frac{\partial\mathcal{L}}{\partial(\partial_j A_i)}
        \right)
    \right]
    =
    0.
    \label{eq:noether_identity_lagrangian}
\end{equation}
By defining the charge density and current density as
\begin{equation*}
    \rho
    \equiv
    -\frac{\partial\mathcal{L}}{\partial\phi},
    \qquad
    J_i
    \equiv
    c
    \left[
        \frac{\partial\mathcal{L}}{\partial A_i}
        -
        \partial_j
        \left(
            \frac{\partial\mathcal{L}}
                 {\partial(\partial_j A_i)}
        \right)
    \right].
\end{equation*}
Eq.~\eqref{eq:noether_identity_lagrangian} becomes the continuity
equation,
\begin{equation*}
    \frac{\partial\rho(\bm{x},t)}{\partial t}
    +
    \bm{\nabla}\cdot\bm{J}(\bm{x},t)
    =
    0.
\end{equation*}
This is the continuity equation implied by gauge invariance when the
Lagrangian depends on both the vector potential and its spatial
gradients.
In the present work, the Lagrangian density is written as
\begin{equation*}    \mathcal{L}
    =
    i\hbar
    \psi^\dagger
    \partial_t\psi
    -
    \mathcal{H}
    \left[
        \psi,\psi^\dagger,\phi,A_i,\partial_j A_i
    \right],
\end{equation*}
where the electromagnetic potentials enter only through the Hamiltonian density $\mathcal{H}$. Therefore,
\begin{equation*}    \frac{\partial\mathcal{L}}{\partial\phi}
    =
    -\frac{\partial\mathcal{H}}{\partial\phi},
    \qquad
    \frac{\partial\mathcal{L}}{\partial A_i}
    =
    -\frac{\partial\mathcal{H}}{\partial A_i},
    \qquad
    \frac{\partial\mathcal{L}}{\partial(\partial_j A_i)}
    =
    -\frac{\partial\mathcal{H}}{\partial(\partial_j A_i)}.
\end{equation*}
In the Hamiltonian representation, the charge density and current density are thus given by
\begin{equation*}    \rho
    =
    \frac{\partial\mathcal{H}}{\partial\phi},
\end{equation*}
and
\begin{equation}    J_i
    =
    -c
    \left[
        \frac{\partial\mathcal{H}}{\partial A_i}
        -
        \partial_j
        \left(
            \frac{\partial\mathcal{H}}{\partial(\partial_j A_i)}
        \right)
    \right].
    \label{eq:current_density_hamiltonian}
\end{equation}
The zero-gradient Hamiltonian depends directly on $A_i$ and therefore contributes through the first term in Eq.~\eqref{eq:current_density_hamiltonian}. By contrast, the first-gradient Hamiltonian depends on $\partial_j A_i$ and contributes through the second term. In the following subsections, we apply this general expression to
$\hat{H}_I^{A1}$,
$\hat{H}_I^{B1}$,
$\hat{H}_I^{A2}$,
and
$\hat{H}_I^{B2}$, and define the corresponding current channels used in the numerical calculation.

\subsection{Local \texorpdfstring{$\bm{A}\cdot\bm{p}$}{A dot p} channel}
\label{sec:current_A1}

We first consider the zero-gradient $\bm{A}\cdot\bm{p}$ interaction,
$\hat{H}_I^{A1}$.
This term is directly proportional to the vector potential
$\bm{A}(\bm{X}_0,t)$ and does not depend on its spatial gradients.
The corresponding current operator is therefore obtained from
\begin{equation*}
    \hat{J}_{i}^{A1}(\bm{X}_0,t)
    =
    -c
    \frac{\partial \hat{H}_I^{A1}}
         {\partial A_i(\bm{X}_0,t)} .
\end{equation*}
Using Eq.~\eqref{eq:summary_HA1}, we obtain
\begin{equation*}
    \hat{J}_{i}^{A1}(\bm{X}_0,t)
    =
    -c
    \sum_{\bm{k}}
    \left[
        d_{\bm{k}}^{i}
        \hat{a}_{\bm{k}}^{\dagger}(\bm{X}_0)
        \hat{b}_{\bm{k}}(\bm{X}_0)
        +
        \mathrm{H.c.}
    \right].
\end{equation*}
Using the time-evolved Bloch state
$\lvert\psi_{\bm{k}}(\bm{X}_0,t)\rangle$, the local
$\bm{A}\cdot\bm{p}$ current is
\begin{equation*}
    J_{i}^{A1}(\bm{X}_0,t)
    =
    -2c\,\mathrm{Re}
    \sum_{\bm{k}}
    d_{\bm{k}}^{i}
    \left[\psi_{\bm{k}}^{A}(\bm{X}_0,t)\right]^*
    \psi_{\bm{k}}^{B}(\bm{X}_0,t).
\end{equation*}
Here, $\lvert\psi_{\bm{k}}(\bm{X}_0,t)\rangle$ is propagated under the
full Hamiltonian, including the $A1$, $B1$, $A2$, and $B2$
interaction sectors.
Accordingly, the label $A1$ specifies the current operator being
resolved and does not imply time evolution under
$\hat{H}_I^{A1}$ alone.
In the spatially uniform-field limit, this channel reduces to the
conventional $\bm{A}\cdot\bm{p}$ current commonly used in
solid-state optical-response and high-harmonic-generation calculations
~\cite{Tamaya2016PRLDiabaticHHG,Tamaya2016PRBEllipticalHHG,
Tamaya2019SubbandHHG,Tamaya2023ShearStrainGrapheneHHG,
Tamaya2021PiezoOpticHHG,Xia2021PRBGaAsHHG,
Yoshikawa2017ScienceGrapheneHHG}.

\subsection{Local \texorpdfstring{$\bm{A}^2$}{A squared} diamagnetic contribution}
\label{sec:current_B1}

We next consider the zero-gradient $A^2$ interaction,
\begin{equation*}
    \hat{H}_I^{B1}
    =
    \frac{e^2 A^2(\bm{X}_0,t)}
         {2m_0c^2}
    \sum_{\bm{k}}
    \left[
        \hat{a}_{\bm{k}}^{\dagger}(\bm{X}_0)
        \hat{a}_{\bm{k}}(\bm{X}_0)
        +
        \hat{b}_{\bm{k}}^{\dagger}(\bm{X}_0)
        \hat{b}_{\bm{k}}(\bm{X}_0)
    \right]
    \equiv
    \frac{e^2 A^2(\bm{X}_0,t)}
         {2m_0c^2}
    \hat{n}(\bm{X}_0),
\end{equation*}
where $\hat{n}(\bm{X}_0)$ is the local particle-number operator.
Since this term depends directly on the vector potential, the
corresponding current operator is
\begin{equation*}
    \hat{J}_{i}^{B1}(\bm{X}_0,t)
    =
    -c
    \frac{\partial \hat{H}_I^{B1}}
         {\partial A_i(\bm{X}_0,t)}
    =
    -\frac{e^2}{m_0c}
    A_i(\bm{X}_0,t)
    \hat{n}(\bm{X}_0).
\end{equation*}
Using the time-evolved Bloch states
$\lvert\psi_{\bm{k}}(\bm{X}_0,t)\rangle$, we define
\begin{equation*}
    n(\bm{X}_0,t)
    =
    \sum_{\bm{k}}
    \left[
        \left|\psi_{\bm{k}}^{A}(\bm{X}_0,t)\right|^2
        +
        \left|\psi_{\bm{k}}^{B}(\bm{X}_0,t)\right|^2
    \right],
\end{equation*}
so that the diamagnetic current is
\begin{equation*}
    J_{i}^{B1}(\bm{X}_0,t)
    =
    -\frac{e^2}{m_0c}
    A_i(\bm{X}_0,t)
    n(\bm{X}_0,t).
\end{equation*}
In the closed two-band model considered here, the norm of each
time-evolved Bloch state is conserved, so that
$n(\bm{X}_0,t)$ is time independent.
Consequently, this contribution is directly proportional to
$A_i(\bm{X}_0,t)$.

\subsection{Gradient \texorpdfstring{$\bm{A}\cdot\bm{p}$}{A dot p} channel}
\label{sec:current_A2}

We now consider the first-gradient $\bm{A}\cdot\bm{p}$ interaction,
\begin{equation*}
    \hat{H}_I^{A2}
    =
    \sum_{\bm{k}}
    \sum_{\lambda,j=x,y}
    \left[
        d_{\bm{k}}^{\lambda j}
        \partial_{X_{\lambda}}A_j(\bm{X}_0,t)
        \hat{a}_{\bm{k}}^{\dagger}(\bm{X}_0)
        \hat{b}_{\bm{k}}(\bm{X}_0)
        +
        \mathrm{H.c.}
    \right].
\end{equation*}
Here, $j=x,y$ denotes the Cartesian component of the vector potential,
whereas $\lambda=x,y$ labels the microscopic coordinate direction
associated with the gradient expansion and the corresponding
coarse-grained derivative $\partial_{X_\lambda}$.
We use the notation $X_x\equiv X$ and $X_y\equiv Y$.
Since $\hat{H}_I^{A2}$ depends on
$\partial_{X_{\lambda}}A_j$ rather than on $A_j$ itself, the
corresponding current is obtained from the gradient term in
Eq.~\eqref{eq:current_density_hamiltonian}:
\begin{equation*}
    \hat{J}_{i}^{A2}(\bm{X}_0,t)
    =
    c
    \sum_{\lambda=x,y}
    \partial_{X_{\lambda}}
    \left[
        \frac{\partial \hat{H}_I^{A2}}
             {\partial(\partial_{X_{\lambda}}A_i)}
    \right].
\end{equation*}
Using the time-evolved Bloch state
$\lvert\psi_{\bm{k}}(\bm{X}_0,t)\rangle$, we define
\begin{equation*}
    D^{\lambda j}(\bm{X}_0,t)
    =
    \sum_{\bm{k}}
    d_{\bm{k}}^{\lambda j}
    \left[\psi_{\bm{k}}^{A}(\bm{X}_0,t)\right]^*
    \psi_{\bm{k}}^{B}(\bm{X}_0,t).
\end{equation*}
The two Cartesian components of the gradient
$\bm{A}\cdot\bm{p}$ current are then
\begin{align*}
    J_x^{A2}(\bm{X}_0,t)
    &=
    2c\,\mathrm{Re}
    \left[
        \partial_X D^{xx}(\bm{X}_0,t)
        +
        \partial_Y D^{yx}(\bm{X}_0,t)
    \right],
    \\
    J_y^{A2}(\bm{X}_0,t)
    &=
    2c\,\mathrm{Re}
    \left[
        \partial_X D^{xy}(\bm{X}_0,t)
        +
        \partial_Y D^{yy}(\bm{X}_0,t)
    \right].
\end{align*}
The first superscript of $D^{\lambda j}$ labels the microscopic
coordinate direction associated with the gradient expansion, whereas
the second superscript denotes the Cartesian component of the vector
potential.

\subsection{\texorpdfstring{$B2$}{B2}-type
\texorpdfstring{$\bm{A}\nabla\bm{A}$}{A grad A} channel}
\label{sec:current_B2}

Finally, we consider the first-gradient $A^2$ interaction,
\begin{equation*}
    \hat{H}_I^{B2}
    =
    \sum_{\lambda=x,y}
    \partial_{X_{\lambda}}
    \left[
        \frac{A^2(\bm{X}_0,t)}{2}
    \right]
    \sum_{\bm{k}}
    \left[
        S_{\lambda}(\bm{k})
        \hat{a}_{\bm{k}}^{\dagger}(\bm{X}_0)
        \hat{b}_{\bm{k}}(\bm{X}_0)
        +
        \mathrm{H.c.}
    \right]
    \equiv
    \sum_{\lambda=x,y}
    \partial_{X_{\lambda}}
    \left[
        \frac{A^2(\bm{X}_0,t)}{2}
    \right]
    \hat{\mathcal{Q}}_{\lambda}(\bm{X}_0).
\end{equation*}
Using
\begin{equation*}
    \partial_{X_{\lambda}}
    \left[
        \frac{A^2(\bm{X}_0,t)}{2}
    \right]
    =
    \sum_{j=x,y}
    A_j(\bm{X}_0,t)
    \partial_{X_{\lambda}}A_j(\bm{X}_0,t),
\end{equation*}
we see that $\hat{H}_I^{B2}$ depends on both $A_j$ and
$\partial_{X_{\lambda}}A_j$.
Both terms in Eq.~\eqref{eq:current_density_hamiltonian} therefore
contribute, giving
\begin{align*}
    \hat{J}_i^{B2}(\bm{X}_0,t)
    &=
    -c
    \sum_{\lambda=x,y}
    \left\{
        \left[
            \partial_{X_\lambda}A_i(\bm{X}_0,t)
        \right]
        \hat{\mathcal{Q}}_\lambda(\bm{X}_0)
        -
        \partial_{X_\lambda}
        \left[
            A_i(\bm{X}_0,t)
            \hat{\mathcal{Q}}_\lambda(\bm{X}_0)
        \right]
    \right\}
    \\
    &\qquad =
    c\,A_i(\bm{X}_0,t)
    \sum_{\lambda=x,y}
    \partial_{X_\lambda}
    \hat{\mathcal{Q}}_\lambda(\bm{X}_0).
\end{align*}
where the terms proportional to
$\partial_{X_\lambda}A_i$ cancel.
Using the time-evolved Bloch states
$\lvert\psi_{\bm{k}}(\bm{X}_0,t)\rangle$, we define
\begin{equation*}
    \mathcal{S}_{\lambda}(\bm{X}_0,t)
    =
    \sum_{\bm{k}}
    S_{\lambda}(\bm{k})
    \left[\psi_{\bm{k}}^{A}(\bm{X}_0,t)\right]^*
    \psi_{\bm{k}}^{B}(\bm{X}_0,t),
\end{equation*}
so that
$\langle\hat{\mathcal{Q}}_{\lambda}(\bm{X}_0)\rangle
=
2\,\mathrm{Re}\,\mathcal{S}_{\lambda}(\bm{X}_0,t)$.
The corresponding $B2$ current is therefore
\begin{equation*}
    J_i^{B2}(\bm{X}_0,t)
    =
    2c\,
    A_i(\bm{X}_0,t)
    \mathrm{Re}
    \left[
        \sum_{\lambda=x,y}
        \partial_{X_\lambda}
        \mathcal{S}_{\lambda}(\bm{X}_0,t)
    \right].
\end{equation*}
In the numerical evaluation,
$\partial_{X_\lambda}\mathcal{S}_{\lambda}(\bm{X}_0,t)$ is obtained
from neighboring points on the real-space observation grid.
For the radial profiles shown below, this requires values at multiple
observation radii rather than at a single observation circle.
This contribution originates from the $A\nabla A$-type interaction in
the first-gradient $A^2$ sector and is referred to below as the
$B2$-type $A\nabla A$ channel.

\subsection{Summary of current-channel notation}
\label{sec:current_channel_summary}

The current operators derived above provide an operator-level
decomposition of the total current into the local and gradient sectors
of the light--matter interaction.
For the numerical analysis, we use the notation
\begin{equation*}
    J_1 \equiv J^{A1},
    \qquad
    J_2 \equiv J^{A2},
    \qquad
    J_3 \equiv J^{B1},
    \qquad
    J_4 \equiv J^{B2}.
\end{equation*}
Here, $J_1$ denotes the local $\bm{A}\cdot\bm{p}$ channel, $J_2$ the
gradient $\bm{A}\cdot\bm{p}$ channel, $J_3$ the local $\bm{A}^2$
diamagnetic contribution, and $J_4$ the $B2$-type
$\bm{A}\nabla\bm{A}$ channel.
In all cases, the electronic state is propagated under the full
Hamiltonian, and each channel is evaluated from the corresponding
current operator defined above.

\section{Additional details on winding selection rules and numerical results}
\label{sec:winding_selection_rules}

In this section, we apply the theoretical framework developed in the
preceding sections to the numerical calculations presented in the main
text and provide additional derivations and numerical results supporting
the sector-resolved winding selection rules.
In Sec.~\ref{sec:structured_light_field}, we specify the LG-beam vector
potential and the field parameters used in the calculations.
In Sec.~\ref{sec:winding_spectrum_definition}, we define the azimuthal
current and the corresponding winding spectrum constructed from the
time-dependent Cartesian current components
$J_x(\phi,t)$ and $J_y(\phi,t)$.
We then clarify the role of the local $\bm{A}^2$ diamagnetic
contribution $J_3$ in Sec.~\ref{sec:J3_role} and derive the angular
origin of the local- and gradient-sector winding selection rules in
Sec.~\ref{sec:sector_resolved_selection_rules}.
Section~\ref{sec:additional_numerical_results} presents additional
numerical results, including current-channel-resolved dc winding
spectra, helicity-resolved channel decompositions, the field-amplitude
dependence of the selected $m=0$ component, the observation-radius
dependence of the winding spectra, and the radial profile of the
$B2$-type channel.
Unless otherwise stated, these spectra are evaluated at the maximum
electric-field amplitude
$E_0^{\max}=5.6~\mathrm{MV/cm}$; the field-amplitude dependence is
examined separately in
Sec.~\ref{sec:field_amplitude_dependence_m0_current}.
Finally, finite-$\bm{q}_{\parallel}$ effects associated with in-plane
optical momentum transfer are examined in
Sec.~\ref{sec:finite_q_numerical_validation}, and the axial magnetic
field reconstructed from the dc current distribution is presented in
Sec.~\ref{sec:magnetic_field_reconstruction}.

\subsection{Structured-light field used in the numerical calculation}
\label{sec:structured_light_field}

In this subsection, we specify the structured-light vector potential used in the numerical calculations. 
In the preceding sections, we derived the Hamiltonian and current operators for a general spatially dependent vector potential \(\bm{A}(\bm{X}_0,t)\). 
Here, as a concrete example, we consider an LG beam
\cite{Siegman1986Lasers,Yao2011OAMReview,Allen1992OAMLG}.
An LG beam is characterized by the azimuthal phase factor
$e^{i\ell\phi}$, where $\ell$ is an integer topological charge associated
with the orbital angular momentum (OAM) of light.
Because of this azimuthal phase, the optical field carries a spatial winding structure, which can also be imprinted on the current response induced in graphene. 
To examine how this winding structure appears in the local and gradient sectors, we incorporate the spatial dependence of the LG beam through the vector potential and its spatial gradients.

We place the graphene sheet at the focal plane of the beam and describe the in-plane position by the polar coordinates \((R,\phi)\). 
Although an LG beam generally has a three-dimensional propagation structure, the present calculation focuses on the in-plane vector potential and its in-plane gradients felt by the electrons in graphene at the focal plane \(z=0\). 
This treatment is justified because the thickness of monolayer graphene is much smaller than the beam length scale, and because the longitudinal component and the \(z\)-dependence of the paraxial beam give higher-order corrections. 
We therefore use the two-dimensional in-plane structure of the LG beam.
For the parameters used below, the beam waist is much larger than the graphene lattice constant, which justifies the microscopic gradient expansion.
At the same time, the paraxial approximation is used only to specify the transverse structured-light profile.
Possible nonparaxial corrections may modify the quantitative field distribution, but they are not expected to change the sector-resolved winding selection rules derived from the in-plane angular structure.

The spatial envelope of the LG beam is defined as
\begin{equation*}
    u_{p\ell}(R)
    =
    \left(
        \frac{\sqrt{2}R}{W_0}
    \right)^{|\ell|}
    L_p^{|\ell|}
    \left(
        \frac{2R^2}{W_0^2}
    \right)
    \exp
    \left(
        -\frac{R^2}{W_0^2}
    \right),
\end{equation*}
where \(W_0\) is the beam waist, \(p\) is the radial index, \(\ell\) is the topological charge, and \(L_p^{|\ell|}\) is the associated Laguerre polynomial. 
In the numerical calculations, we mainly use the $p=0$ LG beam with a
beam waist of $W_0=10~\mu\mathrm{m}$.
The overall normalization factor is absorbed into the amplitude $A_0$
of the vector potential defined below.
For \(p=0\), the \(\ell=0\) beam gives a Gaussian reference beam, whereas beams with \(\ell\neq 0\) have a doughnut-shaped intensity profile with a zero at the beam center.
For the \(p=0\) mode, the radial intensity profile is proportional to
\begin{equation*}
    |u_{0\ell}(R)|^2
    \propto
    \left(
        \frac{2R^2}{W_0^2}
    \right)^{|\ell|}
    \exp
    \left(
        -\frac{2R^2}{W_0^2}
    \right).
\end{equation*}
The intensity reaches its maximum at
\begin{equation}
    R_\ell^{\max}
    =
    W_0\sqrt{\frac{|\ell|}{2}}.
    \label{eq:LG_intensity_maximum_radius}
\end{equation}
Hereafter, \(R_\ell^{\max}\) denotes the intensity-maximum radius of
the \(p=0\) LG beam with OAM index \(\ell\).
For \(\ell=0\), Eq.~\eqref{eq:LG_intensity_maximum_radius} gives
\(R_\ell^{\max}=0\).
At \(R=R_\ell^{\max}\), the radial derivative of the \(p=0\) LG
envelope vanishes.

The vector potential used in the numerical calculation is given by
\begin{equation}
    \bm{A}(R,\phi,t)
    =
    A_0 f(t) u_{p\ell}(R)
    \left[
        \bm{e}_x \cos(\omega_0 t+\ell\phi)
        +
        \epsilon \bm{e}_y \sin(\omega_0 t+\ell\phi)
    \right].
    \label{eq:LG_vector_potential}
\end{equation}
This vector potential and its spatial gradients are substituted into the coarse-grained Hamiltonian \(\hat{H}(\bm{X}_0,t)\) derived in the preceding sections, and the electronic state is propagated in time at each observation point
\begin{equation*}
    \bm{X}_0
    =
    R_0(\cos\phi,\sin\phi).
\end{equation*}
Here, $A_0$ sets the overall amplitude of the vector potential,
$f(t)$ is the temporal envelope, $\omega_0$ is the carrier frequency,
and $\epsilon$ is the polarization ellipticity.
In the numerical calculations, we use
$\omega_0/2\pi=90~\mathrm{THz}$.
The case $\epsilon=0$ corresponds to linearly polarized light along the
$x$ direction, whereas $|\epsilon|=1$ corresponds to circularly
polarized light.
For circular polarization, we define the helicity by
\begin{equation*}
    \sigma=\mathrm{sgn}(\epsilon),
\end{equation*}
so that the two choices \(\epsilon=\pm1\) correspond to the two helicities \(\sigma=\pm1\). 
With this convention, \(\sigma=+1\) and \(\sigma=-1\) label the two opposite senses of rotation of the optical field in the graphene plane. 
A different sign convention would interchange the two helicity labels without changing the physical correspondence between the polarization sense and the selected winding branch.

In comparing different polarization states, the field amplitude \(A_0\) is chosen such that the maximum electric-field amplitude \(E_0^{\max}\) is fixed. 
Thus, the differences between the linearly and circularly polarized cases discussed below originate from the polarization and helicity structure of the field, rather than from a trivial change in the overall field strength. 
The temporal envelope is taken to be
\begin{equation*}
f(t)
=
\exp\left[
-\frac{(t-t_0)^2}{\tau_p^2}
\right],
\qquad
t_0=\frac{24\pi}{\omega_0},
\qquad
\tau_p=\frac{4\pi}{\omega_0}.
\end{equation*}
In Eq.~\eqref{eq:LG_vector_potential}, the azimuthal phase \(\ell\phi\) of the LG beam is added to the carrier phase. 
Thus, the temporal oscillation of the optical field and its spatial winding structure are combined in the light--matter coupling.

\begin{figure}[t]
    \centering
    \includegraphics[width=0.8\linewidth]{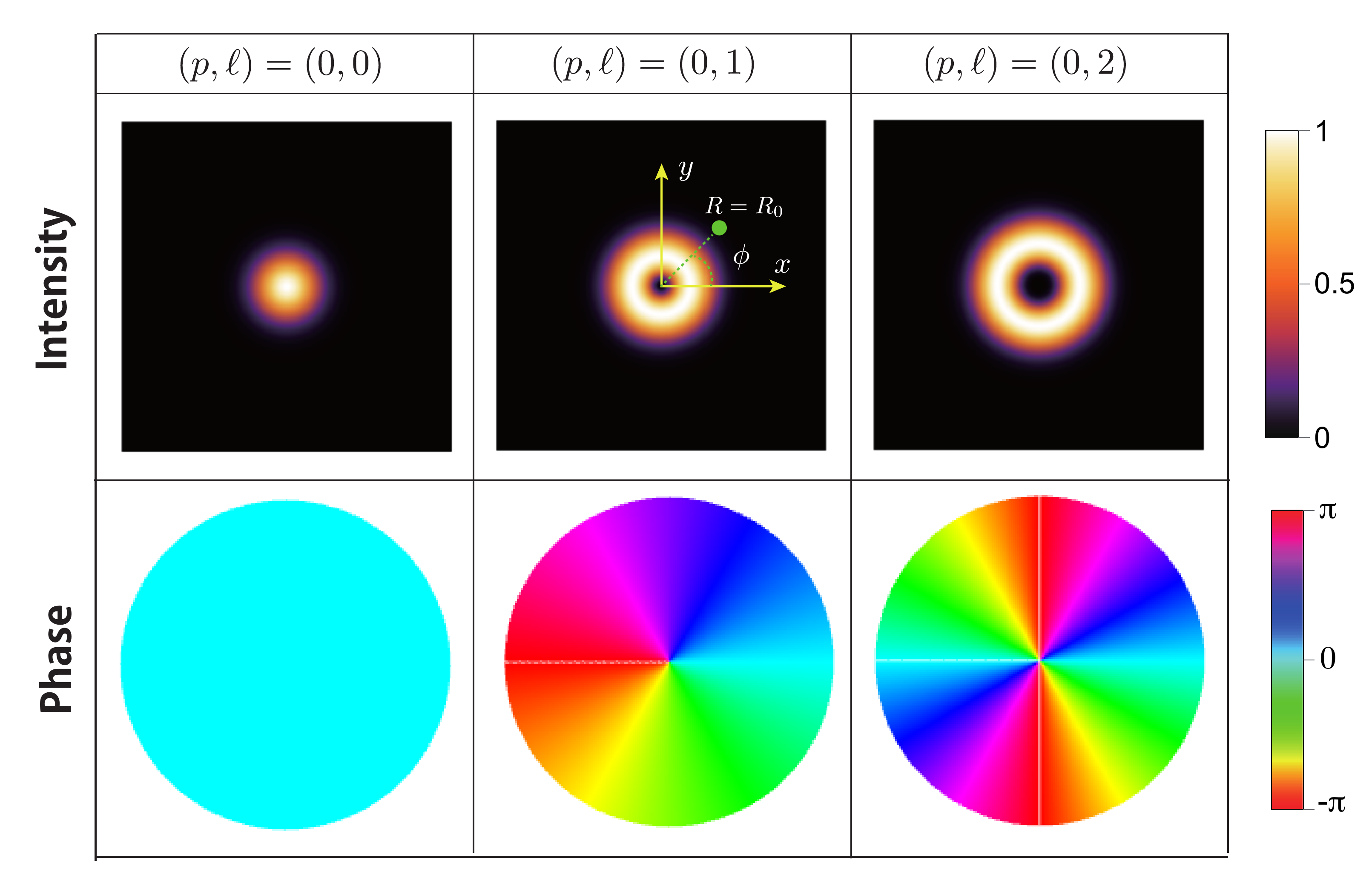}
    \caption{
    Spatial distributions of the LG optical fields used in the numerical calculations.
    The upper panels show the normalized intensity distributions, and the lower panels show the corresponding phase distributions.
    The three columns correspond to \((p,\ell)=(0,0)\), \((0,1)\), and \((0,2)\), respectively.
    The \(\ell=0\) field is a Gaussian reference field without azimuthal phase winding, whereas the \(\ell=1\) and \(\ell=2\) fields carry singly and doubly wound azimuthal phase structures, respectively.
    The schematic in the upper middle panel indicates the polar coordinates $(R,\phi)$ of the coarse-grained position $\bm{X}_0$ in the graphene plane and the observation circle $R=R_0$, on which the azimuthal dependence of the current is evaluated.
    Along this observation circle, the time-dependent current components \(J_x(\phi,t)\) and \(J_y(\phi,t)\) are computed and subsequently converted into the azimuthal current and the winding spectrum defined below.
    }
    \label{fig:LG_beam_profiles}
\end{figure}

Figure~\ref{fig:LG_beam_profiles} illustrates the spatial intensity
and phase profiles of the $p=0$ LG beams with $\ell=0,1,2$.
The observation circle $R=R_0$ is used to sample the azimuthal
dependence of the time-dependent current.
In the next subsection, we construct the azimuthal current from
$J_x(\phi,t)$ and $J_y(\phi,t)$ and define the corresponding winding
spectrum.

\subsection{Azimuthal current and winding spectrum}
\label{sec:winding_spectrum_definition}

In this subsection, we describe how the time-dependent current obtained
from the numerical calculation with the LG beam defined in
Sec.~\ref{sec:structured_light_field} is converted into the azimuthal
current and the corresponding winding spectrum.
The time-dependent Cartesian current components
\(J_x(\phi,t)\) and \(J_y(\phi,t)\) are evaluated at each
coarse-grained position on the observation circle \(R=R_0\), with
\begin{equation*}
    \bm{X}_0
    =
    R_0(\cos\phi,\sin\phi).
\end{equation*}

The radial and azimuthal unit vectors are
\begin{equation*}
    \bm{e}_R
    =
    \cos\phi\,\bm{e}_x
    +
    \sin\phi\,\bm{e}_y,
    \qquad
    \bm{e}_\theta
    =
    -\sin\phi\,\bm{e}_x
    +
    \cos\phi\,\bm{e}_y,
\end{equation*}
and the corresponding current components are
\begin{align*}
    J_R(\phi,t)
    &=
    J_x(\phi,t)\cos\phi
    +
    J_y(\phi,t)\sin\phi,
    \\
    J_\theta(\phi,t)
    &=
    -J_x(\phi,t)\sin\phi
    +
    J_y(\phi,t)\cos\phi.
\end{align*}
We focus on \(J_\theta(\phi,t)\), which represents the circulating
current component along the observation circle and is directly related
to the dc circulating current and its axial magnetic readout.

We extract the frequency components of \(J_\theta(\phi,t)\) by a
window-normalized temporal Fourier transform. For an integer \(q\), we define
\begin{equation*}
    \mathcal{J}_\theta(\phi,q\omega_0)
    =
    \frac{1}{\mathcal{N}_W}
    \int dt\,
    W(t)J_\theta(\phi,t)e^{iq\omega_0t},
\end{equation*}
where
\begin{equation*}
    \mathcal{N}_W
    =
    \int dt\,W(t).
\end{equation*}
This normalization ensures that
\(\mathcal{J}_\theta(\phi,q\omega_0)\) has the same physical dimension
as the time-domain current \(J_\theta(\phi,t)\).
In particular, the \(q=0\) component,
\(\mathcal{J}_\theta(\phi,0)\), is the Gaussian-window-weighted
time-averaged dc current.
We employ the Gaussian window
\begin{equation*}
W(t)
=
\exp\left[
-\frac{(t-t_0)^2}{\tau_w^2}
\right],
\qquad
t_0=\frac{24\pi}{\omega_0},
\qquad
\tau_w=\frac{4\pi}{\omega_0}.
\end{equation*}

The frequency-resolved azimuthal current is then Fourier decomposed
with respect to \(\phi\).
Discretizing the observation circle as
\begin{equation*}
    \phi_n
    =
    \frac{2\pi n}{N_\phi},
    \qquad
    n=0,1,\ldots,N_\phi-1,
\end{equation*}
we define the signed azimuthal Fourier coefficient
\begin{equation*}
    \Gamma_{q,\mu}
=
\frac{2\pi R_0}{N_\phi}
\sum_{n=0}^{N_\phi-1}
\mathcal{J}_\theta(\phi_n,q\omega_0)e^{-i\mu\phi_n},
\qquad
\mu\in\mathbb{Z}.
\end{equation*}
Here, \(q\) is the temporal Fourier index and \(\mu\) is the signed
azimuthal harmonic.
Thus, \(\Gamma_{q,\mu}\) retains the signed angular phase information
of the azimuthal current on the observation circle.

In the following, we focus on the zero-frequency component and use the
abbreviation
\begin{equation*}
    \Gamma_\mu
    \equiv
    \Gamma_{q=0,\mu}.
\end{equation*}
Because $\mathcal{J}_\theta(\phi,0)$ is real,
\begin{equation*}
    \Gamma_{-\mu}
    =
    \Gamma_\mu^*,
    \qquad
    |\Gamma_{-\mu}|^2
    =
    |\Gamma_\mu|^2.
\end{equation*}
We therefore define the nonnegative winding order
\begin{equation*}
    m\equiv|\mu|\geq0
\end{equation*}
and denote the corresponding winding power by
\begin{equation*}
    |\Gamma_m|^2
    \equiv
    |\Gamma_{\mu=m}|^2
    =
    |\Gamma_{\mu=-m}|^2.
\end{equation*}
Accordingly, the winding-power spectra shown below resolve the order
\(m\), whereas the signed harmonic \(\mu\) is used in the analytical
angular-factor tracking.

The same decomposition is applied to each current channel.
For $a=A1,A2,B1,B2$, the channel-resolved azimuthal current is
\begin{equation*}
J_{\theta}^{(a)}(\phi,t)
=
-J_x^{(a)}(\phi,t)\sin\phi
+
J_y^{(a)}(\phi,t)\cos\phi.
\end{equation*}
Applying the window-normalized temporal Fourier transform defined above,
the corresponding signed zero-frequency coefficient is
\begin{equation*}
\Gamma_{\mu}^{(a)}
=
\frac{2\pi R_0}{N_\phi}
\sum_{n=0}^{N_\phi-1}
\mathcal{J}_{\theta}^{(a)}(\phi_n,0)
e^{-i\mu\phi_n}.
\end{equation*}
The corresponding channel-resolved winding power is denoted by
$|\Gamma_m^{(a)}|^2$.
This operator-resolved spectrum identifies the winding orders carried
by the current channels belonging to the local and gradient sectors.

\subsection{Role of the local \texorpdfstring{$\bm{A}^2$}{A squared}
diamagnetic contribution}
\label{sec:J3_role}

We now clarify the role of the local $\bm{A}^2$ diamagnetic
contribution $J_3$.
The corresponding $B1$ interaction is
\begin{equation*}
    \hat{H}_I^{B1}
    =
    \frac{e^2}{2m_0c^2}
    \bm{A}^2(\bm{X}_0,t)\,
    \hat{n}(\bm{X}_0),
\end{equation*}
where $\hat{n}(\bm{X}_0)$ is the local particle-number operator.
In the present momentum-diagonal formulation, different Bloch momenta
are not mixed, and the unitary time evolution preserves the total
occupation at each $\bm{k}$.
Consequently, the $B1$ interaction acts only as a scalar energy shift
within each $\bm{k}$ sector, and its associated current is proportional
to the vector potential,
\begin{equation*}
    J_i^{B1}(\bm{X}_0,t)
    \propto
    A_i(\bm{X}_0,t),
    \qquad i=x,y.
\end{equation*}
Thus, $J_3$ follows the driving field linearly and does not generate an
independent nonlinear dc response.
Accordingly, it is not regarded as an independent channel governing
the dc winding selection rules.
The channel-resolved numerical calculation likewise shows no relevant
$J_3$ contribution to the dc winding spectrum.

\subsection{Sector-resolved winding selection rules}
\label{sec:sector_resolved_selection_rules}

In this subsection, we derive the sector-resolved winding selection
rules from the azimuthal dependence of the current components.
We first relate the azimuthal harmonics of the Cartesian current
components $J_x$ and $J_y$ to those of the projected azimuthal current
$J_\theta$.
We then analyze the carrier-envelope-phase (CEP) dependence of $J_x$
and $J_y$ before the azimuthal projection as an additional diagnostic
of the finite-pulse response.
Using the channel classification summarized in
Sec.~\ref{sec:current_channel_summary}, we refer to the local
$\bm{A}\cdot\bm{p}$ contribution $J_1$ as the local sector, whereas
the gradient sector comprises the $A2$-type gradient
$\bm{A}\cdot\bm{p}$ contribution $J_2$ and the $B2$-type
$\bm{A}\nabla\bm{A}$ contribution $J_4$.
In the following, we take $\ell>0$ and track the component proportional
to $e^{+i\ell\phi}$; its reality-related $e^{-i\ell\phi}$ counterpart
gives the same nonnegative winding order.

\subsubsection{Derivation from azimuthal harmonics}
\label{sec:azimuthal_harmonics}

At a fixed observation radius $R_0$, solving the time-dependent electronic
dynamics yields the Cartesian current components $J_x(\phi,t)$ and
$J_y(\phi,t)$ as functions of the azimuthal angle $\phi$.
Using the same window-normalized temporal Fourier convention as in
Sec.~III B, we denote their zero-frequency components by
$\mathcal{J}_\alpha(\phi,0)$ and decompose them into azimuthal Fourier
harmonics,
\begin{equation}
\mathcal{J}_\alpha(\phi,0)
=
\sum_{N\in\mathbb{Z}}
\mathcal{J}_{\alpha,N}e^{iN\phi},
\qquad
\alpha=x,y.
\label{eq:cartesian_fourier}
\end{equation}
We refer to $N$ as the signed pre-projection azimuthal harmonic.
Thus, $N$ labels the angular Fourier components of the Cartesian
current obtained from the electronic dynamics, rather than an
additional quantum number introduced independently.
The relations associated with the azimuthal projection and spatial
derivatives are derived algebraically below, whereas the remaining
properties of the pre-projection current are obtained from the
time-dependent electronic dynamics.
The following analysis determines how these pre-projection harmonics
$N$ are mapped onto the winding harmonics of the azimuthal current
$\mathcal{J}_\theta$.
The Cartesian current is projected onto the azimuthal direction as

\begin{equation}
 \mathcal{J}_\theta(\phi,0)
=
-\mathcal{J}_x(\phi,0)\sin\phi
+
\mathcal{J}_y(\phi,0)\cos\phi.
 \label{eq:azimuthal_projection}
\end{equation}
Since $\sin\phi$ and $\cos\phi$ contain angular harmonics
$e^{\pm i\phi}$, each pre-projection harmonic $N$ contributes to the
signed projected harmonics

\begin{equation*}
 \mu=N\pm1,
\end{equation*}
with the corresponding winding order

\begin{equation*}
 m=|\mu|.
\end{equation*}
We first consider linearly polarized light.  In the local sector, the
leading Cartesian-current harmonics inherit the azimuthal index of the
LG field, since the local current operator introduces no additional
spatial-derivative angular factor.  The leading pre-projection Fourier
components therefore occur at

\begin{equation*}
 N=\ell .
\end{equation*}
The azimuthal projection then gives

\begin{equation*}
 m_{\rm loc}=|\ell\pm1|.
\end{equation*}
In the gradient sector, by contrast, the current operator contains a
spatial derivative, which introduces an additional angular factor.
In polar coordinates,

\begin{equation*}
 \partial_X
 =
 \cos\phi\,\partial_R
 -
 \frac{\sin\phi}{R}\partial_\phi,
 \qquad
 \partial_Y
 =
 \sin\phi\,\partial_R
 +
 \frac{\cos\phi}{R}\partial_\phi .
\end{equation*}
The explicit $\sin\phi$ and $\cos\phi$ factors shift the azimuthal
index inherited from the LG field by one unit.  The leading
pre-projection Fourier components therefore occur at

\begin{equation}
 N=\ell\pm1 .
 \label{eq:gradient_preprojection_branches}
\end{equation}
The subsequent azimuthal projection produces one further unit of
angular shift, giving

\begin{equation*}
 m_{\rm grad}
 =
 |\ell|,\ |\ell\pm2|.
\end{equation*}
This counting applies to both the first-gradient $A2$ channel $J_2$ and
the $B2$ channel $J_4$.  Although their microscopic origins are
different, the angular structure relevant to the winding classification
contains the same one-unit shift associated with a spatial derivative
in both channels.

We next consider circular polarization.  For the local sector, changing from linear to circular polarization does not alter the pre-projection azimuthal harmonic.  Since no additional
angular shift is introduced before the azimuthal projection, the leading
harmonic remains
\begin{equation*}
 N=\ell .
\end{equation*}
Thus, the difference between linear and circular polarization does not
arise from the pre-projection harmonic $N$, but from the relative phase between the Fourier coefficients
$\mathcal{J}_{x,N}$ and $\mathcal{J}_{y,N}$.
To make this explicit, we introduce

\begin{equation*}
 \mathcal{J}_{+,N}
=
\mathcal{J}_{x,N}+i\mathcal{J}_{y,N},
\qquad
\mathcal{J}_{-,N}
=
\mathcal{J}_{x,N}-i\mathcal{J}_{y,N}.
\end{equation*}
Substituting the Cartesian-current Fourier expansion,
Eq.~\eqref{eq:cartesian_fourier}, into the azimuthal projection,
Eq.~\eqref{eq:azimuthal_projection}, and rewriting the result in terms of
$\mathcal{J}_{+,N}$ and $\mathcal{J}_{-,N}$ gives

\begin{equation*}
 \mathcal{J}_{\theta}(\phi,0)
=
\frac{1}{2i}
\sum_{N\in\mathbb{Z}}
\left[
\mathcal{J}_{+,N}e^{i(N-1)\phi}
-
\mathcal{J}_{-,N}e^{i(N+1)\phi}
\right].
\end{equation*}
This expression shows directly that the
$\mathcal{J}_{+,N}$ and $\mathcal{J}_{-,N}$ components are mapped by the azimuthal projection onto

\begin{equation*}
 \mu=N-1
 \qquad\text{and}\qquad
 \mu=N+1,
\end{equation*}
respectively.
For linear polarization, both components are present at
the leading local harmonic $N=\ell$, giving
$\mu=\ell\pm1$.
For circular polarization with helicity $\sigma=\pm1$, the
pre-projection harmonic remains $N=\ell$, while the corresponding
current obtained from the time-dependent dynamics satisfies

\begin{equation*}
 \mathcal{J}_{y,\ell}
=
-i\sigma\mathcal{J}_{x,\ell}.
\end{equation*}
It follows that

\begin{equation*}
 \mathcal{J}_{+,\ell}
=
(1+\sigma)\mathcal{J}_{x,\ell},
\qquad
\mathcal{J}_{-,\ell}
=
(1-\sigma)\mathcal{J}_{x,\ell}.
\end{equation*}
Thus, only $\mathcal{J}_{+,\ell}$ contributes for $\sigma=+1$,
whereas only $\mathcal{J}_{-,\ell}$ contributes for $\sigma=-1$.  The azimuthal projection
therefore gives

\begin{equation*}
 \mu_{\rm loc}
 =
 \ell-\sigma,
 \qquad
 m_{\rm loc}
 =
 |\ell-\sigma|.
\end{equation*}

For the gradient sector, we introduce the circular derivative

\begin{equation*}
 \partial_\sigma
 =
 \partial_X+i\sigma\partial_Y
 =
 e^{i\sigma\phi}
 \left(
 \partial_R+\frac{i\sigma}{R}\partial_\phi
 \right),
 \qquad
 \sigma=\pm1 .
\end{equation*}
Acting on the LG angular factor gives

\begin{equation*}
 \partial_\sigma
 \left[
 F(R)e^{i\ell\phi}
 \right]
 =
 e^{i(\ell+\sigma)\phi}
 \left(
 \partial_R-\frac{\sigma\ell}{R}
 \right)F(R).
\end{equation*}
Taking $\sigma=\pm1$ reproduces the two derivative-generated
pre-projection branches $N=\ell\pm1$ obtained in
Eq.~\eqref{eq:gradient_preprojection_branches}.
For circular polarization, $\sigma$ is fixed by the physical helicity, and the
helicity-resolved time-dependent dynamics selects the corresponding
branch,

\begin{equation*}
 N=\ell+\sigma .
\end{equation*}
At this selected harmonic, the helicity-resolved gradient current
obtained from the time-dependent dynamics satisfies

\begin{equation*}
 \mathcal{J}_{y,N}^{\rm grad}
=
i\sigma\mathcal{J}_{x,N}^{\rm grad}.
\end{equation*}
It follows that

\begin{equation*}
 \mathcal{J}_{+,N}^{\rm grad}
=
(1-\sigma)\mathcal{J}_{x,N}^{\rm grad},
\qquad
\mathcal{J}_{-,N}^{\rm grad}
=
(1+\sigma)\mathcal{J}_{x,N}^{\rm grad}.
\end{equation*}
Thus, only $\mathcal{J}^{\rm grad}_{-,N}$ contributes for $\sigma=+1$,
whereas only $\mathcal{J}^{\rm grad}_{+,N}$ contributes for $\sigma=-1$. Using the
azimuthal-projection mapping discussed above, the selected gradient
component therefore gives

\begin{equation*}
 \mu_{\rm grad}
 =
 N+\sigma
 =
 \ell+2\sigma,
 \qquad
 m_{\rm grad}
 =
 |\ell+2\sigma|.
\end{equation*}

The leading sector-resolved winding rules are therefore

\begin{equation*}
\begin{aligned}
 \text{linear polarization:}\qquad
 & m_{\rm loc}=|\ell\pm1|,
 &
 & m_{\rm grad}=|\ell|,\ |\ell\pm2|,
 \\[2pt]
 \text{circular polarization:}\qquad
 & m_{\rm loc}=|\ell-\sigma|,
 &
 & m_{\rm grad}=|\ell+2\sigma|.
\end{aligned}
\end{equation*}
The local circular rule also has a simple real-space interpretation.
Using the circular-current components introduced above, the radial and
azimuthal currents can be written as
\begin{equation*}
\mathcal{J}_R
=
\frac{1}{2}
\left(
\mathcal{J}_+e^{-i\phi}
+
\mathcal{J}_-e^{i\phi}
\right),
\qquad
\mathcal{J}_\theta
=
\frac{1}{2i}
\left(
\mathcal{J}_+e^{-i\phi}
-
\mathcal{J}_-e^{i\phi}
\right).
\end{equation*}
For a definite helicity $\sigma$, the selected local component therefore
gives the common azimuthal dependence
\begin{equation*}
\mathcal{J}_R^{\rm loc},
\,
\mathcal{J}_\theta^{\rm loc}
\propto
e^{i(\ell-\sigma)\phi}.
\end{equation*}
Hence, when $\ell=\sigma$, both the radial and azimuthal current
components are independent of $\phi$, so that the local current texture
is rotationally symmetric in the polar basis. This provides a direct
real-space interpretation of the $m=0$ local-sector branch.

More generally, the sector-resolved winding rules derived above follow
from the sector-dependent azimuthal structure of the pre-projection
Cartesian current together with the azimuthal projection. The full
lattice retains these leading winding channels, while its discrete
$C_6$ symmetry can additionally allow symmetry-related angular
sidebands. The winding selection rules are therefore not determined by
the optical OAM alone, but by its coupling to the angular structure of
the current in each interaction sector.

\subsubsection{CEP-harmonic characterization of the pre-projection components}
\label{sec:cep_dependence_preprojection}

As an additional numerical diagnostic of the finite-pulse response, we analyze the CEP dependence of the signed pre-projection azimuthal harmonics $N$ of the zero-frequency Cartesian current. We restrict this analysis to linearly polarized driving, for which the relation between the CEP harmonic and the pre-projection azimuthal harmonic $N$ can be examined directly.

Let $\delta$ denote the CEP, and write the window-normalized
zero-frequency Cartesian current evaluated at a given CEP as
$J_\alpha^{\rm dc}(\phi;\delta)$.
For each Cartesian component $\alpha=x,y$, the coefficient of the
signed pre-projection azimuthal harmonic $N$ is then

\begin{equation*}
 J_{\alpha,N}^{\rm dc}(\delta)
 =
 \frac{1}{2\pi}
 \int_0^{2\pi}
 J_\alpha^{\rm dc}(\phi;\delta)
 e^{-iN\phi}\,d\phi ,
 \qquad
 \alpha=x,y .
\end{equation*}
Here, $N$ is the signed pre-projection azimuthal harmonic introduced
in Sec.~\ref{sec:azimuthal_harmonics}.
In the gradient sector, the angular factors associated with the
spatial derivatives entering the definition of the gradient-sector
current are already contained in the resulting Cartesian current and
are therefore encoded in the pre-projection harmonic $N$.
We next resolve each pre-projection component with respect to the CEP.
The corresponding CEP-Fourier coefficient is defined as

\begin{equation*}
 C_{\alpha,N,\nu}
 =
 \frac{1}{2\pi}
 \int_0^{2\pi}
 J_{\alpha,N}^{\rm dc}(\delta)
 e^{-i\nu\delta}\,d\delta .
\end{equation*}
Here, $\nu$ is the Fourier index conjugate to the CEP $\delta$.
It is distinct from the pre-projection azimuthal harmonic $N$, the
projected signed harmonic $\mu$, and the winding order $m$.

The CEP-Fourier power associated with a given pair $(N,\nu)$ is
defined by

\begin{equation*}
 P_{N,\nu}
 =
 |C_{x,N,\nu}|^2
 +
 |C_{y,N,\nu}|^2 .
\end{equation*}
Because $J_\alpha^{\rm dc}(\phi;\delta)$ is real, the Fourier
coefficients satisfy

\begin{equation*}
 C_{\alpha,-N,\nu}
 =
 \left[
 C_{\alpha,N,-\nu}
 \right]^* ,
\end{equation*}
and hence

\begin{equation*}
 P_{-N,\nu}
 =
 P_{N,-\nu}.
\end{equation*}
Accordingly, we present the spectra in terms of the nonnegative
indices $|N|$ and $|\nu|$ by folding the symmetry-related sign
combinations.  Each $|N|$ spectrum is normalized independently to
its maximum over $|\nu|$.

\begin{figure*}[ht!]
  \centering
  \includegraphics[width=\textwidth]
  {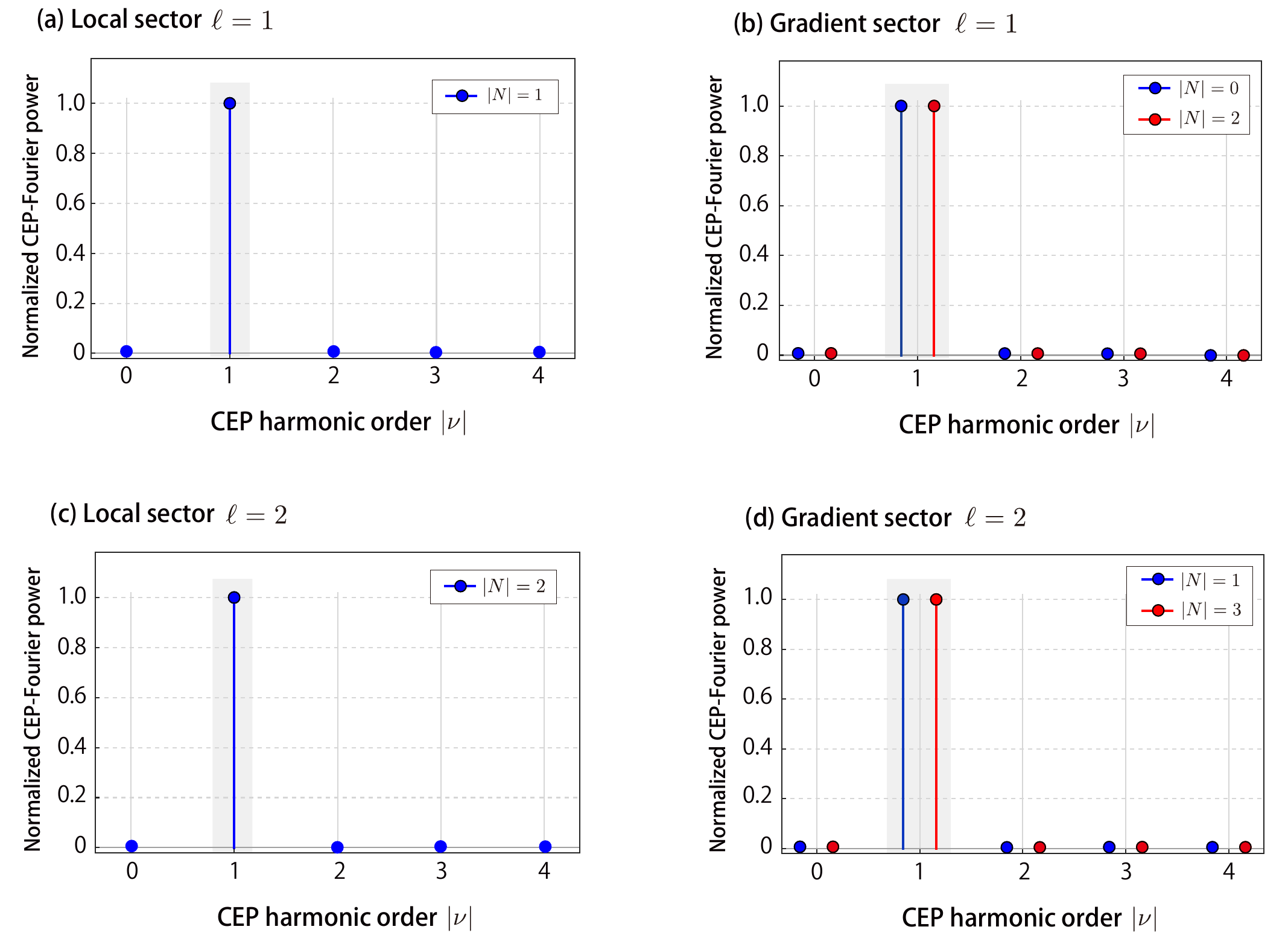}
  \caption{
CEP-Fourier spectra of the selected zero-frequency pre-projection
Cartesian-current components for linearly polarized LG pulses.
The displayed components are the pre-projection branches identified
by the leading angular classification:
$|N|=|\ell|$ in the local sector and
$|N|\in\{|\ell-1|,|\ell+1|\}$ in the gradient sector.
The spectra are evaluated at
$E_0^{\rm max}=5.6$~MV/cm and at the intensity-maximum radius
$R_\ell^{\rm max}$.
Panels (a) and (b) show the local and gradient sectors for
$\ell=1$, respectively, while panels (c) and (d) show the
corresponding sectors for $\ell=2$.
For each selected $|N|$ component, contributions from the different
sign combinations of $N$ and $\nu$ are folded together, and the
resulting CEP-Fourier power is normalized independently by its
maximum over $|\nu|$.
The gray shaded region indicates the fundamental CEP harmonic,
$|\nu|=1$.
In panels containing two pre-projection components, the sticks are
displaced slightly along the horizontal direction solely for
visibility; both correspond to the same integer CEP-harmonic order.
All calculated CEP harmonics in the displayed range
$0\le|\nu|\le4$ are shown.
In every case examined here, the spectrum is dominated by the
fundamental CEP harmonic $|\nu|=1$.
}
  \label{fig:cepft_preprojection}
\end{figure*}

The pre-projection angular classification derived above gives

\begin{equation*}
 |N|=|\ell|
 \qquad
 \text{in the local sector},
\end{equation*}
and

\begin{equation*}
 |N|
 \in
 \left\{
 |\ell-1|,
 |\ell+1|
 \right\}
 \qquad
 \text{in the gradient sector}.
\end{equation*}
Keeping the signs of both Fourier indices, the corresponding
CEP--azimuthal correlations are

\begin{equation*}
 N=-\nu\ell
 \qquad
 \text{in the local sector},
\end{equation*}
and

\begin{equation*}
 N=-\nu\ell\pm1
 \qquad
 \text{in the gradient sector}.
\end{equation*}

Figure~\ref{fig:cepft_preprojection} shows the corresponding
CEP-Fourier spectra for $\ell=1$ and $2$.  
For $\ell=1$, the local-sector response is examined in the $|N|=1$ component, whereas
the gradient-sector response is resolved into the $|N|=0$ and $2$
components.  
For $\ell=2$, the corresponding components are $|N|=2$ in the local sector and $|N|=1$ and $3$ in the gradient sector.
These are precisely the pre-projection angular branches obtained
from the angular classification in Sec.~\ref{sec:azimuthal_harmonics}.
For every selected component shown in Fig.~\ref{fig:cepft_preprojection}, the CEP-Fourier power
is dominated by the fundamental harmonic $|\nu|=1$, while the higher CEP harmonics remain close to the baseline on the displayed linear scale.  
This behavior is found in both the local and gradient sectors and for both incident OAM indices considered here.  Although the calculation includes the full nonlinear electronic dynamics at
$E_0^{\rm max}=5.6$~MV/cm, the CEP dependence of the selected pre-projection components is governed predominantly by the fundamental CEP harmonic.
The dominant $|\nu|=1$ response therefore provides a numerical check
of the pre-projection angular structure identified in
Sec.~\ref{sec:azimuthal_harmonics} and is consistent with the signed
relations $N=-\nu\ell$ in the local sector and
$N=-\nu\ell\pm1$ in the gradient sector.

\subsection{Additional numerical results}
\label{sec:additional_numerical_results}

In this section, we present additional numerical results that further support the sector-resolved winding selection rules discussed in the main text.  
Specifically, we analyze the dc winding spectra resolved by current channel, the helicity-resolved channel decomposition, the field-amplitude dependence of the selected $m=0$ zero-frequency
component, the dependence of the selected $m=0$ component on the observation radius $R_0/W_0$, and the comparison of the $B2$-type channel with an intensity-gradient reference profile.

For the current-channel and helicity-resolved decompositions, we fix
the observation radius at $R_0=R_\ell^{\max}$, where
$R_\ell^{\max}$ is the intensity-maximum radius defined in
Eq.~\eqref{eq:LG_intensity_maximum_radius}.
For the $p=0$ LG mode considered here, the radial derivative of the
beam envelope vanishes at $R_0=R_\ell^{\max}$.
As a result, the radial-gradient contribution is suppressed, and the
gradient-sector response predominantly reflects the azimuthal-gradient
contribution.
Figures~\ref{fig:current_channel_decomposition} and
\ref{fig:helicity_resolved_channel_decomposition} therefore provide
a clean diagnostic of the angular structure underlying the winding
selection rules in the local and gradient sectors.
The role of radial-gradient contributions away from
$R_0=R_\ell^{\max}$ is examined separately below through the
radial-position dependence.  Away from the intensity maximum,
radial-gradient contributions can mix with the azimuthal-gradient
response and quantitatively modify the selected winding components.
As shown below, however, these contributions remain secondary and do
not alter the leading winding selection.  Thus, the results at
$R_0=R_\ell^{\max}$ should be regarded as a clean diagnostic of the
angular selection rules rather than as a restriction on the possible
observation geometry.

The results presented below numerically support the central conclusion
that the local and gradient sectors obey different winding selection
rules, while the two gradient-sector channels, \(H_I^{A2}\) and \(H_I^{B2}\),
share the same leading winding selection rule.

\subsubsection{Current-channel decomposition of dc winding spectra}
\label{sec:additional_channel_decomposition}

\begin{figure}[t]
    \centering
    \includegraphics[width=1.0\linewidth]{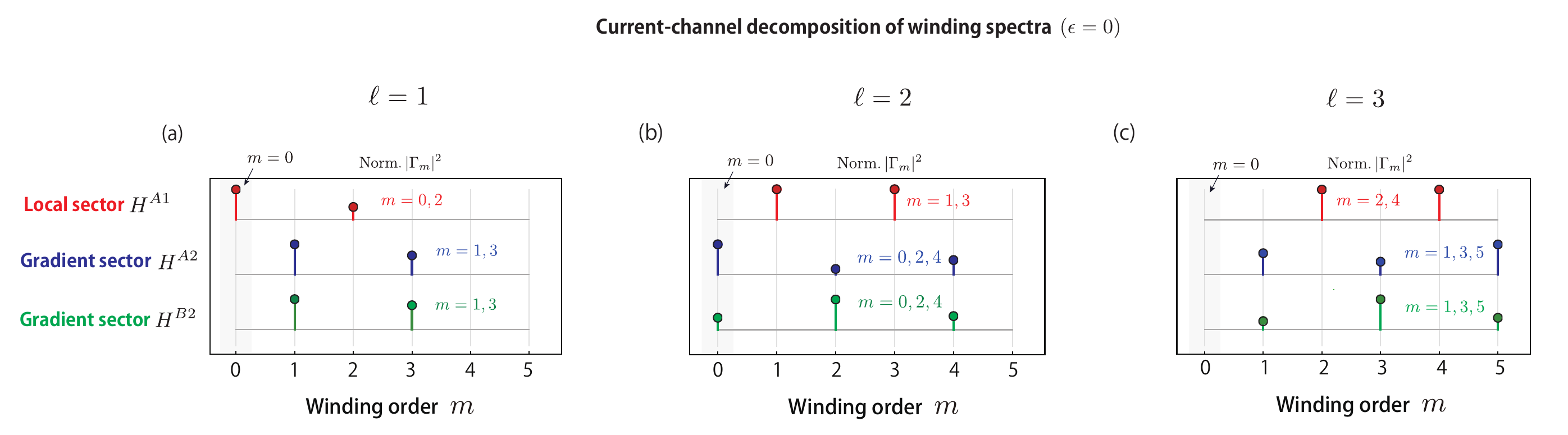}
\caption{
    Current-channel decomposition of the winding spectra under linearly polarized excitation with \(\epsilon=0\).
    The three panels show the results for \(\ell=1\), \(\ell=2\), and \(\ell=3\), respectively.
    The spectra are evaluated at \(E_0^{\max}=5.6~\mathrm{MV/cm}\) and at the
    intensity-maximum radius \(R_0=R_\ell^{\max}\).
    The red bars correspond to the local-sector \(H_I^{A1}\) channel,
    the blue bars to the gradient-sector \(H_I^{A2}\) channel,
    and the green bars to the gradient-sector \(H_I^{B2}\) channel.
    For each channel, the normalized winding spectrum \(|\Gamma_m|^2\) is plotted as a function of the winding order \(m\).
    The local and gradient sectors activate different winding orders, whereas the \(H_I^{A2}\) and \(H_I^{B2}\) channels obey the same gradient-sector selection rule.
    The light-gray shaded region marks the \(m=0\) channel.
}
    \label{fig:current_channel_decomposition}
\end{figure}

Here, we decompose the dc winding spectrum into the contributions from
the individual current channels.  This decomposition allows us to
examine whether the local and gradient sectors obey distinct winding
selection rules and whether the two channels within the gradient sector
follow the same rule.
Figure~\ref{fig:current_channel_decomposition} shows the winding spectra
for LG beams with $\ell=1,2,3$ under linearly polarized excitation,
$\epsilon=0$.  The spectra are decomposed into the
local-sector $H_I^{A1}$ channel, the gradient-sector $H_I^{A2}$ channel,
and the gradient-sector $H_I^{B2}$ channel.

The result can be understood in terms of the winding selection rules for linearly polarized light. 
For the local-sector channel \(H_I^{A1}\), the selected winding orders are given by
\begin{equation*}
    m=|\ell\pm1|.
\end{equation*}
In contrast, for the gradient-sector channels \(H_I^{A2}\) and \(H_I^{B2}\), the selected winding orders are given by
\begin{equation*}
    m=|\ell|,\ |\ell\pm2|.
\end{equation*}
The numerical results in
Fig.~\ref{fig:current_channel_decomposition} are consistent with these
selection rules.  In the local sector, the dominant components occur
at $m=0,2$ for $\ell=1$, at $m=1,3$ for $\ell=2$, and at $m=2,4$
for $\ell=3$.  In the gradient sector, the corresponding dominant
components occur at $m=1,3$ for $\ell=1$, at $m=0,2,4$ for $\ell=2$,
and at $m=1,3,5$ for $\ell=3$.

A particularly important point is that \(H_I^{A2}\) and \(H_I^{B2}\) select the same winding orders.
These two channels have distinct microscopic origins:
\(H_I^{A2}\) arises from the gradient correction to the
\(\bm{A}\cdot\bm{p}\)-type interaction, whereas \(H_I^{B2}\) originates
from the \(B2\)-type \(\bm{A}\nabla\bm{A}\) interaction.
Despite this difference, both channels are governed by spatial
gradients of the optical field and therefore belong to the same
gradient sector.
Consequently, they obey the same gradient-sector winding selection rule.
Thus, Fig.~\ref{fig:current_channel_decomposition} demonstrates that
the winding selection rule is organized primarily by whether a given
current channel belongs to the local or gradient sector, rather than
by its detailed microscopic origin.  The local and gradient sectors
select different winding components, whereas the two gradient-sector
channels exhibit the same leading winding selection rule.  This result
supports the sector-resolved winding classification discussed in the
main text.

\subsubsection{Helicity-resolved channel decomposition}
\label{sec:additional_helicity_resolved_decomposition}

\begin{figure}[t]
    \centering
    \includegraphics[width=0.95\linewidth]{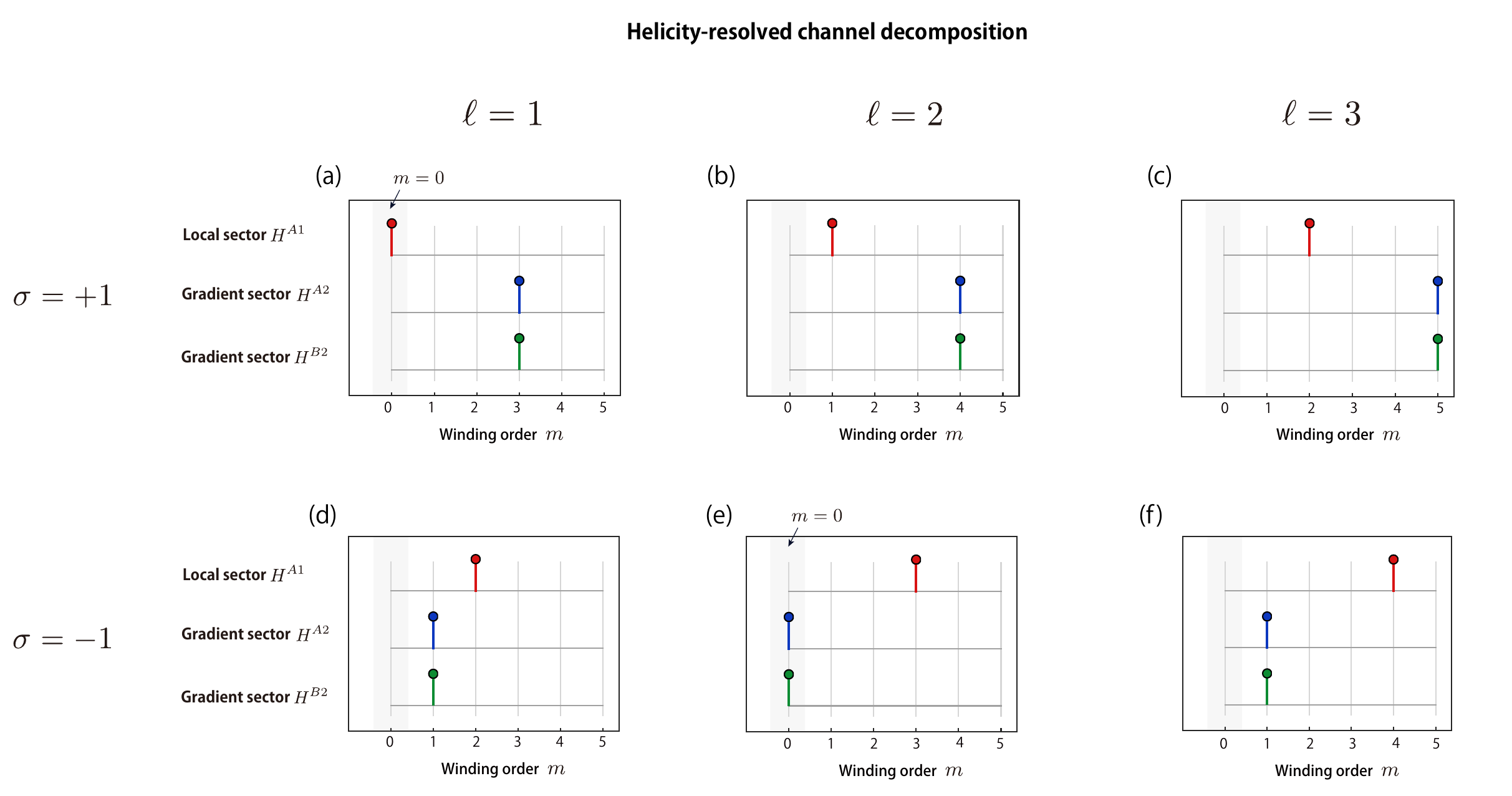}
\caption{
    Current-channel decomposition of the helicity-resolved winding spectra.
    The upper row shows the results for \(\sigma=+1\), and the lower row shows the results for \(\sigma=-1\).
    The three columns correspond to \(\ell=1\), \(\ell=2\), and \(\ell=3\), respectively.
    The spectra are evaluated at \(E_0^{\max}=5.6~\mathrm{MV/cm}\) and at the
    intensity-maximum radius \(R_0=R_\ell^{\max}\).
    The red bars represent the local-sector \(H_I^{A1}\) channel,
    the blue bars represent the gradient-sector \(H_I^{A2}\) channel,
    and the green bars represent the gradient-sector \(H_I^{B2}\) channel.
    Upon reversing the helicity, the selected winding orders in the local and gradient sectors change according to different selection rules.
    In particular, the \(H_I^{A2}\) and \(H_I^{B2}\) channels select the same winding order for a given helicity, demonstrating that channels belonging to the same gradient sector obey a common selection rule.
    The light-gray shaded region marks the \(m=0\) channel.
}
    \label{fig:helicity_resolved_channel_decomposition}
\end{figure}

Here, we examine how the winding selection rules of the local and
gradient sectors differ when the response is resolved with respect to
the helicity of circularly polarized light.
Figure~\ref{fig:helicity_resolved_channel_decomposition} shows the
current-channel-resolved winding spectra for $\ell=1,2,3$, separately
for the two helicities $\sigma=+1$ and $\sigma=-1$.
The red bars represent the local-sector $H_I^{A1}$ channel, the blue
bars represent the gradient-sector $H_I^{A2}$ channel, and the green
bars represent the gradient-sector $H_I^{B2}$ channel.

The key observation from
Fig.~\ref{fig:helicity_resolved_channel_decomposition} is that the
local and gradient sectors obey different helicity-dependent winding
selection rules.  For the local sector, the selected winding order is
given by

\begin{equation*}
    \mathrm{local\ sector}:
    \qquad
    m=|\ell-\sigma|.
\end{equation*}
Thus, for $\sigma=+1$, the selected winding orders are
$m=0,1,2$ for $\ell=1,2,3$, respectively, whereas for
$\sigma=-1$, they are $m=2,3,4$.  This behavior shows that, in the
local sector, the helicity selects one of the two branches present
under linearly polarized excitation.
In contrast, the gradient sector follows the selection rule

\begin{equation*}
    \mathrm{gradient\ sector}:
    \qquad
    m=|\ell+2\sigma|.
\end{equation*}
As discussed in Sec.~\ref{sec:azimuthal_harmonics}, this rule reflects
the combined effect of the derivative-induced angular shift in the
pre-projection current and the subsequent azimuthal projection.
Accordingly, for $\sigma=+1$, the selected winding orders are
$m=3,4,5$ for $\ell=1,2,3$, respectively, whereas for
$\sigma=-1$, they are $m=1,0,1$.  The gradient sector therefore
exhibits a helicity-dependent winding selection rule distinct from
that of the local sector.

A particularly important point is that $H_I^{A2}$ and $H_I^{B2}$
select the same winding order for a fixed helicity.
These two channels have different microscopic origins: the former
originates from the $A2$-type gradient $\bm{A}\cdot\bm{p}$
interaction, whereas the latter originates from the $B2$-type
$\bm{A}\nabla\bm{A}$ interaction.
Nevertheless, their angular structures relevant to the winding
classification contain the same derivative-induced shift and therefore
place both channels in the gradient sector.
Consequently, even in the helicity-resolved case, they obey the same
gradient-sector selection rule,
$m=|\ell+2\sigma|$.
Thus, Fig.~\ref{fig:helicity_resolved_channel_decomposition}
demonstrates that the helicity-dependent winding selection is organized
primarily by the sector structure of the response rather than by the
microscopic details of the individual channels.  The local and gradient
sectors obey distinct selection rules, whereas the two gradient-sector
channels, $H_I^{A2}$ and $H_I^{B2}$, follow the same leading rule.
This result further supports the sector-resolved winding classification
discussed in the main text.

\subsubsection{\texorpdfstring{
Field-amplitude reference for interpreting the radial profiles of the
selected $m=0$ zero-frequency components
}{
Field-amplitude reference for interpreting the radial profiles of the
selected m=0 zero-frequency components
}}
\label{sec:field_amplitude_dependence_m0_current}

Here, we examine the field-amplitude dependence of the selected
$m=0$ zero-frequency components at the intensity-maximum radius
$R_0=R_\ell^{\max}$.  This provides a reference for interpreting the
radial profiles of $\Gamma_{m=0}(R_0)$ shown in
Figs.~3(b) and 3(c) of the main text for the local and gradient
sectors, respectively.
At fixed incident peak amplitude $E_0^{\max}$, varying the observation
radius changes the local field amplitude sampled along the observation
circle.
In the local sector, this provides the primary connection between the
field-amplitude dependence and the radial profile.
In the gradient sector, the same field-amplitude dependence provides a
reference for the radial variation of the azimuthal-gradient
contribution.
Away from $R_0=R_\ell^{\max}$, an additional radial-gradient
contribution appears, and the sum of the two gradient contributions
determines the physical gradient-sector current profile and the
associated axial magnetic field.

\begin{figure}[t]
    \centering
    \includegraphics[width=0.95\linewidth]{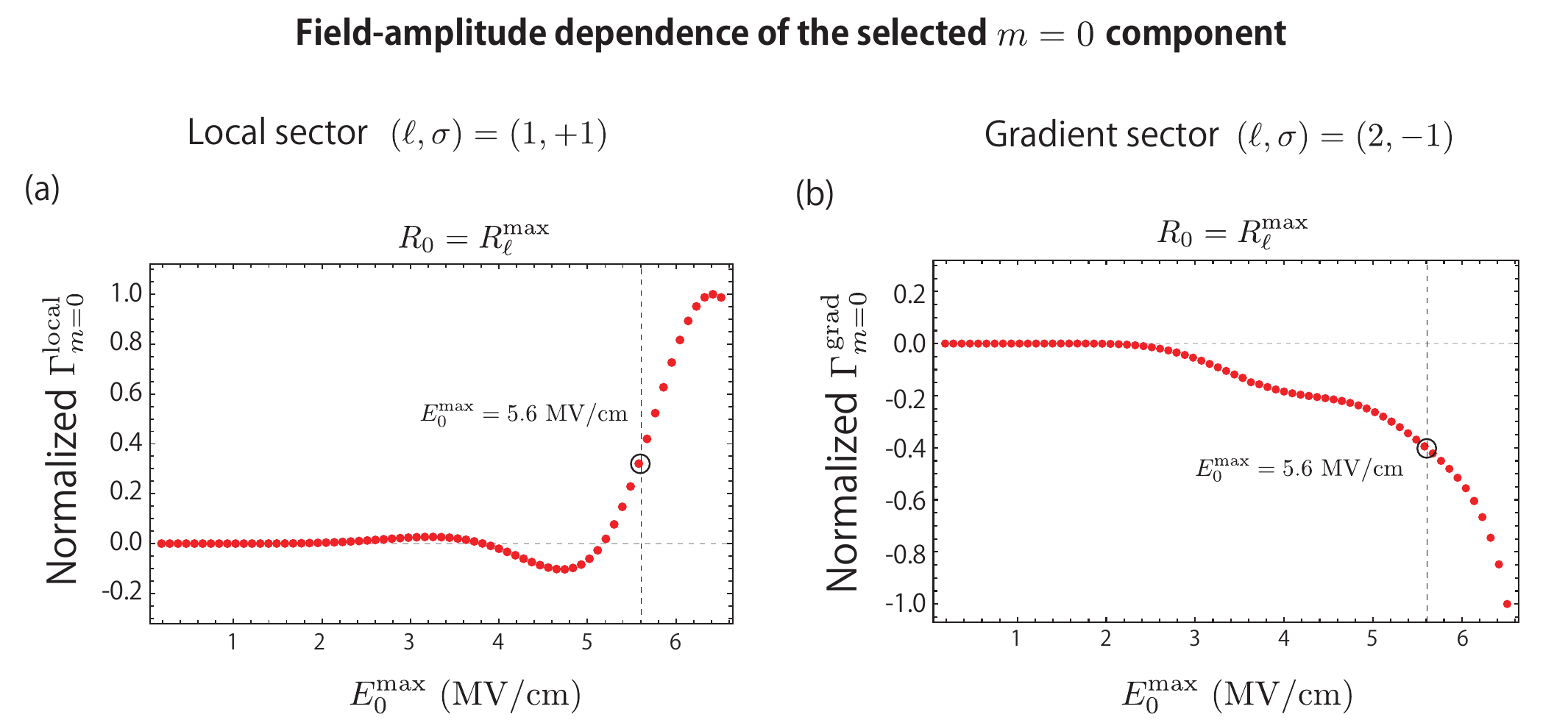}
    \caption{
    Field-amplitude dependence of the selected \(m=0\)
    zero-frequency components at \(R_0=R_\ell^{\max}\).
    (a) Local-sector branch for
    \((\ell,\sigma)=(1,+1)\).
    (b) Gradient-sector branch for
    \((\ell,\sigma)=(2,-1)\).
    Red symbols show the numerically extracted normalized signed
    coefficient \(\Gamma_{m=0}\).
    The open circles indicate
    \(E_0^{\max}=5.6~\mathrm{MV/cm}\), used for the radial-profile and
    magnetic-field analyses in the main text.
    The coefficient varies nonlinearly with the driving strength,
    whereas the selected winding order remains \(m=0\).
    At \(R_0=R_\ell^{\max}\), the radial derivative of the \(p=0\)
    Laguerre--Gaussian envelope vanishes, so that panel (b)
    characterizes the field-amplitude dependence of the
    azimuthal-gradient contribution.
    The two panels are normalized independently.
    }
    \label{fig:field_amplitude_dependence_m0}
\end{figure}

Figure~\ref{fig:field_amplitude_dependence_m0} shows the normalized
signed coefficient $\Gamma_{m=0}$ as a function of the peak
electric-field amplitude $E_0^{\max}$.
We consider the two representative $m=0$ branches discussed in the
main text: the local-sector branch
$(\ell,\sigma)=(1,+1)$ shown in Fig.~3(b) and the gradient-sector
branch $(\ell,\sigma)=(2,-1)$ shown in Fig.~3(c).
In both sectors, $\Gamma_{m=0}$ varies nonlinearly with the driving
strength.
The selected winding order remains $m=0$, whereas the magnitude and
sign of its coefficient are determined by the driven electronic
dynamics.
Accordingly, the present scan characterizes the field-amplitude
dependence of the finite-pulse zero-frequency response without
assuming a particular perturbative scaling.

In the local sector,
Fig.~\ref{fig:field_amplitude_dependence_m0}(a) shows that the selected
$m=0$ coefficient changes sign and varies nonmonotonically with
increasing $E_0^{\max}$.
This behavior provides a qualitative reference for interpreting the
sign-changing radial profile in Fig.~3(b) of the main text.
At fixed incident $E_0^{\max}$, changing the observation radius changes
the locally sampled field amplitude through the LG beam envelope.
A radial scan therefore probes different parts of the nonlinear
field-amplitude dependence shown in
Fig.~\ref{fig:field_amplitude_dependence_m0}(a).
Consequently, $\Gamma_{m=0}(R_0)$ can change sign even though the optical
intensity profile remains nonnegative.
The local-sector radial profile in Fig.~3(b) is therefore not expected
to follow the intensity profile directly; rather, its magnitude and
sign reflect the nonlinear electronic response to the locally sampled
field amplitude.

The gradient-sector radial profile shown in Fig.~3(c) can be
decomposed into azimuthal- and radial-gradient contributions.
At $R_0=R_\ell^{\max}$, the radial derivative of the $p=0$
LG envelope vanishes, and the radial-gradient contribution is therefore
absent.
Figure~\ref{fig:field_amplitude_dependence_m0}(b) consequently
characterizes the nonlinear field-amplitude dependence of the
azimuthal-gradient contribution.

As the observation radius is varied, the locally sampled field
amplitude changes according to the LG envelope.
Using the field-amplitude dependence in
Fig.~\ref{fig:field_amplitude_dependence_m0}(b) as a reference, this
variation qualitatively reproduces the calculated radial profile of
the azimuthal-gradient contribution in Fig.~3(c).
This agreement indicates that the radial dependence of the
azimuthal-gradient contribution is governed primarily by the nonlinear
response to the locally sampled field amplitude.
In particular, the predominantly negative response over the relevant
field-amplitude range accounts for the broad negative region of the
azimuthal-gradient profile in Fig.~3(c).

The radial-gradient contribution, by contrast, cannot be inferred from
the local field amplitude alone, because it also depends on the radial
derivative of the LG envelope.
Away from $R_\ell^{\max}$, this radial derivative becomes nonzero and
generates a contribution with its own radial structure and sign
changes.
This contribution adds to the azimuthal-gradient component and can
locally reinforce or cancel it, thereby shaping the total
gradient-sector profile in Fig.~3(c).
The physical gradient-sector circulating current is determined by the
sum of the azimuthal- and radial-gradient contributions, and it is this
total radial profile that is used to reconstruct the axial magnetic
field in the main text.

These results clarify the connection between the field-amplitude
dependence and the radial profiles in Figs.~3(b) and 3(c).
In the local sector, the radial variation primarily reflects the
nonlinear response to the locally sampled field amplitude.
In the gradient sector, the same mapping qualitatively captures the
azimuthal-gradient contribution, whereas the radial-gradient
contribution introduces an additional spatial dependence that cannot
be inferred from the local field amplitude alone.
Thus, the magnitude and sign of $\Gamma_{m=0}(R_0)$ can vary with radius
without changing the sector-selected winding order.
We verify the robustness of this winding selection explicitly in the
next subsection by varying $R_0/W_0$.

\subsubsection{Observation-radius dependence of the selected winding spectra}
\label{sec:radial_position_dependence}

Here, we examine how the selected $m=0$ winding spectra depend on the
observation radius $R_0/W_0$.
In Figs.~\ref{fig:current_channel_decomposition} and
\ref{fig:helicity_resolved_channel_decomposition}, the winding spectra
were evaluated at the intensity-maximum radius
$R_0=R_\ell^{\max}$ defined in
Eq.~\eqref{eq:LG_intensity_maximum_radius}.
At this radius, the radial derivative of the $p=0$ LG envelope
vanishes, suppressing the radial-gradient contribution and providing a
clean diagnostic of the angular structure underlying the winding
selection rules.
It is therefore important to confirm that the selected $m=0$ component
is not an artifact of choosing this particular observation radius.

Away from $R_\ell^{\max}$, radial-gradient contributions become finite.
To interpret the resulting additional winding components, we decompose
the spatial gradient of each Cartesian component $A_j$ of the vector
potential as
\begin{equation}
\nabla A_j
=
\mathbf e_R \partial_R A_j
+
\mathbf e_\phi \frac{1}{R}\partial_\phi A_j,
\qquad j=x,y,
\label{eq:radial_azimuthal_gradient_decomposition}
\end{equation}
where $\mathbf e_R=(\cos\phi,\sin\phi)$ and
$\mathbf e_\phi=(-\sin\phi,\cos\phi)$.
We refer to the first term in
Eq.~\eqref{eq:radial_azimuthal_gradient_decomposition} as the
radial-gradient contribution and to the second term as the
azimuthal-gradient contribution.
The azimuthal-gradient contribution contains the angular structure
associated with the leading winding selection, whereas the
radial-gradient contribution mainly modifies the amplitude and can
generate weak auxiliary winding components when the observation radius
is displaced from $R_\ell^{\max}$.

\begin{figure}[t]
\centering
\includegraphics[width=0.95\linewidth]{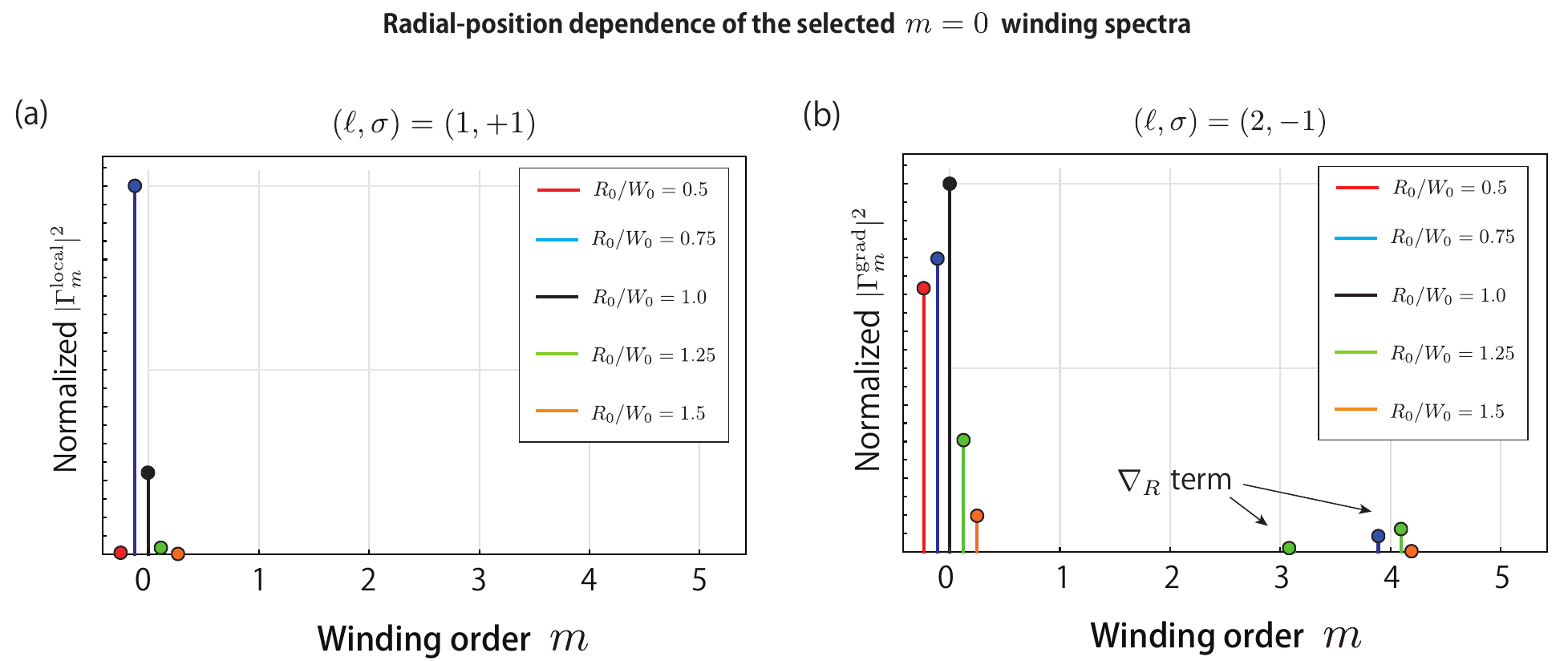}
\caption{
Observation-radius dependence of the selected $m=0$ winding spectra.
The horizontal axis denotes the winding order $m$, and different colors correspond to different observation radii $R_0/W_0$.
Panel (a) shows the local-sector response for $(\ell,\sigma)=(1,+1)$, while panel (b) shows the gradient-sector response for $(\ell,\sigma)=(2,-1)$.
Even when the observation radius is varied, the dominant selected winding order remains $m=0$, whereas its magnitude changes depending on $R_0/W_0$.
The higher-order winding components observed in panel (b), indicated by the $\nabla_R$ label, are auxiliary contributions associated with the radial-gradient part of the structured-light field and remain much smaller than the selected $m=0$ component.
}
\label{fig:radial_position_dependence}
\end{figure}

Figure~\ref{fig:radial_position_dependence}(a) shows the result for the
local sector under the condition $(\ell,\sigma)=(1,+1)$.
For this condition, the local-sector selection rule predicts the
$m=0$ component.
As shown in the figure, the dominant winding component remains at
$m=0$ even when the observation radius $R_0/W_0$ is varied.
At the same time, its magnitude depends strongly on the observation
radius, reflecting the nonlinear response to the locally sampled field
amplitude discussed in the preceding subsection.

Figure~\ref{fig:radial_position_dependence}(b) shows the result for the
gradient sector under the condition $(\ell,\sigma)=(2,-1)$.
For this condition, the gradient-sector selection rule predicts the
$m=0$ component.
As in the local-sector case, the dominant winding component remains at
$m=0$ over the range of observation radii considered.
Weak higher-order winding components also appear at $m=3$ and $m=4$
when $R_0$ is displaced from $R_\ell^{\max}$.
These components are associated with radial-gradient contributions
that become finite away from the intensity-maximum radius and remain
much smaller than the selected $m=0$ component.
Thus, even in the gradient sector, the dominant winding response
remains concentrated at $m=0$, in agreement with the selection rule.

These results demonstrate that the emergence of the $m=0$ winding
channel is not an artifact of choosing a particular observation
radius.
The selected winding order is fixed by the local- or gradient-sector
selection rule, whereas its magnitude and the weak auxiliary winding
components depend on the observation radius through the radial field
profile and, in the gradient sector, the additional radial-gradient
contribution.

\subsubsection{Intensity-gradient-like radial dependence of the B2-type radial current}
\label{sec:intensity_gradient_like_B2_channel}

Here, we compare the radial dependence of the $B2$-channel radial current
with a reference profile derived from the radial gradient of the optical
intensity.
As discussed above, the $B2$ channel originates from an
$\bm{A}\nabla\bm{A}$-type coupling.
For the fixed-polarization Gaussian beam considered below, the
underlying B2 interaction contains the radial field-gradient structure
$A\partial_R A=(1/2)\partial_R(A^2)$, which motivates a comparison
with an intensity-gradient reference profile.

\begin{figure}[t!]
\centering
\includegraphics[width=0.55\linewidth]{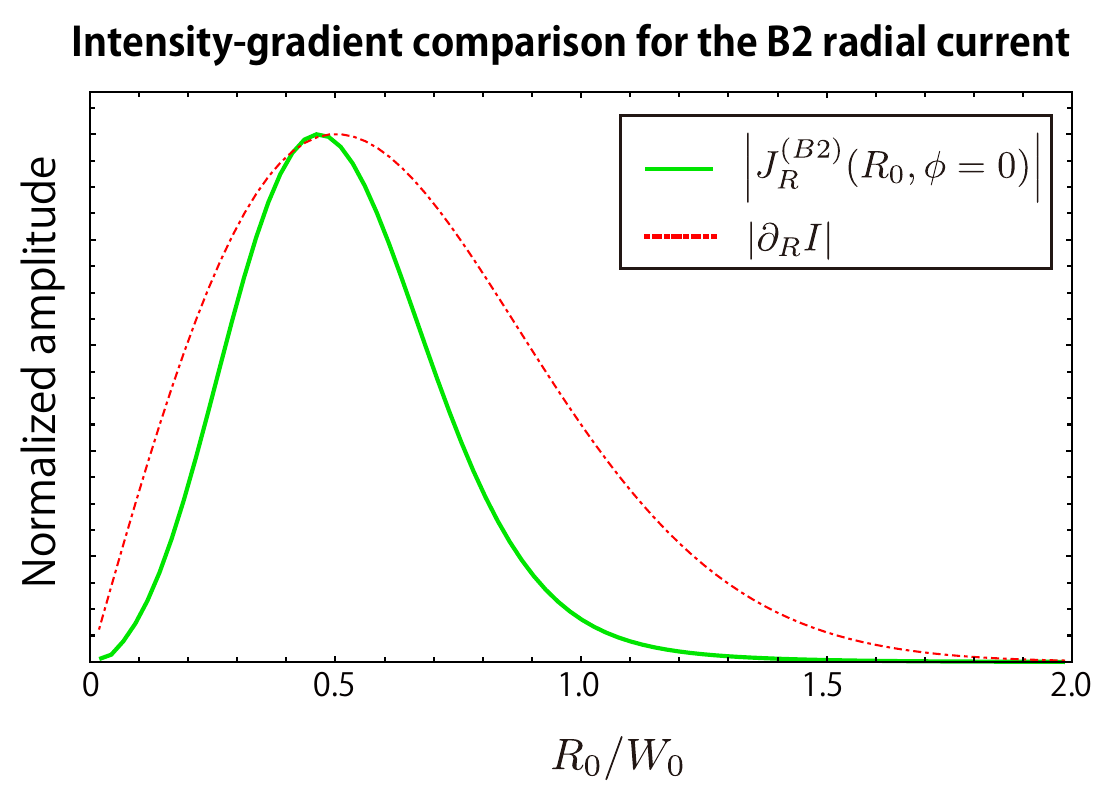}
\caption{
Intensity-gradient-like radial dependence of the $B2$-type radial current.
The green curve shows the radial dependence of the radial-current
component $|J_R^{(B2)}(R_0,\phi=0)|$ along the polarization axis for a
linearly polarized Gaussian beam with $\ell=0$.
It is compared with the reference intensity-gradient profile
$|\partial_R I(R)|$.
Both curves are independently normalized to their respective maxima
for comparison.
The reference profile is proportional to
$R\exp[-2(R/W_0)^2]$.
The similar radial dependence demonstrates an intensity-gradient-like
structure of the $B2$-channel radial response, without implying that the
microscopic current is identical to a ponderomotive-force response.
}
\label{fig:intensity_gradient_like_B2_channel}
\end{figure}

To examine this point, we consider a linearly polarized Gaussian beam,
corresponding to $\ell=0$, and compare the magnitude of the $B2$
radial-current component along the polarization axis,
$|J_R^{(B2)}(R_0,\phi=0)|$, with an intensity-gradient reference profile.
Denoting the optical intensity profile by $I(R)$, we use the magnitude
of its radial gradient, $|\partial_R I(R)|$, as the reference.
For a Gaussian beam, this profile takes the form
\begin{equation*}
|\partial_R I(R)|
\propto
R
\exp\left[
-2\left(\frac{R}{W_0}\right)^2
\right].
\label{eq:intensity_gradient_gaussian}
\end{equation*}
This profile vanishes at the beam center and reaches a maximum at a
finite radius.
The numerical result in
Fig.~\ref{fig:intensity_gradient_like_B2_channel} shows that
$|J_R^{(B2)}(R_0,\phi=0)|$ exhibits a radial profile similar to the
intensity-gradient reference: it is small near the beam center and
reaches a maximum at a finite value of $R_0/W_0$.
This qualitative agreement is consistent with an
intensity-gradient-like radial structure associated with the
$\bm{A}\nabla\bm{A}$-type character of the $B2$ channel.
Here, $J_R^{(B2)}(R_0,\phi=0)$ is a direction-resolved radial-current
component and should not be interpreted as an $m=0$ azimuthal-current
winding amplitude.

A force proportional to the optical-intensity gradient is characteristic
of a ponderomotive response
~\cite{Kibble1966MutualRefraction,Kruer1988LaserPlasma}.
The similarity of the radial profiles may therefore suggest a connection
to such a picture.
However, the $B2$ radial current should not be identified directly with
the ponderomotive force itself.
The quantity $|\partial_R I(R)|$ is used here only as a reference
profile for comparing the radial dependence.
The actual $J_R^{(B2)}$ is a microscopic coherent current response
that includes the effects of the band structure, nonlinear time
evolution, interband coherence, current-channel decomposition, and
projection onto the radial direction.
It is therefore not expected to coincide quantitatively with the
intensity-gradient reference.

Thus, the $B2$ channel is not equivalent to a ponderomotive-force
response, although its radial-current profile exhibits a clear
intensity-gradient-like structure.
This behavior is consistent with the role of the $B2$-type
$\bm{A}\nabla\bm{A}$ channel as a gradient-sector contribution to the
dc current response.

\subsection{Numerical validation of
\texorpdfstring{finite-$\bm{q}_{\parallel}$}{finite-q-parallel} effects}
\label{sec:finite_q_numerical_validation}

In this section, we quantitatively assess the validity of the
momentum-diagonal approximation used in the main calculations.
We compare the finite-$q_{\parallel}$ formulation with the
momentum-diagonal $\bm q_{\parallel}=0$ reference, focusing on the
selected local- and gradient-sector winding branches under circular
polarization.
Here, $\bm{q}_{\parallel}$ denotes the in-plane optical wave vector
associated with the azimuthal phase gradient of the LG beam.
As derived in Secs.~\ref{sec:finite_q_hamiltonian} and
\ref{sec:finite_q_bloch_projection}, finite
$\bm{q}_{\parallel}$ introduces the explicit momentum transfer
$\bm{k}\leftrightarrow\bm{k}\pm\bm{q}_{\parallel}$ through the
$A1$ and $A2$ interactions, whereas the
$\bm{q}_{\parallel}=0$ formulation is momentum diagonal.
The corresponding current operators retain the momentum-transfer
structure associated with finite $\bm{q}_{\parallel}$.
The comparison therefore quantifies the correction to the selected
winding responses arising from the explicit in-plane optical momentum.
We consider two circular-polarization configurations chosen to probe
the characteristic branch selection.
For $(\ell,\sigma)=(1,+1)$, the selected local and gradient branches
are $m=0$ and $m=3$, respectively, whereas for
$(\ell,\sigma)=(2,-1)$ they are $m=3$ and $m=0$.
Thus, the two configurations probe the selected $m=0$ response in both
the local and gradient sectors.

At the beam waist, the spatial phase of the LG beam defined in
Eq.~\eqref{eq:LG_vector_potential} is
$\Phi(R,\phi)=\ell\phi$.
The local in-plane optical wave vector defined in
Eq.~\eqref{eq:local_q_parallel} is therefore
\begin{equation*}
    \bm{q}_{\parallel}
    =
    \bm{\nabla}_{\parallel}\Phi
    =
    \bm{e}_R
    \frac{\partial\Phi}{\partial R}
    +
    \frac{\bm{e}_{\phi}}{R}
    \frac{\partial\Phi}{\partial\phi}
    =
    \frac{\ell}{R}\bm{e}_{\phi}.
\end{equation*}
On the observation circle $R=R_0$, this becomes
\begin{equation*}
    \bm{q}_{\parallel}(\phi)
    =
    \frac{\ell}{R_0}\bm{e}_{\phi}
    =
    \frac{\ell}{R_0}
    \left(
        -\sin\phi,\,
        \cos\phi
    \right).
\end{equation*}
For the $p=0$ LG beam, we choose the observation radius at the radial
intensity maximum given in
Eq.~\eqref{eq:LG_intensity_maximum_radius},
\begin{equation*}
    R_0
    =
    R_{\ell}^{\max}
    =
    W_0
    \sqrt{\frac{|\ell|}{2}}.
\end{equation*}
Since $\bm{X}_0=R_0\bm{e}_R$ on this circle,
$\bm{q}_{\parallel}(\bm{X}_0)\cdot\bm{X}_0=0$.
Consequently, the phase factor
$e^{-is\bm{q}_{\parallel}\cdot\bm{X}_0}$ appearing in
Eqs.~\eqref{eq:finite_q_HA1} and \eqref{eq:finite_q_HA2}
is unity for this geometry.
Thus, the finite $\bm{q}_{\parallel}$ used in the validation is fixed
by the LG field and the observation radius rather than introduced as
an independent parameter.
At each $\bm X_0$, the finite-$\bm q_{\parallel}$ Hamiltonian couples
Bloch states whose wave vectors differ by
$\pm\bm q_{\parallel}$ through the $A1$ and $A2$ interactions.
Repeated transitions generated by these couplings are included
simultaneously in the numerical time evolution.
For evaluating the current, it is sufficient to resolve the
one-particle density matrix according to the momentum transfer
selected by each current operator.

To evaluate the finite-$\bm q_{\parallel}$ current, we first introduce
the momentum-transfer-resolved one-particle density-matrix elements.
We define the intersublattice coherence associated with a momentum
transfer $\Delta\bm k$ as
\begin{equation*}
    \mathcal C_{\bm k}
    (\Delta\bm k;\bm X_0,t)
    =
    \left\langle
        \hat a_{\bm k+\Delta\bm k}^{\dagger}
        \hat b_{\bm k}
    \right\rangle_t .
\end{equation*}
Here, the expectation value denotes the corresponding one-particle
density-matrix element obtained by propagating all initially occupied
valence-band states under the full finite-$\bm q_{\parallel}$
Hamiltonian.
Using the microscopic coefficient functions introduced in
Secs.~\ref{sec:A_dot_p_interaction} and \ref{sec:A2_interaction},
we define
\begin{align*}
    D_i(\Delta\bm k;\bm X_0,t)
    &=
    \sum_{\bm k}
    d_{\bm k}^{\,i}\,
    \mathcal C_{\bm k}
    (\Delta\bm k;\bm X_0,t),
    \\
    D_{\lambda i}(\Delta\bm k;\bm X_0,t)
    &=
    \sum_{\bm k}
    d_{\bm k}^{\,\lambda i}\,
    \mathcal C_{\bm k}
    (\Delta\bm k;\bm X_0,t),
    \\
    \mathcal S_{\lambda}(\Delta\bm k;\bm X_0,t)
    &=
    \sum_{\bm k}
    S_{\lambda}(\bm k)\,
    \mathcal C_{\bm k}
    (\Delta\bm k;\bm X_0,t).
\end{align*}
For the quadratic density channel, we similarly define
\begin{equation*}
    N(\Delta\bm k;\bm X_0,t)
    =
    \sum_{\bm k}
    \left[
        \left\langle
            \hat a_{\bm k+\Delta\bm k}^{\dagger}
            \hat a_{\bm k}
        \right\rangle_t
        +
        \left\langle
            \hat b_{\bm k+\Delta\bm k}^{\dagger}
            \hat b_{\bm k}
        \right\rangle_t
    \right].
\end{equation*}
The momentum sums are evaluated using the same Brillouin-zone
normalization as in the momentum-diagonal calculation.

The finite-$\bm q_{\parallel}$ current is evaluated from the same
source-current principle as in
Eq.~\eqref{eq:current_density_hamiltonian}.
The two Fourier amplitudes
$\mathcal{A}_i^{(+)}$ and $\mathcal{A}_i^{(-)}$
are treated as independent variables when taking the generalized
derivatives, and the real-field condition
$\mathcal{A}_i^{(-)}=[\mathcal{A}_i^{(+)}]^*$
is imposed afterward.
Thus,
\begin{equation*}
    J_i(\bm X_0,t)
    =
    -c
    \sum_{s=\pm1}
    \left[
        \frac{\partial H_I}
        {\partial \mathcal{A}_i^{(s)}}
        -
        \sum_{\lambda=x,y}
        \partial_{X_\lambda}
        \frac{\partial H_I}
        {\partial
        \left(
            \partial_{X_\lambda}\mathcal{A}_i^{(s)}
        \right)}
    \right].
\end{equation*}
The sum over the two conjugate Fourier components reconstructs the
physical real current at $\bm X_0$.
For the $A1$ channel, the current operator carries the momentum
transfers $\Delta\bm k=\pm\bm q_{\parallel}$.
Its expectation value is therefore
\begin{equation*}
    J_i^{A1}(\bm X_0,t)
    =
    -2c\,
    \mathrm{Re}
    \left[
        D_i(+\bm q_{\parallel};\bm X_0,t)
        +
        D_i(-\bm q_{\parallel};\bm X_0,t)
    \right].
\end{equation*}
For the $B1$ channel, the corresponding contribution to the
circularly polarized Hamiltonian is momentum diagonal, as shown in
Sec.~\ref{sec:finite_q_B_projection}.
The current operator is nevertheless obtained by differentiating the
general quadratic interaction before imposing circular polarization.
It therefore contains the momentum transfers
$\Delta\bm k=0,\pm2\bm q_{\parallel}$,
\begin{align*}
    J_i^{B1}(\bm X_0,t)
    =
    -\frac{e^2}{m_0c}\,
    \mathrm{Re}
    \Big[
        &A_i(\bm X_0,t)
        N(\bm 0;\bm X_0,t)
\\
        &+
        \mathcal{A}_i^{(+)}(\bm X_0,t)
        N(+2\bm q_{\parallel};\bm X_0,t)
\\
        &+
        \mathcal{A}_i^{(-)}(\bm X_0,t)
        N(-2\bm q_{\parallel};\bm X_0,t)
    \Big],
\end{align*}
where
\begin{equation*}
    A_i(\bm X_0,t)
    =
    \mathcal{A}_i^{(+)}(\bm X_0,t)
    +
    \mathcal{A}_i^{(-)}(\bm X_0,t).
\end{equation*}
For the $A2$ channel, the same
$\Delta\bm k=\pm\bm q_{\parallel}$ transfers appear.
Defining
\begin{equation*}
    \widetilde D_{\lambda i}(\bm X_0,t)
    =
    D_{\lambda i}
    (+\bm q_{\parallel};\bm X_0,t)
    +
    D_{\lambda i}
    (-\bm q_{\parallel};\bm X_0,t),
\end{equation*}
we obtain
\begin{equation*}
    J_i^{A2}(\bm X_0,t)
    =
    2c\,
    \mathrm{Re}
    \sum_{\lambda=x,y}
    \partial_{X_\lambda}
    \widetilde D_{\lambda i}(\bm X_0,t).
\end{equation*}
The $B2$ current is obtained from the general first-gradient quadratic
interaction before imposing the circular-polarization condition.
As in the momentum-diagonal derivation, the terms proportional to
$\partial_{X_\lambda}\mathcal{A}_i^{(s)}$
cancel between the two terms of the generalized current definition.
The remaining contribution contains
$\Delta\bm k=0,\pm2\bm q_{\parallel}$,
\begin{align*}
    J_i^{B2}(\bm X_0,t)
    =
    2c\,
    \mathrm{Re}
    \sum_{\lambda=x,y}
    \Big[
        &A_i(\bm X_0,t)
        \partial_{X_\lambda}
        \mathcal S_{\lambda}(\bm 0;\bm X_0,t)
\\
        &+
        \mathcal{A}_i^{(+)}(\bm X_0,t)
        \partial_{X_\lambda}
        \mathcal S_{\lambda}
        (+2\bm q_{\parallel};\bm X_0,t)
\\
        &+
        \mathcal{A}_i^{(-)}(\bm X_0,t)
        \partial_{X_\lambda}
        \mathcal S_{\lambda}
        (-2\bm q_{\parallel};\bm X_0,t)
    \Big].
\end{align*}
The momentum-transfer structure of the four current operators is therefore
\begin{equation}
    A1,\ A2:
    \qquad
    \Delta\bm k=\pm\bm q_{\parallel},
    \qquad
    B1,\ B2:
    \qquad
    \Delta\bm k=0,\pm2\bm q_{\parallel}.
    \label{eq:finite_q_current_transfer_summary}
\end{equation}
The $B1$ and $B2$ contributions to the circularly polarized
Hamiltonian remain momentum diagonal during the time evolution.
The $\pm2\bm q_{\parallel}$ components in
Eq.~\eqref{eq:finite_q_current_transfer_summary} arise only in the
corresponding current operators and therefore do not represent
additional finite-momentum couplings in the Hamiltonian. 
For the numerical comparison, we evaluate the currents at
$R_0=R_\ell^{\max}$, where the radial derivative of the $p=0$ LG
envelope vanishes. The validation therefore focuses on the
azimuthal phase-gradient contribution along the observation circle.
The local and gradient currents used in the comparison are
\begin{equation*}
    \bm J_{\mathrm{loc}}
    =
    \bm J^{A1}
    +
    \bm J^{B1},
    \qquad
    \bm J_{\mathrm{grad}}
    =
    \bm J^{A2}
    +
    \bm J^{B2}.
\end{equation*}
The resulting Cartesian currents are projected onto the azimuthal
direction, and their zero-frequency winding components are evaluated
using the same temporal and azimuthal Fourier definitions as in
Sec.~\ref{sec:winding_spectrum_definition}.

We take the momentum-diagonal calculation used in the main
calculations as the $\bm q_{\parallel}=0$ reference, with all other
parameters unchanged.
For each selected branch, we quantify the finite-$q_{\parallel}$
correction by defining
\begin{equation*}
    \mathcal R_m
    =
    \frac{
        \left|
            \Gamma_m^{\mathrm{finite}\text{-}q}
        \right|
    }{
        \left|
            \Gamma_m^{q=0}
        \right|
    },
    \qquad
    \Delta_m
    =
    \left|
        \mathcal R_m-1
    \right|
    \times100\%.
\end{equation*}
The resulting values are summarized in
Table~\ref{tab:finite_q_validation}.
\begin{table}[t]
    \centering
    \caption{
Comparison of the selected zero-frequency winding amplitudes obtained
from the finite-$\bm q_{\parallel}$ calculation and the
momentum-diagonal $\bm q_{\parallel}=0$ reference.
All other parameters are kept unchanged.
The values of $\Delta_m$ are evaluated from the unrounded ratios.
}
    \label{tab:finite_q_validation}
    \begin{tabular}{ccccc}
        \hline\hline
        $(\ell,\sigma)$
        & Sector
        & $m$
        & $\mathcal R_m$
        & $\Delta_m$ \\
        \hline
        $(1,+1)$
        & Local
        & $0$
        & $0.999992$
        & $0.000791\%$ \\
        $(1,+1)$
        & Gradient
        & $3$
        & $1.000061$
        & $0.00606\%$ \\
        $(2,-1)$
        & Local
        & $3$
        & $1.000006$
        & $0.000579\%$ \\
        $(2,-1)$
        & Gradient
        & $0$
        & $1.000184$
        & $0.0184\%$ \\
        \hline\hline
    \end{tabular}
\end{table}
For all four selected branches,
$\mathcal R_m$ remains very close to unity.
The largest relative change is below $0.02\%$ and occurs for the
gradient-sector $m=0$ branch at $(\ell,\sigma)=(2,-1)$.
The winding orders selected in the finite-$q_{\parallel}$ calculation
are unchanged from those in the $\bm q_{\parallel}=0$ reference.
Thus, including the explicit finite in-plane optical momentum in both
the Hamiltonian and the associated current operators produces only
negligible corrections to the selected winding amplitudes for the
cases examined here, while leaving the selected winding orders
unchanged.
These results support the momentum-diagonal approximation used in the
main calculations within the parameter regime examined here.

\FloatBarrier
\subsection{Magnetic field reconstructed from the dc current distribution}
\label{sec:magnetic_field_reconstruction}

Here, we relate the dc winding current to the axial magnetic field.
As discussed above, the $m=0$ winding channel is selected under
specific conditions in both the local and gradient sectors.
Because the $m=0$ component represents an azimuthally uniform
circulating current, it contributes directly to the on-axis magnetic
field $B_z(z)$ normal to the graphene plane.

Before applying the Biot--Savart law, we calibrate the physical
current scale using an independent weak-field calculation.
For a spatially uniform field in the weak-field limit, the calculated
linear optical response at the carrier frequency is fitted to the
universal optical conductivity of graphene
~\cite{Nair2008FineStructureGraphene,
Mak2008OpticalConductivityGraphene,
Kuzmenko2008UniversalOpticalConductance,
Stauber2008OpticalConductivityGrapheneVisible},
\begin{equation*}
    \sigma_0=\frac{e^2}{4\hbar}.
\end{equation*}
The current scale determined from this weak-field fit is then kept
fixed in the high-intensity structured-light calculations.
The nonlinear dc current and the corresponding winding coefficient
$\Gamma_{m=0}(R)$ are extracted from the resulting time-dependent current
without any further fitting. Specifically, the dc current is obtained
from the window-normalized $q=0$ component defined in Sec.~\ref{sec:winding_spectrum_definition}, so that
it has the same physical dimension as the time-domain current.
The same window-normalized temporal Fourier convention is used for the
weak-field calibration at the carrier frequency and for the nonlinear
dc-current extraction.

We next relate the $m=0$ winding coefficient to the azimuthally
averaged dc sheet current.
Let $\Gamma_{m=0}(R)$ denote the signed $m=0$ coefficient constructed
from this window-normalized dc current at radius $R$. The corresponding averaged azimuthal sheet current is
\begin{equation}
    J_{\theta}^{\mathrm{dc}}(R)
    =
    \frac{\Gamma_{m=0}(R)}{2\pi R}.
    \label{eq:average_azimuthal_current}
\end{equation}
The radial profile $\Gamma_{m=0}(R)$ is obtained by evaluating the
winding coefficient at multiple observation radii and is used to
reconstruct the axial magnetic field.
The azimuthally averaged sheet current can then be regarded as a
continuum of circular current loops distributed in the radial
direction. Using the Biot--Savart law, the axial magnetic field at
height $z$ above the graphene plane is
\begin{equation}
    B_z(z)
    =
    \frac{\mu_0}{2}
    \int \mathrm{d}R\,
    \frac{
        R^2 J_{\theta}^{\mathrm{dc}}(R)
    }{
        \left(z^2+R^2\right)^{3/2}
    }.
    \label{eq:Biot_Savart_Jtheta}
\end{equation}
Substituting Eq.~\eqref{eq:average_azimuthal_current} into
Eq.~\eqref{eq:Biot_Savart_Jtheta} gives
\begin{equation}
    B_z(z)
    =
    \frac{\mu_0}{4\pi}
    \int \mathrm{d}R\,
    \frac{
        R\,\Gamma_{m=0}(R)
    }{
        \left(z^2+R^2\right)^{3/2}
    }.
    \label{eq:Biot_Savart_Gamma}
\end{equation}
Thus, the axial magnetic field is determined by the radial profile of
the calibrated $m=0$ winding coefficient $\Gamma_{m=0}(R)$, with its
sign set by the corresponding weighted radial integral.
In the actual calculation, we evaluate $\Gamma_{m=0}(R)$ at multiple
observation radii and numerically perform the radial integral in
Eq.~\eqref{eq:Biot_Savart_Gamma}.
Winding components with nonzero $m$ have zero azimuthal average and therefore do not contribute to the on-axis axial field \(B_z\).
Consequently, the axial magnetic field on the symmetry axis is
determined by the $m=0$ component.

This procedure connects the sector-resolved winding spectrum to an
axial magnetic readout of the $m=0$ dc circulating current.
For the parameters considered here, the reconstructed magnetic field
lies in the $\mathrm{nT}$--$\mu\mathrm{T}$ range.
The estimated magnitude corresponds to the ideal coherent limit;
dephasing, dissipation, and finite-temperature effects may reduce the
absolute field strength, whereas the symmetry-based winding selection
rules are expected to be more robust.

\bibliography{references}